\documentclass[aps,prd,twocolumn,amsmath,amssymb,showpacs,footnote,reprint]{revtex4-1}

\usepackage{graphicx}
\usepackage{dcolumn}
\usepackage{bm}

\usepackage{epstopdf} 

\usepackage[colorlinks,hyperindex]{hyperref}  
\hypersetup
{
  colorlinks,%
  citecolor=blue,%
  linkcolor=blue,%
  urlcolor=blue,%
}

\usepackage{tensor}

\begin{document}

\title{
Stueckelberg massive electromagnetism in curved spacetime: \\
Hadamard renormalization of the stress-energy tensor and the Casimir effect
}

\author{Andrei Belokogne}
\email{belokogne.andrei@gmail.com}

\affiliation{Equipe Physique
Th\'eorique - Projet COMPA, \\ SPE, UMR 6134 du CNRS
et de l'Universit\'e de Corse,\\
Universit\'e de Corse, BP 52, F-20250 Corte,
France}

\author{Antoine Folacci}
\email{folacci@univ-corse.fr}

\affiliation{Equipe Physique
Th\'eorique - Projet COMPA, \\ SPE, UMR 6134 du CNRS
et de l'Universit\'e de Corse,\\
Universit\'e de Corse, BP 52, F-20250 Corte,
France}

\date{\today}

\begin{abstract}

We discuss Stueckelberg massive electromagnetism on an arbitrary four-dimensional curved spacetime (gauge invariance of the classical theory and covariant quantization; wave equations for the massive spin-1 field $A_\mu$, for the auxiliary Stueckelberg scalar field $\Phi$ and for the ghost fields $C$ and $C^\ast$; Ward identities; Hadamard representation of the various Feynman propagators and covariant Taylor series expansions of the corresponding coefficients). This permits us to construct, for a Hadamard quantum state, the expectation value of the renormalized stress-energy tensor associated with the Stueckelberg theory. We provide two alternative but equivalent expressions for this result. The first one is obtained by removing the contribution of the ``Stueckelberg ghost'' $\Phi$ and only involves state-dependent and geometrical quantities associated with the massive vector field $A_\mu$. The other one involves contributions coming from both the massive vector field and the auxiliary Stueckelberg scalar field, and it has been constructed in such a way that, in the zero-mass limit, the massive vector field contribution reduces smoothly to the result obtained from Maxwell's theory. As an application of our results, we consider the Casimir effect outside a perfectly conducting medium with a plane boundary. We discuss the results obtained using Stueckelberg but also de Broglie-Proca electromagnetism and we consider the zero-mass limit of the vacuum energy in both theories. We finally compare the de Broglie-Proca and Stueckelberg formalisms and highlight the advantages of the Stueckelberg point of view, even if, in our opinion, the de Broglie-Proca and Stueckelberg approaches of massive electromagnetism are two faces of the same field theory.

\end{abstract}


\maketitle


\section{Introduction}
\label{Sec.I}

It is generally assumed that the electromagnetic interaction is mediated by a massless photon. This seems largely justified (i) by the countless theoretical and practical successes of Maxwell's theory of electromagnetism and of its extension in the framework of quantum field theory as well as (ii) by the stringent upper limits on the photon mass (see p.~559 of Ref.~\cite{Agashe:2014kda} and references therein) which have been obtained by various terrestrial and extraterrestrial experiments (currently, one of the most reliable results provides for the photon mass $m$ the limit $m \le 10^{-18}~\mathrm{eV} \approx 2 \times 10^{-54}~\mathrm{kg}$ \cite{Ryutov:2007zz}).

Despite this, physicists are seriously considering the possibility of a massive but, of course, ultralight photon and are very interested by the associated non-Maxwellian theories of electromagnetism (for recent reviews on the subject, see Refs.~\cite{Tu:2005ge,Goldhaber:2008xy}). Indeed, the incredibly small value mentioned above does not necessarily imply that the photon mass is exactly zero, and from a theoretical point of view, massive electromagnetism can be rather easily included in the Standard Model of particle physics. Moreover, in order to test the masslessness of the photon or, more precisely, to impose experimental constraints on its mass, it is necessary to have a good understanding of the various massive non-Maxwellian theories. Among these, two theories are particularly important, and we intend to discuss them at more length in our article:
\begin{itemize}

  \item[(i)] The most popular one, which is the simplest generalization of Maxwell's electromagnetism, is mainly due to de Broglie (note that the idea of an ultralight massive photon is already present in de Broglie's doctoral thesis \cite{deBroglie1924,deBroglie:1922zz} and has been developed by him in modern terms in a series of works \cite{deBroglie1934,deBroglie1936,deBroglie1940} where he has considered the theory from a Lagrangian point of view and has explicitly shown the modifications induced by the photon mass for Maxwell's equations) but is attributed in the literature to its ``PhD student'' Proca (for the series of his original articles dating from 1930 to 1938 which led him to introduce in Ref.~\cite{Proca:1900nv} the so-called Proca equation for a massive vector field, see Ref.~\cite{ProcaOeuvres}, but note, however, that the main aim of Proca was the description of spin-$1/2$ particles inspired by the neutrino theory of light due to de Broglie). Here, it is worth pointing out that, due to the mass term, the de Broglie-Proca theory is not a gauge theory, and this has some important consequences when we compare, in the limit $m^2 \rightarrow 0$, the results obtained via the de Broglie-Proca theory with those derived from Maxwell's electromagnetism. It is also important to recall that, in general, it is the de Broglie-Proca theory that is used to impose experimental constraints on the photon mass \cite{Ryutov:2007zz,Tu:2005ge,Goldhaber:2008xy}.

  \item[(ii)] The most aesthetically appealing one which, contrarily to the de Broglie-Proca theory preserves the local $U(1)$ gauge invariance of Maxwell's electromagnetism, has been proposed by Stueckelberg (see Refs.~\cite{Stueckelberg:1900zz,Stueckelberg:1938zz} for the original articles on the subject and also Ref.~\cite{Ruegg:2003ps} for a nice recent review). The construction of such a massive gauge theory can be achieved by coupling appropriately an auxiliary scalar field to the massive spin-$1$ field. This theory is unitary and renormalizable and can be included in the Standard Model of particle physics \cite{Ruegg:2003ps}. Moreover, it is interesting to note that extensions of the Standard Model based on string theory predict the existence of a hidden sector of particles which could explain the nature of dark matter. Among these exotic particles, there exists in particular a dark photon, the mass of which arises also via the Stueckelberg mechanism (see, e.g., Ref.~\cite{Jaeckel:2010ni}). This ``heavy'' photon may be detectable in low energy experiments (see, e.g., Refs~\cite{Essig:2013lka,An:2013yua,Balewski:2014pxa,Eigen:2015rea}). It is also worth pointing out that the Stueckelberg procedure is not limited to vector fields. It has been recently extended to ``restore'' the gauge invariance of various massive field theories (see, e.g., Refs.~\cite{Kuzmin:2002gn,Buchbinder:2008jf} which discuss the case of massive antisymmetric tensor fields and, e.g., Ref.~\cite{deRham:2014zqa} where massive gravity is considered).

\end{itemize}
\vspace{-2.0mm}
In the two following paragraphs, we shall briefly review these two theories at the classical level.

De Broglie-Proca massive electromagnetism is described by a vector field $A_\mu$, and its action $S=S[A_\mu, g_{\mu \nu}]$, which is directly obtained from the original Maxwell Lagrangian by adding a mass contribution, is given by
\begin{equation}
\label{Action_dBProca}
S = \int_\mathcal{M} d^4 x \, \sqrt{-g} \, \left[ - \frac{1}{4} \, F^{\mu \nu} F_{\mu \nu} - \frac{1}{2} \, m^2 A^\mu A_\mu \right] .
\end{equation}
Here, $m$ is the mass of the vector field $A_\mu$, and the associated field strength $F{}_\mu{}_\nu $ is defined as usual by
\begin{equation}
\label{Def_F}
F_{\mu \nu} = \nabla_\mu A_\nu - \nabla_\nu A_\mu = \partial_\mu A_\nu - \partial_\nu A_\mu .
\end{equation}
Let us note that, while Maxwell's theory is invariant under the gauge transformation
\begin{eqnarray}
\label{invGauge_Maxwell}
A_\mu \rightarrow A'_\mu = A_\mu + \nabla_\mu \Lambda
\end{eqnarray}
for an arbitrary scalar field $\Lambda$, this gauge invariance is broken for the de Broglie-Proca theory due to the mass term. The extremization of \eqref{Action_dBProca} with respect to $A_\mu$ leads to the Proca equation $\nabla^\nu F_{\mu \nu} + m^2 A_\mu = 0$. Applying $\nabla^\mu$ to this equation, we obtain the Lorenz condition
\begin{subequations} \label{LCetWEQ_dBProca}
\begin{equation} \label{LorenzCondition_dBProca}
\nabla^\mu A_\mu = 0
\end{equation}
which is here a dynamical constraint (and not a gauge condition) as well as the wave equation
\begin{equation} \label{WEQ_dBProca}
\Box A_\mu - m^2 A_\mu - R_{\mu}^{\phantom{\mu} \nu} A_\nu = 0.
\end{equation}
\end{subequations}
\vspace{-1.0mm}

\noindent It should be noted that the action \eqref{Action_dBProca} is also directly relevant at the quantum level because the de Broglie-Proca theory is not a gauge theory.

Stueckelberg massive electromagnetism is described by a vector field $A_\mu$ and an auxiliary scalar field $\Phi$, and its action $S_\mathrm{Cl}=S_\mathrm{Cl}[A_\mu,\Phi, g_{\mu \nu}]$, which can be constructed from the de Broglie-Proca action \eqref{Action_dBProca} by using the substitution
\begin{equation}
\label{Action_dBProca_to_Stueck}
A_\mu  \to A_\mu + \frac{1}{m} \nabla_\mu \Phi,
\end{equation}
is given by
\begin{widetext}
\begin{subequations}
\label{Action_Stueck}
\begin{eqnarray}
\label{Action_Stueck_1}
&& S_\mathrm{Cl} = \int_{\cal M} d^4 x \sqrt{-g} \, \left[ - \frac{1}{4} F^{\mu \nu} F_{\mu \nu} - \frac{1}{2} \, m^2 \left(A^\mu + \frac{1}{m} \, \nabla^\mu \Phi \right) \left(A_\mu + \frac{1}{m} \, \nabla_\mu \Phi \right) \right]
\\
\label{Action_Stueck_2}
&& \phantom{S_\mathrm{Cl}} = \int _{\cal M} d^4 x \sqrt{-g} \, \left[ - \frac{1}{2} \, \nabla^\mu A^\nu \nabla_\mu A_\nu + \frac{1}{2} \, (\nabla^\mu A_\mu)^2 - \frac{1}{2} \, m^2 A^\mu A_\mu - \frac{1}{2} \, R_{\mu \nu} A^\mu A^\nu \right. \nonumber
\\
&& \qquad\qquad\qquad\qquad\qquad\qquad\qquad\qquad \left. - \frac{1}{2} \, \nabla^\mu \Phi \nabla_\mu \Phi - m A^\mu \nabla_\mu \Phi \right] .
\end{eqnarray}
\end{subequations}
\end{widetext}
It should be noted that, at the classical level, the vector field $A_\mu$ and the scalar field $\Phi$ are coupled [see, in Eq.~\eqref{Action_Stueck_2}, the last term $- m A^\mu \nabla_\mu \Phi$]. Here, it is important to note that Stueckelberg massive electromagnetism is invariant under the gauge transformation
\begin{subequations}
\label{invGauge_Stueck}
\begin{eqnarray}
\label{invGauge_Stueck_A}
&& A_\mu \to  A'_\mu = A_\mu + \nabla_\mu \Lambda ,
\\
\label{invGauge_Stueck_Phi}
&& \Phi \to \Phi' = \Phi - m \Lambda,
\end{eqnarray}
\end{subequations}
for an arbitrary scalar field $\Lambda $, so the local $U(1)$ gauge symmetry of Maxwell's electromagnetism remains unbroken for the spin-$1$ field of the Stueckelberg theory. As a consequence, in order to treat this theory at the quantum level (see below), it is necessary to add to the action \eqref{Action_Stueck} a gauge-breaking term and the compensating ghost contribution.

Here, it seems important to highlight some considerations which will play a crucial role in this article. Let us note that the de Broglie-Proca theory can be obtained from Stueckelberg electromagnetism by taking
\begin{equation}
\label{Action_Stueck_to_dBProca}
\Phi =0.
\end{equation}
We can therefore consider that the de Broglie-Proca theory is nothing other than the Stueckelberg gauge theory in the particular gauge \eqref{Action_Stueck_to_dBProca}. However, it is worth noting that this is a ``bad'' choice of gauge leading to some complications. In particular:
\begin{itemize}
  \item[(i)] Due to the constraint \eqref{LorenzCondition_dBProca}, the Feynman propagator associated with the vector field $A_\mu$ does not admit a Hadamard representation (see below), and, as a consequence, the quantum states of the de Broglie-Proca theory are not of Hadamard type. This complicates the regularization and renormalization procedures.
  \item[(ii)] In the limit $m^2 \rightarrow 0$, singularities occur, and a lot of physical results obtained in the context of the de Broglie-Proca theory do not coincide with the corresponding results obtained with Maxwell's theory.
\end{itemize}

In this article, we intend to focus on the Stueckelberg theory at the quantum level, and we shall analyze its energetic content with possible applications to the Casimir effect (in this paper) and to cosmology of the very early universe (in a next paper) in mind. More precisely, we shall develop the formalism permitting us to construct, for a normalized Hadamard quantum state $|\psi\rangle$ of the Stueckelberg theory, the quantity $\langle \psi | \widehat{T}_{\mu \nu} | \psi \rangle_\mathrm{ren}$ which denotes the renormalized expectation value of the stress-energy-tensor operator. It is well known that such an expectation value is of fundamental importance in quantum field theory in curved spacetime (see, e.g., Refs.~\cite{DeWitt:1975ys,Birrell:1982ix,Fulling:1989nb,Wald:1995yp,Parker:2009uva}). Indeed, it permits us to analyze the quantum state $|\psi \rangle$ without any reference to its particle content, and, moreover, it acts as a source in the semiclassical Einstein equations $G_{\mu \nu} = 8\pi \, \langle \psi | \widehat{T}_{\mu \nu} | \psi \rangle_\mathrm{ren}$ which govern the backreaction of the quantum field theory on the spacetime geometry.

Let us recall that the stress-energy tensor $\widehat{T}_{\mu \nu}$ is an operator quadratic in the quantum fields which is, from the mathematical point of view, an operator-valued distribution. As a consequence, this operator is ill defined and the associated expectation value $\langle \psi | \widehat{T}_{\mu \nu} |\psi \rangle$ is formally infinite. In order to extract from this expectation value a finite and physically acceptable contribution which could act as the source in the semiclassical Einstein equations, it is necessary to regularize it and then to renormalize all the coupling constants. For a description of the various techniques of regularization and renormalization in the context of quantum field theory in curved spacetime (adiabatic regularization method, dimensional regularization method, $\zeta$-function approach, point-splitting methods, \dots), see Refs.~\cite{DeWitt:1975ys,Birrell:1982ix,Fulling:1989nb,Wald:1995yp,Parker:2009uva} and references therein.

In this paper, we shall deal with Stueckelberg electromagnetism by using the so-called Hadamard renormalization procedure (for a rigorous axiomatic presentation of this approach, we refer to the monographs of Wald \cite{Wald:1995yp} and Fulling \cite{Fulling:1989nb}). Here, we just recall that it is an extension of the point-splitting method \cite{DeWitt:1975ys,Christensen:1976vb,Christensen:1978yd} which has been developed in connection with the Hadamard representation of the Green functions (see, e.g., Refs.~\cite{Wald:1977up,Wald:1978pj,Adler:1976jx,Adler:1977ac,Castagnino:1984mk,Brown:1986tj,Bernard:1986vc,Tadaki:1987dq,Allen:1987bn,Folacci:1990eb,Decanini:2005eg,Moretti:2001qh,Hack:2012qf} and, more particularly, Refs.~\cite{Adler:1976jx,Adler:1977ac,Brown:1986tj,Allen:1987bn,Folacci:1990eb} where gauge theories are considered).

Our article is organized as follows. In Sec.~\ref{Sec.II}, we review the covariant quantization of Stueckelberg massive electromagnetism on an arbitrary four-dimensional curved spacetime (gauge-breaking action and associated ghost contribution; wave equations for the massive spin-$1$ field $A_\mu$, for the auxiliary Stueckelberg scalar field $\Phi$ and for the ghost fields $C$ and $C^\ast$; Feynman propagators and Ward identities). In Sec.~\ref{Sec.III}, we focus on the particular gauge for which the various Feynman propagators and the associated Hadamard Green functions admit Hadamard representation, or, in other words, we consider quantum states of Hadamard type. We also construct the covariant Taylor series expansions of the geometrical and state-dependent coefficients involved in the Hadamard representation of the Green functions. In Sec.~\ref{Sec.IV}, we obtain, for a Hadamard quantum state, the renormalized expectation value of the stress-energy-tensor operator, and we discuss carefully its geometrical ambiguities. In fact, we provide two alternative but equivalent expressions for this renormalized expectation value. The first one is obtained by removing the contribution of the auxiliary scalar field $\Phi$ (here, it plays the role of a kind of ghost field) and only involves state-dependent and geometrical quantities associated with the massive vector field $A_\mu$. The other one involves contributions coming from both the massive vector field and the auxiliary Stueckelberg scalar field, and it has been constructed in such a way that, in the zero-mass limit, the massive vector field contribution reduces smoothly to the result obtained from Maxwell's theory. In Sec.~\ref{Sec.V}, as an application of our results, we consider in the Minkowski spacetime the Casimir effect outside of a perfectly conducting medium with a plane boundary wall separating it from free space. We discuss the results obtained using Stueckelberg but also de Broglie-Proca electromagnetism, and we consider the zero-mass limit of the vacuum energy in both theories. Finally, in a conclusion (Sec.~\ref{Sec.VI}), we provide a step-by-step guide for the reader wishing to use our formalism, we briefly discuss and compare the de Broglie-Proca and Stueckelberg approaches in the light of the results obtained in our paper, and we highlight the advantages of the latter. In a short Appendix, we have gathered some important results which are helpful to do the calculations of Secs.~\ref{Sec.III} and \ref{Sec.IV}, and, in particular, (i) we define the geodetic interval $\sigma(x,x')$, the Van Vleck-Morette determinant $\Delta(x,x')$ and the bivector of parallel transport $g_{\mu \nu'}(x,x')$ which play a crucial role along our article, and (ii) we discuss the concept of covariant Taylor series expansions.

It should be noted that, in this paper, we consider a four-dimensional curved spacetime $(\mathcal{M}, g_{\mu \nu})$ with no boundary ($\partial \mathcal{M} = \varnothing$), and we use units with $\hbar=c=G=1$ and the geometrical conventions of Hawking and Ellis \cite{HawkingEllis} concerning the definitions of the scalar curvature $R$, the Ricci tensor $R_{\mu \nu}$ and the Riemann tensor $R_{\mu \nu \rho \sigma}$ as well as the commutation of covariant derivatives. It is moreover important to note that we provide the covariant Taylor series expansions of the Hadamard coefficients in irreducible form by using the algebraic proprieties of the Riemann tensor (and more particularly the cyclicity relation and its consequences) as well as the Bianchi identity.

\section{Quantization of Stueckelberg electromagnetism}
\label{Sec.II}

In this section, we review the covariant quantization of Stueckelberg electromagnetism on an arbitrary four-dimensional curved spacetime. The gauge-breaking term considered includes an arbitrary gauge parameter $\xi$, and all the results concerning the wave equations for the massive vector field $A_\mu$, for the auxiliary scalar field $\Phi$, for the ghost fields $C$ and $C^\ast$ and for all the associated Feynman propagators as well as the Ward identities are expressed in terms of $\xi$.

\subsection{Quantum action}
\label{Sec.IIa}

At the quantum level, the action defining Stueckelberg massive electromagnetism is given by (see, e.g., Ref.~\cite{Ruegg:2003ps})
\begin{eqnarray}
\label{Action_Stueck_Quant_v1}
&& S[A_\mu, \Phi, C, C^\ast, g_{\mu \nu} ] =
  S_\mathrm{Cl}[A_\mu, \Phi, g_{\mu \nu}]
\nonumber \\ && \qquad
+ S_\mathrm{GB}[A_\mu, \Phi, g_{\mu \nu}]
+ S_\mathrm{Gh}[C, C^\ast, g_{\mu \nu}] ,
\end{eqnarray}
where we have added to the classical action \eqref{Action_Stueck} the gauge-breaking term
\begin{equation}
\label{Action_Stueck_Quant_v1_GB}
S_\mathrm{GB} = \int_{\cal M} d^4 x \, \sqrt{-g} \, \left[ -\frac{1}{2\xi} (\nabla^\mu A_\mu + \xi m \Phi)^2 \right]
\end{equation}
and the compensating ghost action
\begin{equation}
\label{Action_Stueck_Quant_v1_Ghosts}
S_\mathrm{Gh} = \int_{\cal M} d^4 x \, \sqrt{-g} \, \left[ \nabla^\mu C^\ast \nabla_\mu C + \xi m^2 C^\ast C \right] .
\end{equation}
By collecting the fields in the explicit expression \eqref{Action_Stueck_Quant_v1}, the quantum action can be written in the form
\begin{eqnarray}
\label{Action_Stueck_Quant_v2}
&& S\left[A_\mu, \Phi, C, C^\ast, g_{\mu \nu} \right] =
  S_A \left[A_\mu, g_{\mu \nu} \right]
\nonumber \\ && \qquad
+ S_\Phi \left[\Phi, g_{\mu \nu} \right]
+ S_\mathrm{Gh} \left[C, C^\ast, g_{\mu \nu} \right] ,
\end{eqnarray}
where
\begin{subequations}
\label{Action_Stueck_Quant_v2_A}
\begin{eqnarray}
\label{Action_Stueck_Quant_v2_A_1}
&& S_A = \int_{\cal M} d^4 x \, \sqrt{-g} \, \left[
- \frac{1}{4} F^{\mu \nu} F_{\mu \nu}
\right. \nonumber \\ && \left. \qquad \qquad
- \frac{1}{2} \, m^2 A^\mu A_\mu
- \frac{1}{2\xi} \, \left( \nabla^\mu A_\mu \right)^2
\right]
\\
\label{Action_Stueck_Quant_v2_A_2}
&& \phantom{S_A} = \int_{\cal M} d^4 x \, \sqrt{-g} \, \left[
- \frac{1}{2} \, \nabla^\mu A^\nu \nabla_\mu A_\nu
- \frac{1}{2} \, R_{\mu \nu} A^\mu A^\nu
\right. \nonumber \\ && \left. \qquad \qquad
- \frac{1}{2} \, m^2 A^\mu A_\mu
+ \frac{1}{2} \, \left( 1 - \frac{1}{\xi} \right) \left( \nabla^\mu A_\mu \right)^2
\right]
\end{eqnarray}
\end{subequations}
and
\begin{eqnarray}
\label{Action_Stueck_Quant_v2_Phi}
&& \hspace{-5mm} S_\Phi = \int_{\cal M} d^4 x \, \sqrt{-g} \, \left[
- \frac{1}{2} \, \nabla^\mu \Phi \nabla_\mu \Phi
- \frac{1}{2} \, \xi m^2 \Phi^2
\right] ,
\end{eqnarray}
$S_\mathrm{Gh}$ remaining unchanged and still given by Eq.~\eqref{Action_Stueck_Quant_v1_Ghosts}. It is worth noting that the term $- m A^\mu \nabla_\mu \Phi$ coupling the fields $A_\mu$ and $\Phi$ in the classical action \eqref{Action_Stueck_2} has disappeared; because spacetime is assumed with no boundary, it is neutralized by the term $- m \Phi \nabla^\mu A_\mu$ in the gauge-breaking action \eqref{Action_Stueck_Quant_v1_GB}.

The functional derivatives with respect to the fields $A_\mu$, $\Phi$, $C$ and $C^\ast$ of the quantum action \eqref{Action_Stueck_Quant_v1} or \eqref{Action_Stueck_Quant_v2} will allow us to obtain, in Sec.~\ref{Sec.IIb}, the wave equations for all the fields and to discuss, in Sec.~\ref{Sec.IVa}, the conservation of the stress-energy tensor associated with Stueckelberg electromagnetism. They are given by
\begin{eqnarray}
\label{DerivFunct_Stueck_champA_Quant}
&& \frac{1}{\sqrt{-g}} \frac{\delta S}{\delta A_\mu}
= \left[ g^{\mu \nu} \Box - \left( 1 - 1/\xi \right) \nabla^\mu \nabla^\nu
\nonumber \right. \\ && \left. \hphantom{\frac{1}{\sqrt{-g}} \frac{\delta S}{\delta A_\mu} \equiv}
- R^{\mu \nu} - m^2 g^{\mu \nu} \right] A_\nu
\end{eqnarray}
for the vector field $A_\mu$,
\begin{equation}
\label{DerivFunct_Stueck_champPhi_Quant}
\frac{1}{\sqrt{-g}} \frac{\delta S}{\delta \Phi}
= \left[ \Box - \xi m^2 \right] \Phi
\end{equation}
for the auxiliary scalar field $\Phi$, as well as
\begin{equation}
\label{DerivFunct_Stueck_ghosts_Quant_a}
\frac{1}{\sqrt{-g}} \frac{\delta_\mathrm{R} S}{\delta C}
= - \left[ \Box - \xi m^2 \right] C^\ast
\end{equation}
and
\begin{equation}
\label{DerivFunct_Stueck_ghosts_Quant_b}
\frac{1}{\sqrt{-g}} \frac{\delta_\mathrm{L} S}{\delta C^\ast}
= - \left[ \Box - \xi m^2 \right] C
\end{equation}
for the ghost fields $C$ and $C^\ast$. It should be noted that, due to the fermionic behavior of the ghost fields, we have introduced in Eq.~\eqref{DerivFunct_Stueck_ghosts_Quant_a} the right functional derivative and in Eq.~\eqref{DerivFunct_Stueck_ghosts_Quant_b} the left functional derivative.

\subsection{Wave equations}
\label{Sec.IIb}

The extremization of the quantum action \eqref{Action_Stueck_Quant_v1} or \eqref{Action_Stueck_Quant_v2} permits us to obtain the wave equations for the fields $A_\mu$, $\Phi$, $C$ and $C^\ast$. The vanishing of the functional derivatives \eqref{DerivFunct_Stueck_champA_Quant}--\eqref{DerivFunct_Stueck_ghosts_Quant_b} provides
\begin{equation}
\label{DerivFunct_WEQ_Stueck_champA_Quant}
\left[ g^{\mu \nu} \Box - \left( 1 - 1/\xi \right) \nabla^\mu \nabla^\nu - R^{\mu \nu} - m^2 g^{\mu \nu} \right] A_\nu = 0
\end{equation}
for the vector field $A_\mu$,
\begin{equation}
\label{DerivFunct_WEQ_Stueck_champPhi_Quant}
\left[ \Box - \xi m^2 \right] \Phi = 0
\end{equation}
for the auxiliary scalar field $\Phi$, as well as
\begin{eqnarray}
\label{DerivFunct_WEQ_Stueck_ghosts_Quant}
&& \left[ \Box - \xi m^2 \right] C = 0
\quad \text{and} \quad
\left[ \Box - \xi m^2 \right] C^\ast = 0
\end{eqnarray}
for the ghost fields $C$ and $C^\ast$.

\subsection{Feynman propagators and Ward identities}
\label{Sec.IIc}

From now on, we shall assume that the Stueckelberg field theory previously described has been quantized and is in a normalized quantum state $|\psi \rangle$. The Feynman propagator
\begin{equation}
\label{FeynmanProp_A}
G^{A}_{\mu \nu'} (x,x') = i \langle \psi | T A_\mu(x) A_{\nu'} (x') | \psi \rangle
\end{equation}
associated with the field $A_\mu$ (here, $T$ denotes time ordering) is, by definition, a solution of
\begin{eqnarray}
\label{WEQ_GFA}
&& \left[ g^{\phantom{\mu} \nu}_\mu \Box_x - \left( 1 - 1/\xi \right) \nabla^\nu \nabla_\mu
- R^{\phantom{\mu} \nu}_\mu - m^2 g^{\phantom{\mu} \nu}_\mu \right]
G^{A}_{\nu \rho'} (x,x')
\nonumber \\ && \qquad
= - g_{\mu \rho'} \, \delta^4(x,x')
\end{eqnarray}
with $\delta^4 (x,x') = [-g(x)]^{-1/2} \, \delta^4 (x-x')$. Similarly, the Feynman propagator
\begin{equation}
\label{FeynmanProp_Phi}
G^{\Phi} (x,x') = i \langle \psi | T \Phi(x) \Phi(x') | \psi \rangle
\end{equation}
associated with the scalar field $\Phi$ satisfies
\begin{equation}
\label{WEQ_GFPhi}
\left[ \Box_x  - \xi m^2 \right] G^{\Phi} (x,x') = - \delta^4 (x,x') ,
\end{equation}
and the Feynman propagator
\begin{equation}
\label{FeynmanProp_Gh}
G^\mathrm{Gh} (x,x') = i \langle \psi | T C^\ast(x) C(x') | \psi \rangle
\end{equation}
associated with the ghost fields $C$ and $C^\ast$ satisfies
\begin{equation}
\label{WEQ_GFGh}
\left[ \Box_x  - \xi m^2 \right] G^\mathrm{Gh} (x,x') = - \delta^4 (x,x') .
\end{equation}

The three propagators are related by two Ward identities. The first one is a nonlocal relation linking the propagators $G^{A}_{\mu \nu'} (x,x')$ and $G^\mathrm{Gh} (x,x')$. It can be obtained by extending the approach of DeWitt and Brehme in Ref.~\cite{DeWitt:1960fc} as follows: we take the covariant derivative $\nabla^\mu$ of Eq.~\eqref{WEQ_GFA} and the covariant derivative $\nabla_{\rho'}$ of Eq.~\eqref{WEQ_GFGh}; then, by commuting suitably the various covariant derivatives involved and by using the relation $\nabla^\mu [ g_{\mu \rho'} \, \delta^4(x,x') ] = - \nabla_{\rho'} \delta^4(x,x')$, we obtain the formal relation
\begin{eqnarray}
\label{WardId_GFA}
&& (1/\xi) \, \nabla^\mu G^{A}_{\mu \nu'} (x,x') + \nabla_{\nu'} G^\mathrm{Gh} (x,x')
\nonumber \\ && \hspace{-5mm}
= \left( 1 - 1/\xi \right) \left[ \Box_x  - \xi m^2 \right]^{-1} \left[ \nabla^\mu \left\{ R^{\phantom{\mu} \rho}_\mu G^{A}_{\rho \nu'} (x,x') \right\} \right] .
\end{eqnarray}
It should be noted that the nonlocal term in the right-hand side of this equation is associated with the nonminimal term $\left( 1 - 1/\xi \right) \nabla^\nu \nabla_\mu$ appearing in the wave equation \eqref{WEQ_GFA} and includes appropriate boundary conditions. The second Ward identity can be obtained directly from the wave equations \eqref{WEQ_GFPhi} and \eqref{WEQ_GFGh} by using arguments of uniqueness. We have
\begin{equation}
\label{WardId_GFPhi}
G^{\Phi} (x,x') - G^\mathrm{Gh} (x,x') = 0 .
\end{equation}

\section{Hadamard expansions of the Green functions of Stueckelberg electromagnetism}
\label{Sec.III}

From now on, we assume that $\xi=1$. (For $\xi \neq 1$, the various Feynman propagators cannot be represented in the Hadamard form.) For this choice of gauge parameter, the wave equations \eqref{WEQ_GFA}, \eqref{WEQ_GFPhi} and \eqref{WEQ_GFGh} for the Feynman propagators $G^{A}_{\mu \nu'} (x,x')$, $G^{\Phi} (x,x')$ and $G^\mathrm{Gh} (x,x')$ reduce to
\begin{eqnarray}
\label{WEQ_GFA_1}
&& \left[ g^{\phantom{\mu} \nu}_\mu \Box_x - R^{\phantom{\mu} \nu}_\mu - m^2 g^{\phantom{\mu} \nu}_\mu \right] G^{A}_{\nu \rho'} (x,x')
\nonumber \\ && \qquad
= - g_{\mu \rho'} \, \delta^4(x,x') ,
\end{eqnarray}
\begin{equation}
\label{WEQ_GFPhi_1}
\left[ \Box_x  - m^2 \right] G^{\Phi} (x,x') = - \delta^4 (x,x')
\end{equation}
and
\begin{equation}
\label{WEQ_GFGh_1}
\left[ \Box_x  - m^2 \right] G^\mathrm{Gh} (x,x') = - \delta^4 (x,x') .
\end{equation}
As far as the Ward identity \eqref{WardId_GFA} is concerned, it takes now the local form
\begin{equation}
\label{WardId_GFA_1}
\nabla^\mu G^{A}_{\mu \nu'} (x,x') + \nabla_{\nu'} G^\mathrm{Gh} (x,x') = 0 ,
\end{equation}
while the Ward identity \eqref{WardId_GFPhi} remains unchanged. Because this last relation expresses the equality of the Feynman propagators associated with the auxiliary scalar field and the ghost fields, we shall often use a generic form for these propagators (and for their Hadamard representation discussed below) where the labels $\Phi$ and $\mathrm{Gh}$ are omitted, and we shall write
\begin{equation}
\label{GFPhi=GFGh}
G (x,x') = G^\Phi (x,x') = G^\mathrm{Gh} (x,x') .
\end{equation}

For $\xi=1$ the nonminimal term in the wave equation for $G^{A}_{\mu \nu'} (x,x')$ has disappeared [compare Eq.~\eqref{WEQ_GFA_1} with Eq.~\eqref{WEQ_GFA}]. As consequence, we can consider a Hadamard representation for this propagator as well as for the propagators $G^\Phi (x,x')$ and $G^\mathrm{Gh} (x,x')$. In other words, we can assume that all fields of Stueckelberg theory are in a normalized quantum state $|\psi\rangle$ of Hadamard type.

\subsection{Hadamard representation of the Feynman propagators}
\label{Sec.IIIa}

The Feynman propagator $G^A_{\mu \nu'} (x,x')$ associated with the vector field $A_\mu$ can be now represented in the Hadamard form
\begin{eqnarray}
\label{HadamardRep_GAF}
&& G^A_{\mu \nu'} (x,x') = \frac{i}{8\pi^2} \, \left\{
\frac{\Delta^{1/2}(x,x')}{\sigma(x,x') + i \epsilon} \, g_{\mu \nu'}(x,x')
\right. \nonumber \\ && \left. \quad \vphantom{\frac{\Delta^{1/2}(x,x')}{\sigma(x,x') + i \epsilon}}
+ V^A_{\mu \nu'}(x,x') \ln[\sigma(x,x') + i \epsilon] + W^A_{\mu \nu'}(x,x')
\right\} ,
\end{eqnarray}
where the bivectors $V^A_{\mu \nu'}(x,x')$ and $W^A_{\mu \nu'}(x,x')$ are symmetric in the sense that $V^A_{\mu \nu'}(x,x') = V^A_{\nu' \mu}(x',x)$ and $W^A_{\mu \nu'}(x,x') = W^A_{\nu' \mu}(x',x)$ and are regular for $x' \to x$. Furthermore, these bivectors have the following expansions
\begin{subequations}
\label{Expansion_VA_WA}
\begin{eqnarray}
\label{Expansion_VA}
&& V^A_{\mu \nu'}(x,x') = \sum_{n=0}^{+\infty} V^A_n{}_{\mu \nu'}(x,x') \, \sigma^n(x,x') ,
\\
\label{Expansion_WA}
&& W^A_{\mu \nu'}(x,x') = \sum_{n=0}^{+\infty} W^A_n{}_{\mu \nu'}(x,x') \, \sigma^n(x,x') .
\end{eqnarray}
\end{subequations}
Similarly, the Hadamard expansion of the Feynman propagator $G(x,x')$ associated with the auxiliary scalar field $\Phi$ or the ghost fields is given by
\begin{eqnarray}
\label{HadamardRep_GF}
&& G(x,x') = \frac{i}{8\pi^2} \, \left\{
\frac{\Delta^{1/2}(x,x')}{\sigma(x,x') + i \epsilon}
\right. \nonumber \\ && \left. \quad \vphantom{\frac{\Delta^{1/2}(x,x')}{\sigma(x,x') + i \epsilon}}
+ V(x,x') \ln[\sigma(x,x') + i \epsilon] + W(x,x')
\right\} ,
\end{eqnarray}
where the biscalars $V(x,x')$ and $W(x,x')$ are symmetric, i.e., $V(x,x') = V(x',x)$ and $W(x,x') = W(x',x)$, regular for $x' \to x$ and possess expansions of the form
\begin{subequations}
\label{Expansion_V_W}
\begin{eqnarray}
\label{Expansion_V}
&& V(x,x') = \sum_{n=0}^{+\infty} V_n(x,x') \, \sigma^n(x,x') ,
\\
\label{Expansion_W}
&& W(x,x') = \sum_{n=0}^{+\infty} W_n(x,x') \, \sigma^n(x,x') .
\end{eqnarray}
\end{subequations}
In Eqs.~\eqref{HadamardRep_GAF} and \eqref{HadamardRep_GF}, the factor $i\epsilon$ with $\epsilon \to 0_+$ ensures the singular behavior prescribed by the time-ordered product introduced in the definition of the Feynman propagators [see Eqs.~\eqref{FeynmanProp_A}, \eqref{FeynmanProp_Phi} and \eqref{FeynmanProp_Gh}].

The Hadamard coefficients $V^A_n{}_{\mu \nu'}(x,x')$ and $W^A_n{}_{\mu \nu'}(x,x')$ introduced in Eq.~\eqref{Expansion_VA_WA} are also symmetric and regular bivector functions. The coefficients $V^A_n{}_{\mu \nu'}(x,x')$ satisfy the recursion relations
\begin{subequations}
\label{Recursion_VA}
\begin{eqnarray}
\label{Recursion_VAn}
&& 2(n+1)(n+2) \, V^A_{n+1}{}_{\mu \nu'} + 2(n+1) \, V^A_{n+1}{}_{\mu \nu' ;a} \sigma^{;a}
\nonumber \\ && \quad
- 2(n+1) \, V^A_{n+1}{}_{\mu \nu'} \Delta^{-1/2} ( \Delta^{1/2} ){}_{;a} \sigma^{;a}
\nonumber \\ && \quad
+ \left[ g^{\phantom{\mu} \rho}_\mu \Box_x - R^{\phantom{\mu} \rho}_\mu - m^2 g^{\phantom{\mu} \rho}_\mu \right] V^A_n{}_{\rho \nu'} = 0
\end{eqnarray}
for $n \in \mathbb{N}$ with the boundary condition
\begin{eqnarray}
\label{Recursion_VA0}
&& 2 \, V^A_0{}_{\mu \nu'} + 2 \, V^A_0{}_{\mu \nu' ;a} \sigma^{;a}
- 2 \, V^A_0{}_{\mu \nu'} \Delta^{-1/2} ( \Delta^{1/2} ){}_{;a} \sigma^{;a}  \nonumber \\ && \quad
+ \left[ g^{\phantom{\mu} \rho}_\mu \Box_x - R^{\phantom{\mu} \rho}_\mu - m^2 g^{\phantom{\mu} \rho}_\mu \right] ( g_{\rho \nu'} \Delta^{1/2} ) = 0 ,
\end{eqnarray}
\end{subequations}
while the coefficients $W^A_n{}_{\mu \nu'}(x,x')$ satisfy the recursion relations
\begin{eqnarray}
\label{Recursion_WAn}
&& 2(n+1)(n+2) \, W^A_{n+1}{}_{\mu \nu'} + 2(n+1) \, W^A_{n+1}{}_{\mu \nu' ;a} \sigma^{;a}
\nonumber \\ && \quad
- 2(n+1) \, W^A_{n+1}{}_{\mu \nu'} \Delta^{-1/2} ( \Delta^{1/2} ){}_{;a} \sigma^{;a}
\nonumber \\ && \quad
+ 2(2n+3) \, V^A_{n+1}{}_{\mu \nu'} + 2 \, V^A_{n+1}{}_{\mu \nu' ;a} \sigma^{;a}
\nonumber \\ && \quad
- 2 \, V^A_{n+1}{}_{\mu \nu'} \Delta^{-1/2} ( \Delta^{1/2} ){}_{;a} \sigma^{;a}
\nonumber \\ && \quad
+ \left[ g^{\phantom{\mu} \rho}_\mu \Box_x - R^{\phantom{\mu} \rho}_\mu - m^2 g^{\phantom{\mu} \rho}_\mu \right] W^A_n{}_{\rho \nu'} = 0
\end{eqnarray}
for $n \in \mathbb{N}$. It should be noted that from the recursion relations \eqref{Recursion_VA} and \eqref{Recursion_WAn} we can show that
\begin{eqnarray}
\label{EQ_VA}
&& \left[ g^{\phantom{\mu} \nu}_\mu \Box_x - R^{\phantom{\mu} \nu}_\mu - m^2 g^{\phantom{\mu} \nu}_\mu \right] V^A_{\nu \rho'} = 0
\end{eqnarray}
and
\begin{eqnarray}
\label{EQ_WA}
&& \hspace{-3mm} \sigma \left[ g^{\phantom{\mu} \nu}_\mu \Box_x - R^{\phantom{\mu} \nu}_\mu - m^2 g^{\phantom{\mu} \nu}_\mu \right] W^A_{\nu \rho'} =
\nonumber \\ && \hspace{-3mm} \quad
- \left[ g^{\phantom{\mu} \nu}_\mu \Box_x - R^{\phantom{\mu} \nu}_\mu - m^2 g^{\phantom{\mu} \nu}_\mu \right] ( g_{\nu \rho'} \Delta^{1/2} )
\nonumber \\ && \hspace{-3mm} \quad
- 2 \, V^A_{\mu \rho'} - 2 \, V^A_{\mu \rho' ;a} \sigma^{;a} + 2 \, V^A_{\mu \rho'} \Delta^{-1/2} ( \Delta^{1/2} ){}_{;a} \sigma^{;a} .
\end{eqnarray}
These two ``wave equations'' permit us to prove that the Feynman propagator \eqref{HadamardRep_GAF} solves the wave equation \eqref{WEQ_GFA_1}.

Similarly, the Hadamard coefficients $V_n(x,x')$ and $W_n(x,x')$ are also symmetric and regular biscalar functions. The coefficients $V_n(x,x')$ satisfy the recursion relations
\begin{subequations}
\label{Recursion_V}
\begin{eqnarray}
\label{Recursion_Vn}
&& 2(n+1)(n+2) \, V_{n+1} + 2(n+1) \, V_{n+1}{}_{;a} \sigma^{;a}
\nonumber \\ && \quad
- 2(n+1) \, V_{n+1} \Delta^{-1/2} ( \Delta^{1/2} ){}_{;a} \sigma^{;a}
\nonumber \\ && \quad
+ \left[ \Box_x - m^2 \right] V_n = 0
\end{eqnarray}
for $n \in \mathbb{N}$ with the boundary condition
\begin{eqnarray}
\label{Recursion_V0}
&& 2 \, V_0 + 2 \, V_0{}_{;a} \sigma^{;a}
- 2 \, V_0 \Delta^{-1/2} ( \Delta^{1/2} ){}_{;a} \sigma^{;a}  \nonumber \\ && \quad
+ \left[ \Box_x - m^2 \right] \Delta^{1/2} = 0 ,
\end{eqnarray}
\end{subequations}
while the coefficients $W_n(x,x')$ satisfy the recursion relations
\begin{eqnarray}
\label{Recursion_Wn}
&& 2(n+1)(n+2) \, W_{n+1} + 2(n+1) \, W_{n+1}{}_{;a} \sigma^{;a}
\nonumber \\ && \quad
- 2(n+1) \, W_{n+1} \Delta^{-1/2} ( \Delta^{1/2} ){}_{;a} \sigma^{;a}
\nonumber \\ && \quad
+ 2(2n+3) \, V_{n+1} + 2 \, V_{n+1}{}_{;a} \sigma^{;a}
\nonumber \\ && \quad
- 2 \, V_{n+1} \Delta^{-1/2} ( \Delta^{1/2} ){}_{;a} \sigma^{;a}
\nonumber \\ && \quad
+ \left[ \Box_x - m^2 \right] W_n = 0
\end{eqnarray}
for $n \in \mathbb{N}$. It should be also noted that from the recursion relations \eqref{Recursion_V} and \eqref{Recursion_Wn} we can show that
\begin{eqnarray}
\label{EQ_V}
&& \left[ \Box_x - m^2 \right] V = 0
\end{eqnarray}
and
\begin{eqnarray}
\label{EQ_W}
&& \sigma \left[ \Box_x - m^2 \right] W = - \left[ \Box_x - m^2 \right] \Delta^{1/2}
\nonumber \\ && \quad
- 2 \, V - 2 \, V_{;a} \sigma^{;a} + 2 \, V \Delta^{-1/2} ( \Delta^{1/2} ){}_{;a} \sigma^{;a} .
\end{eqnarray}
These two ``wave equations'' permit us to prove that the Feynman propagator \eqref{HadamardRep_GF} solves the wave equation \eqref{WEQ_GFPhi_1} or \eqref{WEQ_GFGh_1}.

The Hadamard representation of the Feynman propagators permits us to straightforwardly identify their singular and regular parts (when the coincidence limit $x' \to x$ is considered). We can write
\begin{equation}
\label{HadamardRep_GFA_sing+reg}
G^A_{\mu \nu'} (x,x') = G^A_\mathrm{sing}{}_{\mu \nu'} (x,x') + G^A_\mathrm{reg}{}_{\mu \nu'} (x,x')
\end{equation}
with
\begin{subequations}
\label{HadamardRep_GFA_sing_reg}
\begin{eqnarray}
\label{HadamardRep_GFA_sing}
&& G^A_\mathrm{sing}{}_{\mu \nu'} (x,x') = \frac{i}{8\pi^2} \, \left\{  \frac{\Delta^{1/2}(x,x')}{\sigma(x,x') + i \epsilon} \, g_{\mu \nu'}(x,x')
\right. \nonumber \\ && \quad \left. \vphantom{\frac{\Delta^{1/2}(x,x')}{\sigma(x,x') + i \epsilon}}
+ V^A_{\mu \nu'}(x,x') \ln[\sigma(x,x') + i \epsilon]
\right\}
\end{eqnarray}
and
\begin{eqnarray}
\label{HadamardRep_GFA_reg}
&& G^A_\mathrm{reg}{}_{\mu \nu'} (x,x') = \frac{i}{8\pi^2} \, W^A_{\mu \nu'}(x,x')
\end{eqnarray}
\end{subequations}
as well as
\begin{equation}
\label{HadamardRep_GF_sing+reg}
G (x,x') = G_\mathrm{sing} (x,x') + G_\mathrm{reg} (x,x')
\end{equation}
with
\begin{subequations}
\label{HadamardRep_GF_sing_reg}
\begin{eqnarray}
\label{HadamardRep_GF_sing}
&& G_\mathrm{sing} (x,x') = \frac{i}{8\pi^2} \, \left\{
\frac{\Delta^{1/2}(x,x')}{\sigma(x,x') + i \epsilon}
\right. \nonumber \\ && \quad \left. \vphantom{\frac{\Delta^{1/2}(x,x')}{\sigma(x,x') + i \epsilon}}
+ V(x,x') \ln[\sigma(x,x') + i \epsilon]
\right\}
\end{eqnarray}
and
\begin{eqnarray}
\label{HadamardRep_GF_reg}
&& G_\mathrm{reg} (x,x') = \frac{i}{8\pi^2} \, W(x,x') .
\end{eqnarray}
\end{subequations}
Here, it is important to note that, due to the geometrical nature of $\sigma(x,x')$, $g_{\mu \nu'}(x,x')$, $\Delta^{1/2}(x,x')$ (see the Appendix) and of $V^A_{\mu \nu'}(x,x')$ and $V(x,x')$ (see Sec.~\ref{Sec.IIIc}), the singular parts \eqref{HadamardRep_GFA_sing} and \eqref{HadamardRep_GF_sing} are purely geometrical objects. By contrast, the regular parts \eqref{HadamardRep_GFA_reg} and \eqref{HadamardRep_GF_reg} are state dependent (see Sec.~\ref{Sec.IIId}).

\subsection{Hadamard Green functions}
\label{Sec.IIIb}

In the context of the regularization of the stress-energy-tensor operator, instead of working with the Feynman propagators, it is more convenient to use the associated so-called Hadamard Green functions. Their representations can be derived from those of the Feynman propagators by using the formal identities
\begin{equation}
\label{Distribution_1}
\frac{1}{\sigma + i \epsilon} = \mathcal{P} \frac{1}{\sigma} - i \pi \delta(\sigma)
\end{equation}
and
\begin{equation}
\label{Distribution_2}
\ln(\sigma + i \epsilon) = \ln|\sigma| + i \pi \Theta(-\sigma) .
\end{equation}
Here, $\mathcal{P}$ is the symbol of the Cauchy principal value, and $\Theta$ denotes the Heaviside step function. Indeed, these identities permit us to rewrite the expression \eqref{HadamardRep_GAF} of the Feynman propagator associated with the massive vector field $A_\mu$ as
\begin{equation}
\label{GFA_GmA_G1A}
G^A_{\mu \nu'} (x,x') = \overline{G}{}^A_{\mu \nu'} (x,x') + \frac{i}{2} \, G^{(1)A}_{\mu \nu'} (x,x') ,
\end{equation}
where the average of the retarded and advanced Green functions is represented by
\begin{eqnarray}
\label{HadamardRep_GmA}
&& \overline{G}{}^A_{\mu \nu'} (x,x') = \frac{1}{8 \pi} \left\{
  \Delta^{1/2} (x,x') \, g_{\mu \nu'} (x,x') \, \delta[\sigma(x,x')]
\right. \nonumber \\ && \left. \quad
- V^A_{\mu \nu'} (x,x') \, \Theta[-\sigma(x,x')]
\vphantom{\Delta^{1/2}(x,x')} \right\}
\end{eqnarray}
and the Hadamard Green function has the representation
\begin{eqnarray}
\label{HadamardRep_G1A}
&& G^{(1)A}_{\mu \nu'} (x,x') = \frac{1}{4 \pi^2} \left\{
  \frac{\Delta^{1/2}(x,x')}{\sigma(x,x')} \, g_{\mu \nu'}(x,x')
\right. \nonumber \\ && \left. \quad
+ V^A_{\mu \nu'}(x,x') \ln | \sigma(x,x') | + W^A_{\mu \nu'}(x,x')
\vphantom{\frac{\Delta^{1/2}(x,x')}{\sigma(x,x')}} \right\} .
\end{eqnarray}
Similarly, we have for the Feynman propagator \eqref{HadamardRep_GF} associated with the auxiliary scalar field $\Phi$ or the ghost fields
\begin{equation}
\label{GF_Gm_G1}
G (x,x') = \overline{G} (x,x') + \frac{i}{2} \, G^{(1)} (x,x') ,
\end{equation}
where
\begin{eqnarray}
\label{HadamardRep_Gm}
&& \overline{G} (x,x') = \frac{1}{8 \pi} \left\{
  \Delta^{1/2}(x,x') \delta[\sigma(x,x')]
\right. \nonumber \\ && \left. \quad
- V (x,x') \Theta[-\sigma(x,x')]
\vphantom{\Delta^{1/2}(x,x')} \right\}
\end{eqnarray}
and
\begin{eqnarray}
\label{HadamardRep_G1}
&& G^{(1)} (x,x') = \frac{1}{4 \pi^2} \left\{
  \frac{\Delta^{1/2}(x,x')}{\sigma(x,x')}
\right. \nonumber \\ &&  \left. \quad
+ V(x,x') \ln | \sigma(x,x') | + W(x,x')
\vphantom{\frac{\Delta^{1/2}(x,x')}{\sigma(x,x')}} \right\} .
\end{eqnarray}

It is important to recall that the Hadamard Green function associated with the massive vector field $A_\mu$ is defined as the anticommutator
\begin{equation}
\label{HadamardGreenFn_A}
G^{(1)A}_{\mu \nu'} (x,x') = \langle \psi | \left\{ A_\mu (x), A_{\nu'} (x') \right\} |\psi \rangle
\end{equation}
and satisfies the wave equation
\begin{equation}
\label{WEQ_G1A}
\left[ g^{\phantom{\mu} \nu}_\mu \Box_x - R^{\phantom{\mu} \nu}_\mu  - m^2 g^{\phantom{\mu} \nu}_\mu \right] G^{(1)A}_{\nu \rho'} (x,x') = 0 .
\end{equation}
Similarly, the Hadamard Green function associated with the auxiliary scalar field $\Phi$ is defined as the anticommutator
\begin{equation}
\label{HadamardGreenFn_Phi}
G^{(1)\Phi} (x,x') = \langle \psi | \left\{ \Phi (x), \Phi (x') \right\} |\psi \rangle
\end{equation}
which is a solution of
\begin{equation}
\label{WEQ_G1Phi}
\left[ \Box_x  - m^2 \right] G^{(1)\Phi} (x,x') = 0 ,
\end{equation}
while the Hadamard Green function associated with the ghost fields is defined as the commutator
\begin{equation}
\label{HadamardGreenFn_Gh}
G^{(1)\mathrm{Gh}} (x,x') = \langle \psi | \left[ C^\ast (x), C (x') \right] |\psi \rangle
\end{equation}
and satisfies the wave equation
\begin{equation}
\label{WEQ_G1Gh}
\left[ \Box_x  - m^2 \right] G^{(1)\mathrm{Gh}} (x,x') = 0 .
\end{equation}
It should be noted that the expressions \eqref{HadamardGreenFn_A}, \eqref{HadamardGreenFn_Phi} and \eqref{HadamardGreenFn_Gh} can be obtained from the definitions \eqref{FeynmanProp_A}, \eqref{FeynmanProp_Phi} and \eqref{FeynmanProp_Gh} by noting that a Hadamard Green function is twice the imaginary part of the corresponding Feynman propagator [see also Eqs.~\eqref{GFA_GmA_G1A} and \eqref{GF_Gm_G1}]. The transition from \eqref{FeynmanProp_A} to \eqref{HadamardGreenFn_A} or from \eqref{FeynmanProp_Phi} to \eqref{HadamardGreenFn_Phi} is rather immediate but it is not so obvious to obtain \eqref{HadamardGreenFn_Gh} from \eqref{FeynmanProp_Gh}: it is necessary to use the fact that the time-ordered product appearing in Eq.~\eqref{FeynmanProp_Gh} is defined with a minus sign due to the anticommuting nature of the ghost fields [i.e., we have $T C^\ast(x) C(x') = \theta (x^0 - x^{'0}) \, C^\ast (x) C(x') - \theta (x^{'0} - x^0) \, C(x') C^\ast(x)$] and to consider that the ghost field $C$ is hermitian while the anti-ghost field $C^\ast$ is anti-hermitian (see, e.g., Ref.~\cite{Kugo:1977yx}).

The Ward identities \eqref{WardId_GFA_1} and \eqref{WardId_GFPhi} satisfied by the Feynman propagators are also valid for the Hadamard Green functions. We have
\begin{equation}
\label{WardId_G1A}
\nabla^\mu G^{(1)A}_{\mu \nu'} (x,x') + \nabla_{\nu'} G^\mathrm{(1)\mathrm{Gh}} (x,x') = 0
\end{equation}
and
\begin{equation}
\label{WardId_G1Phi}
G^{(1)\Phi} (x,x') - G^{(1)\mathrm{Gh}} (x,x') = 0 .
\end{equation}

Similarly, as it has been previously noted in the case of the Feynman propagators, the Hadamard representaion of the Hadamard Green functions permits us to straightforwardly identify their singular and purely geometrical parts as well as their regular and state-dependent parts (when the coincidence limit $x' \to x$ is considered). We can write
\begin{equation}
\label{HadamardRep_G1A_sing+reg}
G^{(1)A}_{\mu \nu'} (x,x') = G^{(1)A}_\mathrm{sing}{}_{\mu \nu'} (x,x') + G^{(1)A}_\mathrm{reg}{}_{\mu \nu'} (x,x')
\end{equation}
with
\begin{subequations}
\label{HadamardRep_G1A_sing_reg}
\begin{eqnarray}
\label{HadamardRep_G1A_sing}
&& G^{(1)A}_\mathrm{sing}{}_{\mu \nu'} (x,x') = \frac{1}{4\pi^2} \, \left\{  \frac{\Delta^{1/2}(x,x')}{\sigma(x,x')} \, g_{\mu \nu'}(x,x')
\right. \nonumber \\ && \quad \left. \vphantom{\frac{\Delta^{1/2}(x,x')}{\sigma(x,x')}}
+ V^A_{\mu \nu'}(x,x') \ln \left| \sigma(x,x') \right|
\right\}
\end{eqnarray}
and
\begin{eqnarray}
\label{HadamardRep_G1A_reg}
&& G^{(1)A}_\mathrm{reg}{}_{\mu \nu'} (x,x') = \frac{1}{4\pi^2} \, W^A_{\mu \nu'}(x,x')
\end{eqnarray}
\end{subequations}
as well as
\begin{equation}
\label{HadamardRep_G1_sing+reg}
G^{(1)} (x,x') = G^{(1)}_\mathrm{sing} (x,x') + G^{(1)}_\mathrm{reg} (x,x')
\end{equation}
with
\begin{subequations}
\label{HadamardRep_G1_sing_reg}
\begin{eqnarray}
\label{HadamardRep_G1_sing}
&& G^{(1)}_\mathrm{sing} (x,x') = \frac{1}{4\pi^2} \, \left\{
\frac{\Delta^{1/2}(x,x')}{\sigma(x,x')}
\right. \nonumber \\ && \quad \left. \vphantom{\frac{\Delta^{1/2}(x,x')}{\sigma(x,x')}}
+ V(x,x') \ln \left| \sigma(x,x') \right|
\right\}
\end{eqnarray}
and
\begin{eqnarray}
\label{HadamardRep_G1_reg}
&& G^{(1)}_\mathrm{reg} (x,x') = \frac{1}{4\pi^2} \, W(x,x') .
\end{eqnarray}
\end{subequations}
It should be pointed out that the regular part of the Hadamard Green function given by Eq.~\eqref{HadamardRep_G1A_reg} [respectively, by Eq.~\eqref{HadamardRep_G1_reg}] is proportional to that of the Feynman propagator given by Eq.~\eqref{HadamardRep_GFA_reg} [respectively, by Eq.~\eqref{HadamardRep_GF_reg}].

\subsection{Geometrical Hadamard coefficients and associated covariant Taylor series expansions}
\label{Sec.IIIc}

Formally, the Hadamard coefficients $V^A_n{}_{\mu \nu'}(x,x')$ or $V_n(x,x')$ can be determined uniquely by solving the recursion relations \eqref{Recursion_VA} or \eqref{Recursion_V}, i.e., by integrating these recursion relations along the unique geodesic joining $x$ to $x'$ (it is unique for $x'$ near $x$ or more generally for $x'$ in a convex normal neighborhood of $x$). As a consequence, all these coefficients as well as the sums given by Eqs.~\eqref{Expansion_VA} and \eqref{Expansion_V} are of purely geometric nature, i.e., they only depend on the geometry along the geodesic joining $x$ to $x'$.

From the point of view of the practical applications considered in this work, it is sufficient to know the expressions of the two first geometrical Hadamard coefficients. Furthermore, their covariant Taylor series expansions are needed up to order $\sigma^1$ for $n=0$ and $\sigma^0$ for $n=1$.
The covariant Taylor series expansions of the bivector coefficients $V^A_0{}_{\mu \nu}(x,x')$ and $V^A_1{}_{\mu \nu}(x,x')$ are given by [see Eqs.~\eqref{CovTaylorSeries_Vector} and \eqref{VectorTaylorCoef_Sym}]
\begin{subequations}
\label{covTaylorSeries_VA0_VA1}
\begin{eqnarray}
\label{covTaylorSeries_VA0}
&& \hspace{-3mm} V^A_0{}_{\mu \nu} = g_\nu^{\phantom{\nu} \nu'} \, V^A_0{}_{\mu \nu'}
= v^A_0{}_{(\mu \nu)}
\nonumber \\ && \hspace{-3mm} \quad
- \left\{ (1/2) \, v^A_0{}_{(\mu \nu) ;a} + v^A_0{}_{[\mu \nu] a} \right\} \sigma^{;a}
\nonumber \\ && \hspace{-3mm} \quad
+ \frac{1}{2!} \, \left\{ v^A_0{}_{(\mu \nu) a b} + v^A_0{}_{[\mu \nu] a ;b} \right\} \sigma^{;a} \sigma^{;b} + O( \sigma^{3/2} )
\end{eqnarray}
and
\begin{eqnarray}
\label{covTaylorSeries_VA1}
&& V^A_1{}_{\mu \nu} = g{}_\nu^{\phantom{\nu} \nu'} \, V^A_1{}_{\mu \nu'}
= v^A_1{}_{(\mu \nu)} + O( \sigma^{1/2} ).
\end{eqnarray}
\end{subequations}
Here, the explicit expressions of the Taylor coefficients can be determined from the recursion relations \eqref{Recursion_VA}. We have
\begin{subequations}
\allowdisplaybreaks
\label{coefTaylorSeries_vA0_vA1}
\begin{eqnarray}
\label{coefTaylorSeries_vA0()}
&& v^A_0{}_{(\mu \nu)} = (1/2) \, R_{\mu \nu}
+ g_{\mu \nu} \left\{ (1/2) \, m^2 - (1/12) \, R \right\} ,
\\
\label{coefTaylorSeries_vA0[]a}
&& v^A_0{}_{[\mu \nu] a} = (1/6) \, R_{a [\mu ;\nu]} ,
\\
\label{coefTaylorSeries_vA0()ab}
&& v^A_0{}_{(\mu \nu) a b} = (1/6) \, R_{\mu \nu ;(a b)} + (1/12) \, R_{\mu \nu} \, R_{a b}
\nonumber \\ && \quad
+ (1/12) R_{\mu p q (a|} \, R^{\phantom{\nu} p q}_{\nu \phantom{p} \phantom{q} |b)}
\nonumber \\ && \quad
+ g_{\mu \nu} \left\{
  (1/12) \, m^2 \, R_{a b} - (1/40) \, R_{;a b} - (1/120) \, \Box R_{a b}
\vphantom{R^{\phantom{b} p}_b} \right. \nonumber \\ && \qquad \left.
- (1/72) \, R \, R_{a b}
+ (1/90) \, R_{a p} \, R^{\phantom{b} p}_b
\right. \nonumber \\ && \qquad \left.
- (1/180) \, R_{p q} \, R^{\phantom{b} p \phantom{b} q}_{a \phantom{p} b} - (1/180) \, R_{a p q r} \, R^{\phantom{b} p q r}_b
\right\} ,
\\
\label{coefTaylorSeries_vA1()}
&& v^A_1{}_{(\mu \nu)} = (1/4) \, m^2 \, R_{\mu \nu}  - (1/24) \, \Box R_{\mu \nu} - (1/24) \, R \, R_{\mu \nu}
\nonumber \\ && \quad
+ (1/8) \, R_{\mu p} \, R^{\phantom{\nu} p}_{\nu} - (1/48) \, R_{\mu p q r} \, R^{\phantom{\nu} p q r}_\nu
\nonumber \\ && \quad
+ g_{\mu \nu} \left\{
  (1/8) \, m^4 - (1/24) \, m^2 \, R + (1/120) \, \Box R
\right. \nonumber \\ && \qquad \left.
+ (1/288) \, R^2 - (1/720) \, R_{p q} \, R^{p q}
\right. \nonumber \\ && \qquad \left.
+ (1/720) \, R_{p q r s} \, R^{p q r s}
\right\} .
\end{eqnarray}
\end{subequations}
The covariant Taylor series expansions of the biscalar coefficients $V_0(x,x')$ and $V_1(x,x')$ are given by [see Eqs.~\eqref{CovTaylorSeries_Scalar} and \eqref{ScalarTaylorCoef_Sym}]
\begin{subequations}
\label{covTaylorSeries_V0_V1}
\begin{eqnarray}
\label{covTaylorSeries_V0}
&& V_0 = v_0 - \left\{ (1/2) \, v_0{}_{;a} \right\} \sigma^{;a}
\nonumber \\ && \quad
+ \frac{1}{2!} \, v_0{}_{a b} \, \sigma^{;a} \sigma^{;b} + O( \sigma^{3/2} )
\end{eqnarray}
and
\begin{eqnarray}
\label{covTaylorSeries_V1}
&& V_1 = v_1 + O( \sigma^{1/2} ).
\end{eqnarray}
\end{subequations}
Here, the explicit expressions of the Taylor coefficients can be determined from the recursion relations \eqref{Recursion_V}. We have
\begin{subequations}
\allowdisplaybreaks
\label{coefTaylorSeries_v0_v1}
\begin{eqnarray}
\label{coefTaylorSeries_v0}
&& v_0 = (1/2) \, m^2 - (1/12) \, R ,
\\
\label{coefTaylorSeries_v0ab}
&& v_0{}_{a b} =
  (1/12) \, m^2 \, R_{a b} - (1/40) \, R_{;a b} - (1/120) \, \Box R_{a b}
\nonumber \\ && \quad
- (1/72) \, R \, R_{a b}
+ (1/90) \, R_{a p} \, R^{\phantom{b} p}_b
\nonumber \\ && \quad
- (1/180) \, R_{p q} \, R^{\phantom{b} p \phantom{b} q}_{a \phantom{p} b} - (1/180) \, R_{a p q r} \, R^{\phantom{b} p q r}_b ,
\\
\label{coefTaylorSeries_v1}
&& v_1 =
  (1/8) \, m^4 - (1/24) \, m^2 \, R + (1/120) \, \Box R
\nonumber \\ && \quad
+ (1/288) \, R^2 - (1/720) \, R_{p q} \, R^{p q}
\nonumber \\ && \quad
+ (1/720) \, R_{p q r s} \, R^{p q r s} .
\end{eqnarray}
\end{subequations}
In order to obtain the expressions of the Taylor coefficients given by Eqs.~\eqref{coefTaylorSeries_vA0_vA1} and \eqref{coefTaylorSeries_v0_v1}, we have used some of the properties of $\sigma(x,x')$, $g_{\mu \nu'}(x,x')$, $\Delta^{1/2}(x,x')$ mentioned in the Appendix as well as the algebraic properties of the Riemann tensor.

\subsection{State-dependent Hadamard coefficients and associated covariant Taylor series expansions}
\label{Sec.IIId}

\subsubsection{General considerations}
\label{Sec.IIId1}

Unlike the geometrical Hadamard coefficients, the coefficients $W^A_n{}_{\mu \nu'}(x,x')$ and $W_n(x,x')$ are neither uniquely defined nor purely geometrical. Indeed, the coefficient $W^A_0{}_{\mu \nu'}(x,x')$ [respectively, $W_0(x,x')$] is unrestrained by the recursion relations \eqref{EQ_WA} [respectively, by the recursion relations \eqref{EQ_W}]. As a consequence, this is also true for all the coefficients $W^A_n{}_{\mu \nu'}(x,x')$ and $W_n(x,x')$ with $n \ge 1$ and for the sums \eqref{Expansion_WA} and \eqref{Expansion_W}. This arbitrariness is in fact very interesting, and it can be used to encode the quantum state dependence of the theory in the coefficients $W^A_0{}_{\mu \nu'}(x,x')$ and $W_0(x,x')$. Once they have been specified, the coefficients $W^A_n{}_{\mu \nu'}(x,x')$ and $W_n(x,x')$ with $n \ge 1$ as well as the bivector $W^A_{\mu \nu'}(x,x')$ and the biscalar $W(x,x')$ are uniquely determined.

In the following, instead of working with the state-dependent Hadamard coefficients, we shall consider the sums $W^A_{\mu \nu'}(x,x')$ and $W(x,x')$, and, more precisely, we shall use their covariant Taylor series expansions up to order $\sigma^{3/2}$. We have [see Eqs.~\eqref{CovTaylorSeries_Vector} and \eqref{VectorTaylorCoef_Sym}]
\begin{eqnarray}
\label{covTaylorSeries_WA}
&& W^A_{\mu \nu} = g^{\phantom{\nu} \nu' }_{\nu}  W^A_{\mu \nu'}
= s_{\mu \nu}
- \left\{ (1/2) \, s_{\mu \nu ;a} + a_{\mu \nu a} \right\} \sigma^{;a}
\nonumber \\ && \quad
+ \frac{1}{2!} \, \left\{ s_{\mu \nu a b} + a_{\mu \nu a ;b} \right\} \sigma^{;a} \sigma^{;b}
- \frac{1}{3!} \, \left\{ (3/2) \, s_{\mu \nu a b ;c}
\right. \nonumber \\ && \left. \qquad
- (1/4) \, s_{\mu \nu ;a b c} + a_{\mu \nu a b c} \right\} \sigma^{;a} \sigma^{;b} \sigma^{;c}
+ O( \sigma^2 )
\end{eqnarray}
and  [see Eqs.~\eqref{CovTaylorSeries_Scalar} and \eqref{ScalarTaylorCoef_Sym}]
\begin{eqnarray}
\label{covTaylorSeries_W}
&& W = w - \left\{ (1/2) \, w_{;a} \right\} \sigma^{;a} + \frac{1}{2!} \, w_{a b} \, \sigma^{;a} \sigma^{;b}
\nonumber \\ && \quad
- \frac{1}{3!} \, \left\{ (3/2) \, w_{a b ;c}
- (1/4) \, w_{;a b c} \right\} \sigma^{;a} \sigma^{;b}  \sigma^{;c}
+ O( \sigma^2 ) .
\nonumber \\
\end{eqnarray}
In the expansion (\ref{covTaylorSeries_WA}) we have introduced the notations
\begin{subequations}
\label{coefTaylorSeries_s_a}
\begin{eqnarray}
\label{coefTaylorSeries_s}
&& s_{\mu \nu a_1 \cdots a_p} \equiv w^A_{(\mu \nu) a_1 \cdots a_p}
\end{eqnarray}
for the symmetric part of the Taylor coefficients and
\begin{eqnarray}
\label{coefTaylorSeries_a}
&& a_{\mu \nu a_1 \cdots a_p} \equiv w^A_{[\mu \nu] a_1 \cdots a_p}
\end{eqnarray}
\end{subequations}
for their antisymmetric part.

It is important to note that, with practical applications in mind, it is interesting to express some of the Taylor coefficients appearing in Eqs.~\eqref{covTaylorSeries_WA} and \eqref{covTaylorSeries_W} in terms of the bitensors $W^A_{\mu \nu'}(x,x')$ and $W(x,x')$. This can be done by inverting these equations. From Eq.~\eqref{covTaylorSeries_WA}, we obtain
\begin{subequations}
\label{coefTaylorSeries_WA}
\begin{eqnarray}
\label{coefTaylorSeries_WA_1}
&& s_{\mu \nu} (x) = \lim_{x' \to x} W^A_{\mu \nu'} (x,x') ,
\\
\label{coefTaylorSeries_WA_2}
&& a_{\mu \nu a} (x) = \frac{1}{2} \lim_{x' \to x} \left[ W^A_{\mu \nu' ;a'} (x,x') - W^A_{\mu \nu' ;a} (x,x') \right] ,
\\
\label{coefTaylorSeries_WA_3}
&& s_{\mu \nu a b} (x) = \frac{1}{2} \lim_{x' \to x} \left[ W^A_{\mu \nu' ;(a' b')} (x,x') + W^A_{\mu \nu' ;(a b)} (x,x') \right] .
\nonumber \\
\end{eqnarray}
\end{subequations}
(Here, the coefficient $a_{\mu \nu a b c}$ is not relevant because it does not appear in the final expressions of the renormalized stress-energy-tensor operator given in Sec.~\ref{Sec.IVc}). Similarly, from Eq.~\eqref{covTaylorSeries_W}, we straightforwardly establish that
\begin{subequations}
\label{coefTaylorSeries_W}
\begin{eqnarray}
\label{coefTaylorSeries_W_1}
&& w (x) = \lim_{x' \to x} W (x,x') ,
\\
\label{coefTaylorSeries_W_2}
&& w_{a b} (x) = \lim_{x' \to x} W_{;(a' b')} (x,x') .
\end{eqnarray}
\end{subequations}

We shall now rewrite the wave equations \eqref{WEQ_G1A}, \eqref{WEQ_G1Phi} and \eqref{WEQ_G1Gh} as well as the Ward identity \eqref{WardId_G1A} in terms of the Taylor coefficients of $W^A_{\mu \nu}(x,x')$ and $W(x,x')$. To achieve the calculations, we shall use extensively the properties of $\sigma(x,x')$, $g_{\mu \nu'}(x,x')$, $\Delta^{1/2}(x,x')$ mentioned in the Appendix.

\subsubsection{Wave equations}
\label{Sec.IIId2}

By inserting the Hadamard representation \eqref{HadamardRep_G1A} of the Green function $G^{(1)A}_{\mu \nu'} (x,x')$ into the wave equation \eqref{WEQ_G1A}, we obtain a wave equation with source for the state-dependent Hadamard coefficient $W^A_{\mu \nu'}(x,x')$. We have
\begin{eqnarray}
\label{WEQ_WA}
&& g^{\phantom{\rho} \rho'}_{\rho} \left[ g^{\phantom{\mu} \nu}_{\mu} \, \Box_x - R^{\phantom{\mu} \nu}_{\mu} - m^2 g^{\phantom{\mu} \nu}_{\mu} \right] W^A_{\nu \rho'}
\nonumber \\ && \quad
= - 6 \, V^A_1{}_{\mu \rho} - 2 \, g^{\phantom{\rho} \rho'}_{\rho} V^A_1{}_{\mu \rho' ;a} \sigma^{;a} + O( \sigma )
\nonumber \\ && \quad
= - 6 \, v^A_1{}_{(\mu \rho)} + \left( 2 \, v^A_1{}_{(\mu \rho) ;a} + 8 \, v^A_1{}_{[\mu \rho] a} \right) \sigma^{;a} + O( \sigma ) .
\nonumber \\
\end{eqnarray}
Here, we have used the expansions of the geometrical Hadamard coefficients given by Eqs.~\eqref{Expansion_VA} and \eqref{covTaylorSeries_VA0_VA1}. By inserting the expansion \eqref{covTaylorSeries_WA} of $W^A_{\mu \nu}(x,x')$ into the left-hand side of Eq.~\eqref{WEQ_WA}, we find the following relations:
\begin{subequations}
\allowdisplaybreaks
\label{Relation_coefTaylorSeries_WA}
\begin{eqnarray}
\label{Relation_coefTaylorSeries_WA_1}
&& s^{\phantom{\mu \rho \nu} \nu}_{\mu \rho \nu} = R^{p}_{\phantom{p} (\mu}  s_{\rho) p} + m^2 s_{\mu \rho} - 6 \, v^A_1{}_{(\mu \rho)} ,
\\
\label{Relation_coefTaylorSeries_WA_2}
&& s^{\phantom{\mu} \mu \phantom{\nu} \nu}_{\mu \phantom{\mu} \nu} = R^{p q} s_{p q} + m^2 s^{\phantom{p} p}_{p}  - 6 \, v^A_1{}^{\phantom{p} p}_{p} ,
\\
\label{Relation_coefTaylorSeries_WA_3}
&& a^{\phantom{\mu \rho \nu} ;\nu}_{\mu \rho \nu} = - R^{p}_{\phantom{p} [\mu}  s_{\rho] p} ,
\\
\label{Relation_coefTaylorSeries_WA_4}
&& s^{\phantom{\mu \rho \nu a} ;\nu}_{\mu \rho \nu a} =
  (1/4) \, (\Box s_{\mu \rho})_{;a}
+ (1/2) \, R^{p}_{\phantom{p} a ;(\mu} s_{\rho) p}
\nonumber \\ && \quad
- (1/2) \, R^{\phantom{(\mu| a} ;p}_{(\mu| a} s_{|\rho) p}
- (1/2) \, R^{p}_{\phantom{p} (\mu} s_{\rho) p ;a}
\nonumber \\ && \quad
+ (1/2) \, R^{p}_{\phantom{p} a} s_{\mu \rho ;p}
- R^{p \phantom{(\mu|} q}_{\phantom{p} (\mu| \phantom{q} a} s_{|\rho) p ;q}
\nonumber \\ && \quad
- R^{p}_{\phantom{p} (\mu|} a_{p |\rho) a}
- R^{p \phantom{(\mu|} q}_{\phantom{p} (\mu| \phantom{q} a} a_{|\rho) p q}
+ (1/2) \, s^{\phantom{\mu \rho p} p}_{\mu \rho p \phantom{p} ;a}
\nonumber \\ && \quad
- (1/2) \, m^2 s_{\mu \rho ;a}
+ 2 \, v^A_1{}_{(\mu \rho) ;a} ,
\\
\label{Relation_coefTaylorSeries_WA_5}
&& s^{\phantom{\mu} \mu \phantom{\nu a} ;\nu}_{\mu \phantom{\mu} \nu a} =
  (1/4) \, (\Box s^{\phantom{p} p}_{p})_{;a}
+ (1/2) \, R^{q}_{\phantom{q} a} s^{\phantom{p} p}_{p \phantom{p} ;q}
\nonumber \\ && \quad
- (1/2) \, R^{p q} s_{p q ;a}
+ R^{p q r}_{\phantom{p q r} a} a_{p q r}
+ (1/2) \, s^{\phantom{p} p \phantom{q} q}_{p \phantom{p} q \phantom{q} ;a}
\nonumber \\ && \quad
- (1/2) \, m^2 s^{\phantom{p} p}_{p \phantom{p} ;a}
+ 2 \, v^A_1{}^{\phantom{p} p}_{p \phantom{p} ;a} .
\end{eqnarray}
Furthermore, by combining Eq.~\eqref{Relation_coefTaylorSeries_WA_4} with Eq.~\eqref{Relation_coefTaylorSeries_WA_1} and Eq.~\eqref{Relation_coefTaylorSeries_WA_5} with Eq.~\eqref{Relation_coefTaylorSeries_WA_2}, we also establish that
\begin{eqnarray}
\label{Relation_coefTaylorSeries_WA_4_bis}
&& s^{\phantom{\mu \rho \nu a} ;\nu}_{\mu \rho \nu a} =
  (1/4) \, (\Box s_{\mu \rho})_{;a}
+ (1/2) \, R^{p}_{\phantom{p} a ;(\mu} s_{\rho) p}
\nonumber \\ && \quad
- (1/2) \, R^{\phantom{(\mu| a} ;p}_{(\mu| a} s_{|\rho) p}
+ (1/2) \, R^{p}_{\phantom{p} (\mu| ;a} s_{|\rho) p}
\nonumber \\ && \quad
+ (1/2) \, R^{p}_{\phantom{p} a} s_{\mu \rho ;p}
- R^{p \phantom{(\mu|} q}_{\phantom{p} (\mu| \phantom{q} a} s_{|\rho) p ;q}
\nonumber \\ && \quad
- R^{p}_{\phantom{p} (\mu|} a_{p |\rho) a}
- R^{p \phantom{(\mu|} q}_{\phantom{p} (\mu| \phantom{q} a} a_{|\rho) p q}
- v^A_1{}_{(\mu \rho) ;a}
\end{eqnarray}
and
\begin{eqnarray}
\label{Relation_coefTaylorSeries_WA_5_bis}
&& s^{\phantom{\mu} \mu \phantom{\nu a} ;\nu}_{\mu \phantom{\mu} \nu a} =
  (1/4) \, (\Box s^{\phantom{p} p}_{p})_{;a}
+ (1/2) \, R^{q}_{\phantom{q} a} s^{\phantom{p} p}_{p \phantom{p} ;q}
\nonumber \\ & & \quad
+ (1/2) \, R^{p q}_{\phantom{p q} ;a} s_{p q}
+ R^{p q r}_{\phantom{p q r} a} a_{p q r}
- v^A_1{}^{\phantom{p} p}_{p \phantom{p} ;a} .
\end{eqnarray}
\end{subequations}

\textit{Mutatis mutandis}, by inserting the Hadamard representation \eqref{HadamardRep_G1} of the Green function $G^{(1)} (x,x')$ into the wave equation \eqref{WEQ_G1Phi} or \eqref{WEQ_G1Gh}, we obtain a wave equation with source for the state-dependent Hadamard coefficient $W(x,x')$. We have
\begin{eqnarray}
\label{WEQ_W}
&& \left(\Box_x - m^2 \right) W
= - 6 \, V_1 - 2 \, V_1{}_{;a} \sigma^{;a} + O( \sigma )
\nonumber \\ && \hphantom{\left(\Box_x - m^2 \right) W}
= - 6 \, v_1 + 2 \, v_1{}_{;a} \sigma^{;a} + O( \sigma ) .
\end{eqnarray}
Here, we have used the expansions of the geometrical Hadamard coefficients given by  Eqs.~\eqref{Expansion_V} and \eqref{covTaylorSeries_V0_V1}. By inserting the expansion \eqref{covTaylorSeries_W} of $W(x,x')$ into the left-hand side of Eq.~\eqref{WEQ_W}, we find the following relations:
\begin{subequations}
\label{Relation_coefTaylorSeries_W}
\begin{eqnarray}
\label{Relation_coefTaylorSeries_W_1}
&& w^{\phantom{\rho} \rho}_\rho = m^2 w  - 6 \, v_1 ,
\\
\label{Relation_coefTaylorSeries_W_2}
&& w^{\phantom{\rho a} ;\rho}_{\rho a} = (1/4) \, (\Box w)_{;a} + (1/2) \, w^{\phantom{p} p}_{p \phantom{p} ;a}
\nonumber \\ && \quad
+ (1/2) \, R^p_{\phantom{p} a} w_{;p} - (1/2) \, m^2 w_{;a} + 2 \, v_1{}_{;a} .
\end{eqnarray}
Furthermore, by combining Eq.~\eqref{Relation_coefTaylorSeries_W_2} with Eq.~\eqref{Relation_coefTaylorSeries_W_1}, we also establish that
\begin{eqnarray}
\label{Relation_coefTaylorSeries_W_2_bis}
&& w^{\phantom{\rho a} ;\rho}_{\rho a} = (1/4) \, (\Box w)_{;a} + (1/2) \, R^p_{\phantom{p} a} w_{;p} - v_1{}_{;a} .
\end{eqnarray}
\end{subequations}

\subsubsection{Ward identities}
\label{Sec.IIId3}

The first Ward identity given by Eq.~\eqref{WardId_G1A} expressed in terms of the Hadamard representation of the Green functions $G^{(1)A}_{\mu \nu'} (x,x')$ [see Eq.~\eqref{HadamardRep_G1A}] and $G^{(1)} (x,x')$ [see Eq.~\eqref{HadamardRep_G1}] permits us to write a relation between the geometrical Hadamard coefficients $V^A_{\mu \nu'}(x,x')$ and $V(x,x')$ as well as another one between the state-dependent Hadamard coefficients $W^A_{\mu \nu'}(x,x')$ and $W(x,x')$. We obtain
\begin{eqnarray}
\label{WardId_VA_V}
&& g^{\phantom{\nu} \nu'}_{\nu} \left( V^A_{\mu \nu'}{}^{;\mu} + V_{;\nu'} \right) = 0
\end{eqnarray}
which is an identity between the geometrical Taylor coefficients \eqref{coefTaylorSeries_vA0()}--\eqref{coefTaylorSeries_vA1()} and  \eqref{coefTaylorSeries_v0}--\eqref{coefTaylorSeries_v1} and
\begin{eqnarray}
\label{WardId_WA_W}
&& g^{\phantom{\nu} \nu'}_{\nu} \left( W^A_{\mu \nu'}{}^{;\mu} + W_{;\nu'} \right)
\nonumber \\ && \quad
= - {V^A_1}_{\mu \nu} \sigma^{; \mu} + V_1 \sigma_{; \nu} + O \left( \sigma \right)
\nonumber \\ && \quad
= - \left( {v^A_1}_{(\nu a)} - v_1 g_{\nu a} \right) \sigma^{; a} + O \left( \sigma \right) .
\end{eqnarray}
To establish Eq.~\eqref{WardId_WA_W}, we have used the expansions of the geometrical Hadamard coefficients given by Eqs.~\eqref{Expansion_VA}, \eqref{Expansion_V}, \eqref{covTaylorSeries_VA0_VA1} and \eqref{covTaylorSeries_V0_V1}. By inserting the expansions \eqref{covTaylorSeries_WA} of $W^A_{\mu \nu}(x,x')$ and \eqref{covTaylorSeries_W} of $W(x,x')$ into the left-hand side of Eq.~\eqref{WardId_WA_W}, we find the following relations:
\begin{subequations}
\label{Relation_coefTaylorSeries_WA_W}
\begin{eqnarray}
\label{Relation_coefTaylorSeries_WA_W_1}
&& a^{\phantom{\mu \nu} \mu}_{\mu \nu} = (1/2) \, s^{\phantom{p \nu} ;p}_{p \nu} + (1/2) \, w_{;\nu} ,
\\
\label{Relation_coefTaylorSeries_WA_W_2}
&& s^{\phantom{\mu \nu} \mu}_{\mu \nu \phantom{\mu} a} =
  (1/2) \, s^{\phantom{p \nu} ;p}_{p \nu \phantom{;p} a}
+ (1/2) \, R^{p}_{\phantom{p} a} s_{\nu p}
+ a^{p}_{\phantom{p} \nu [a ;p]}
\nonumber \\ && \quad
+ w_{\nu a} - v^A_1{}_{(\nu a)} + v_1 g_{\nu a} .
\end{eqnarray}
Furthermore, by combining Eq.~\eqref{Relation_coefTaylorSeries_WA_W_2} with Eq.~\eqref{Relation_coefTaylorSeries_WA_W_1}, we also establish that
\begin{eqnarray}
\label{Relation_coefTaylorSeries_WA_W_2_bis}
&& s^{\phantom{\mu \nu} \mu}_{\mu \nu \phantom{\mu} a} =
  (1/4) \, s^{\phantom{p \nu} ;p}_{p \nu \phantom{;p} a}
+ (1/2) \, R^{p}_{\phantom{p} a} s_{\nu p}
+ (1/2) \, a^{\phantom{p \nu a} ;p}_{p \nu a}
\nonumber \\ && \quad
- (1/4) \, w_{;\nu a} + w_{\nu a} - {v^A_1}_{(\nu a)} + v_1 g_{\nu a} .
\end{eqnarray}
\end{subequations}

Of course, the second Ward identity given by Eq.~\eqref{WardId_G1Phi} provides trivially the equality of the Taylor coefficients of the Hadamard coefficients associated with the auxiliary scalar field and the ghost fields. We have
\begin{eqnarray}
\label{WardId_VPhi_VGh}
&& V^\Phi = V^\mathrm{Gh}
\end{eqnarray}
and
\begin{eqnarray}
\label{WardId_WPhi_WGh}
&& W^\Phi = W^\mathrm{Gh} .
\end{eqnarray}

\section{Renormalized stress-energy tensor of Stueckelberg electromagnetism}
\label{Sec.IV}

\subsection{Stress-energy tensor}
\label{Sec.IVa}

The functional derivation of the quantum action of the Stueckelberg theory with respect to the metric tensor $g_{\mu \nu}$ permits us to obtain the associated stress-energy tensor $T_{\mu \nu}$. By definition, we have
\begin{equation}
\label{SET_def}
T^{\mu \nu} = \frac{2}{\sqrt{-g}} \, \frac{\delta}{\delta g_{\mu \nu}} \, S[ A_\mu, \Phi, C, C^\ast, g_{\mu \nu} ] ,
\end{equation}
and its explicit expression can be obtain by using that, in the elementary variation
\begin{equation}
\label{varTensMet1}
g_{\mu \nu} \to g_{\mu \nu} + \delta g_{\mu \nu}
\end{equation}
of the metric tensor, we have (see, for example, Ref.~\cite{Barth:1983hb})
\begin{subequations}
\label{varTensMet3}
\begin{eqnarray}
\label{varTensMet3a}
&& g^{\mu \nu } \to g^{\mu \nu } + \delta g^{\mu \nu } ,
\\
\label{varTensMet3b}
&& \sqrt{-g} \to \sqrt{-g} + \delta\sqrt{-g} ,
\\
\label{varTensMet3c}
&& \Gamma^\rho_{\phantom{\rho} \mu \nu} \to \Gamma^\rho_{\phantom{\rho} \mu \nu} + \delta \Gamma^\rho_{\phantom{\rho} \mu \nu}
\end{eqnarray}
with
\begin{eqnarray}
\label{varTensMet4a}
&& \delta g^{\mu \nu} = - g^{\mu \rho} g^{\nu \sigma} \delta g_{\rho \sigma} ,
\\
\label{varTensMet4b}
&&  \delta\sqrt{-g} = \frac{1}{2} \sqrt{-g} \, g^{\mu \nu} \delta g_{\mu \nu} ,
\\
\label{varTensMet4c}
&&  \delta \Gamma^\rho_{\phantom{\rho} \mu \nu} = \frac{1}{2} \left( - \delta g^{\phantom{\mu \nu} ;\rho}_{\mu \nu} + \delta g^\rho_{\phantom{\rho} \mu ;\nu} + \delta g^\rho_{\phantom{\rho} \nu ;\mu} \right) .
\end{eqnarray}
\end{subequations}

The stress-energy tensor derived from the action \eqref{Action_Stueck_Quant_v1} is given by
\begin{equation}
\label{SET_v1}
T^{\mu \nu} = T^{\mu \nu}_\mathrm{cl} + T^{\mu \nu}_\mathrm{GB} + T^{\mu \nu}_\mathrm{Gh} ,
\end{equation}
where the contributions of the classical and gauge-breaking parts take the forms
\begin{subequations}
\allowdisplaybreaks
\label{SET_v1_cl_GB_Gh}
\begin{eqnarray}
\label{SET_v1_cl}
&& T^{\mu \nu}_\mathrm{cl} = F^{\mu}_{\phantom{\mu} \rho} F^{\nu \rho} + m^2 A^\mu A^\nu
\nonumber \\ && \qquad
+ \nabla^\mu \Phi \nabla^\nu \Phi + 2 \, m A^{(\mu} \nabla^{\nu)} \Phi
\nonumber \\ && \qquad
- (1/4) \, g^{\mu \nu} \left\{
F_{\rho \tau}F^{\rho \tau} + 2 \, m^2 A_\rho A^\rho
\right. \nonumber \\ && \left. \qquad \quad
+ 2 \, \nabla_\rho \Phi \nabla^\rho \Phi + 4 \, m  A_\rho \nabla^\rho \Phi
\vphantom{m^2} \right\}
\nonumber \\ && \phantom{T^{\mu \nu}_A}
= \nabla_\rho A^{\mu} \nabla^\rho A^{\nu} - 2 \, \nabla_\rho A^{(\mu} \nabla^{\nu)} A^\rho + \nabla^\mu A_{\rho} \nabla^\nu A^\rho
\nonumber \\ && \qquad
+ m^2 A^\mu A^\nu + \nabla^\mu \Phi \nabla^\nu \Phi + 2 \, m A^{(\mu} \nabla^{\nu)} \Phi
\nonumber \\ && \qquad
- (1/2) \, g^{\mu \nu} \left\{
\nabla_\rho A_\tau \nabla^\rho A^\tau - \nabla_\rho A_\tau \nabla^\tau A^\rho
\vphantom{m^2} \right. \nonumber \\ && \left. \qquad \quad
+ m^2 A_\rho A^\rho + \nabla_\rho \Phi \nabla^\rho \Phi + 2 \, m  A_\rho \nabla^\rho \Phi
\right\}
\end{eqnarray}
and
\begin{eqnarray}
\label{SET_v1_GB}
&& T^{\mu \nu}_\mathrm{GB} = - 2 \, A^{(\mu} \nabla^{\nu)} \nabla_\rho A^\rho - 2 \, m A^{(\mu} \nabla^{\nu)} \Phi
\nonumber \\ && \qquad
- (1/2) \, g^{\mu \nu} \left\{
- 2 \, A_\rho \nabla^\rho \nabla_\tau A^\tau - \left( \nabla_\rho A^\rho \right)^2
\right. \nonumber \\ && \left. \qquad \quad
+ m^2 \Phi^2 - 2 \, m  A_\rho \nabla^\rho \Phi
\vphantom{\left( \nabla_\rho A^\rho \right)^2} \right\} ,
\end{eqnarray}
while the contribution associated with the ghost fields is given by
\begin{eqnarray}
\label{SET_v1_Gh}
&& T^{\mu \nu}_\mathrm{Gh} = - 2 \, \nabla^{(\mu|} C^\ast \nabla^{|\nu)} C
\nonumber \\ && \phantom{T^{\mu \nu}_\mathrm{Gh}}
+ g^{\mu \nu} \left\{ \nabla_\rho C^\ast \nabla^\rho C + m^2 C^\ast C  \right\} .
\end{eqnarray}
\end{subequations}
We can note the existence of terms coupling the fields $A_\mu$ and $\Phi$ in the expression of $T^{\mu \nu}_\mathrm{cl}$ [see Eq.~\eqref{SET_v1_cl}] as well as in the expression of $T^{\mu \nu}_\mathrm{GB}$ [see Eq.~\eqref{SET_v1_GB}].

We also give an alternative expression for the stress-energy tensor which can be derived from the action \eqref{Action_Stueck_Quant_v2} or by summing $T^{\mu \nu}_\mathrm{cl}$ and $T^{\mu \nu}_\mathrm{GB}$. This eliminates any coupling between the fields $A_\mu$ and $\Phi$ and permits us to straightforwardly infer that the stress-energy tensor has three independent contributions corresponding to the massive vector field $A_\mu$, the auxiliary scalar field $\Phi$ and the ghost fields $C$ and $C^\ast$. We can write
\begin{equation}
\label{SET_v2}
T^{\mu \nu} = T^{\mu \nu}_A + T^{\mu \nu}_\Phi + T^{\mu \nu}_\mathrm{Gh}
\end{equation}
with
\begin{subequations}
\allowdisplaybreaks
\label{SET_v2_A_Phi}
\begin{eqnarray}
\label{SET_v2_A}
&& T^{\mu \nu}_A = F^{\mu}_{\phantom{\mu} \rho} F^{\nu \rho} + m^2 A^\mu A^\nu - 2 \, A^{(\mu} \nabla^{\nu)} \nabla_\rho A^\rho
\nonumber \\ && \qquad
- (1/4) \, g^{\mu \nu} \left\{
F_{\rho \tau}F^{\rho \tau} + 2 \, m^2 A_\rho A^\rho
\vphantom{\left( \nabla_\rho A^\rho \right)^2} \right. \nonumber \\ && \left. \qquad \quad
- 4 \, A_\rho \nabla^\rho \nabla_\tau A^\tau - 2 \, \left( \nabla_\rho A^\rho \right)^2
\right\}
\nonumber \\ && \phantom{T^{\mu \nu}_A}
= \nabla_\rho A^{\mu} \nabla^\rho A^{\nu} - 2 \, \nabla_\rho A^{(\mu} \nabla^{\nu)} A^\rho + \nabla^\mu A_{\rho} \nabla^\nu A^\rho
\nonumber \\ && \qquad
+ m^2 A^\mu A^\nu - 2 \, A^{(\mu} \nabla^{\nu)} \nabla_\rho A^\rho
\nonumber \\ && \qquad
- (1/2) \, g^{\mu \nu} \left\{
\nabla_\rho A_\tau \nabla^\rho A^\tau - \nabla_\rho A_\tau \nabla^\tau A^\rho
\vphantom{\left( \nabla_\rho A^\rho \right)^2} \right. \nonumber \\ && \left. \qquad \quad
+ m^2 A_\rho A^\rho - 2 \, A_\rho \nabla^\rho \nabla_\tau A^\tau - \left( \nabla_\rho A^\rho \right)^2
\right\}
\end{eqnarray}
and
\begin{eqnarray}
\label{SET_v2_Phi}
&& \hspace{-9mm} T^{\mu \nu}_\Phi = \nabla^\mu \Phi \nabla^\nu \Phi - (1/2) \, g^{\mu \nu} \left\{ \nabla_\rho \Phi \nabla^\rho \Phi + m^2 \Phi^2 \right\} ,
\end{eqnarray}
\end{subequations}
while the contribution associated with the ghost fields remains unchanged [see Eq.~\eqref{SET_v1_Gh}].

By construction, the stress-energy tensor \eqref{SET_def} [see also its explicit expressions \eqref{SET_v1} and \eqref{SET_v2}] is conserved, i.e.,
\begin{equation}
\label{divSET}
\nabla_\nu T^{\mu \nu} = 0 .
\end{equation}
Indeed, this property is due to the invariance of the action \eqref{Action_Stueck_Quant_v1} or \eqref{Action_Stueck_Quant_v2} under spacetime diffeomorphisms and therefore under the infinitesimal coordinate transformation
\begin{equation}
\label{InfTransfotrmation_Coordinate}
x^\mu \to x^\mu + \epsilon^\mu \quad \mathrm{with} \quad |\epsilon^\mu| \ll 1 .
\end{equation}
Under this transformation, the vector, scalar and ghost fields as well as the background metric transform as
\begin{subequations}
\label{InfTransfotrmation_Fields}
\begin{eqnarray}
\label{InfTransfotrmation_A}
&& A_\mu \to A_\mu + \delta A_\mu ,
\\
\label{InfTransfotrmation_Phi}
&& \Phi \to \Phi + \delta \Phi ,
\\
\label{InfTransfotrmation_C}
&& C \to C + \delta C ,
\\
\label{InfTransfotrmation_C*}
&& C^\ast \to C^\ast + \delta C^\ast ,
\\
\label{InfTransfotrmation_g}
&& g_{\mu \nu} \to g_{\mu \nu} + \delta g_{\mu \nu} .
\end{eqnarray}
\end{subequations}
The variations associated with the field transformations \eqref{InfTransfotrmation_Fields} are obtained by Lie derivation with respect to the vector $-\epsilon^\mu$:
\begin{subequations}
\label{derLie_Fields}
\begin{eqnarray}
\label{derLie_A}
&& \delta A_\mu = \mathcal{L}_{-\epsilon} A_\mu = - \epsilon^\rho \nabla_\rho A_\mu - (\nabla_\mu \epsilon^\rho) A_\rho ,
\\
\label{derLie_Phi}
&& \delta \Phi = \mathcal{L}_{-\epsilon} \Phi = - \epsilon^\rho \nabla_\rho \Phi ,
\\
\label{derLie_C}
&& \delta C = \mathcal{L}_{-\epsilon} C = - \epsilon^\rho \nabla_\rho C ,
\\
\label{derLie_C*}
&& \delta C^\ast = \mathcal{L}_{-\epsilon} C^\ast = - \epsilon^\rho \nabla_\rho C^\ast ,
\\
\label{derLie_g}
&& \delta g_{\mu \nu} = \mathcal{L}_{-\epsilon} g_{\mu \nu} = - \nabla_\mu \epsilon_\nu - \nabla_\nu \epsilon_\mu .
\end{eqnarray}
\end{subequations}
The invariance of the action \eqref{Action_Stueck_Quant_v1} or \eqref{Action_Stueck_Quant_v2} leads to
\begin{eqnarray}
\label{InvAction}
&& \int_\mathcal{M} d^4 x \sqrt{-g} \, \left[
  \left( \frac{1}{\sqrt{-g}} \, \frac{\delta S} {\delta A_\mu} \right) \delta A_\mu
+ \left( \frac{1}{\sqrt{-g}} \, \frac{\delta S} {\delta \Phi} \right) \delta \Phi
\right. \nonumber \\ && \left. \qquad
+ \left( \frac{1}{\sqrt{-g}} \, \frac{\delta_\mathrm{R} S} {\delta C} \right) \delta C
+ \delta C^\ast \left( \frac{1}{\sqrt{-g}} \, \frac{\delta_\mathrm{L} S} {\delta C^\ast} \right)
\right. \nonumber \\ && \left. \qquad
+ \frac{1}{2} \, \left( \frac{2}{\sqrt{-g}} \, \frac{\delta S} {\delta g_{\mu \nu }} \right) \delta g_{\mu \nu}
\right] = 0
\end{eqnarray}
which implies
\begin{eqnarray}
\label{divSET_Expr}
&& \nabla_\nu T^{\mu \nu} =
\nonumber \\ && \quad \phantom{+}
  \left[ \nabla^\mu A_\alpha - \nabla_\alpha A^\mu -A^\mu \nabla_\nu \right] \left( \Box A^\nu  - R^{\nu}_{\phantom{\nu} \rho} A^\rho - m^2 A^\nu \right)
\nonumber \\ && \quad
+ \left[ \nabla^\mu \Phi \right] \left( \Box \Phi - m^2 \Phi \right)
\nonumber \\ && \quad
- \left( \Box C^\ast - m^2 C^\ast \right) \left[ \nabla^\mu C \right]
- \left[ \nabla^\mu C^\ast \right] \left( \Box C - m^2 C \right)
\end{eqnarray}
by using Eq.~\eqref{derLie_Fields} as well as Eqs.~\eqref{DerivFunct_Stueck_champA_Quant}--\eqref{DerivFunct_Stueck_ghosts_Quant_b}. From the wave equations associated with the massive vector field $A_\mu$ [see Eq.~\eqref{DerivFunct_WEQ_Stueck_champA_Quant}], the auxiliary scalar field $\Phi$ [see Eq.~\eqref{DerivFunct_WEQ_Stueck_champPhi_Quant}] and the ghost fields $C$ and $C^\ast$ [see Eq.~\eqref{DerivFunct_WEQ_Stueck_ghosts_Quant}], we then obtain Eq.~\eqref{divSET}.

\subsection{Expectation value of the stress-energy tensor}
\label{Sec.IVb}

At the quantum level, all the fields involved in the Stueckelberg theory as well as  the associated stress-energy tensor [see Eqs.~\eqref{SET_v1} and \eqref{SET_v2}] are operators. From now on, we shall denote the stress-energy-tensor operator by $\widehat{T}_{\mu \nu}$ and we shall focus on the quantity $\langle \psi | \widehat{T}_{\mu \nu} | \psi \rangle$ which denotes its expectation value with respect to the Hadamard quantum state $|\psi \rangle$ discussed in Sec.~\ref{Sec.III}.

The expectation value $\langle \psi | \widehat{T}_{\mu \nu} | \psi \rangle$ corresponding to the expression \eqref{SET_v1} of the stress-energy tensor is decomposed as follows:
\begin{equation}
\label{qSET_v1}
\langle \psi | \widehat{T}_{\mu \nu} | \psi \rangle =
  \langle \psi | \widehat{T}^{\mathrm{cl}}_{\mu \nu} | \psi \rangle
+ \langle \psi | \widehat{T}^\mathrm{GB}_{\mu \nu} | \psi \rangle
+ \langle \psi | \widehat{T}^\mathrm{Gh}_{\mu \nu} | \psi \rangle .
\end{equation}
The three terms in the right-hand side of this equation are explicitly given by
\begin{subequations}
\allowdisplaybreaks
\label{qSET_v1_cl_GB_Gh}
\begin{eqnarray}
\label{qSET_v1_cl}
&& \langle \psi | \widehat{T}^{\mathrm{cl}}_{\mu \nu}(x) | \psi \rangle =
  \frac{1}{2} \lim_{x' \to x} \mathcal{T}^{\mathrm{cl}_A}_{\mu \nu}{}^{\rho \sigma'} (x,x') \left[ G^{(1)A}_{\rho \sigma'} (x,x') \right]
\nonumber \\ && \phantom{\langle \psi | \widehat{T}^{\mathrm{cl}}_{\mu \nu} | \psi \rangle}
+ \frac{1}{2} \lim_{x' \to x} \mathcal{T}^{\mathrm{cl}_\Phi}_{\mu \nu} (x,x') \left[ G^{(1)\Phi} (x,x') \right] ,
\end{eqnarray}
\begin{eqnarray}
\label{qSET_v1_GB}
&& \langle \psi | \widehat{T}^\mathrm{GB}_{\mu \nu}(x) | \psi \rangle =
  \frac{1}{2} \lim_{x' \to x} \mathcal{T}^{\mathrm{GB}_A}_{\mu \nu}{}^{\rho \sigma'} (x,x') \left[ G^{(1)A}_{\rho \sigma'} (x,x') \right]
\nonumber \\ && \phantom{\langle \psi | \widehat{T}^\mathrm{GB}_{\mu \nu} | \psi \rangle}
+ \frac{1}{2} \lim_{x' \to x} \mathcal{T}^{\mathrm{GB}_\Phi}_{\mu \nu} (x,x') \left[ G^{(1)\Phi} (x,x') \right]
\end{eqnarray}
and
\begin{eqnarray}
\label{qSET_v1_Gh}
&& \hspace{-10mm} \langle \psi | \widehat{T}^\mathrm{Gh}_{\mu \nu}(x) | \psi \rangle = \frac{1}{2} \lim_{x' \to x} \mathcal{T}^{\mathrm{Gh}}_{\mu \nu} (x,x') \left[ G^{(1)\mathrm{Gh}} (x,x') \right] ,
\end{eqnarray}
\end{subequations}
where $\mathcal{T}^{\mathrm{cl}_A}_{\mu \nu}{}^{\rho \sigma'}$, $\mathcal{T}^{\mathrm{cl}_\Phi}_{\mu \nu}$, $\mathcal{T}^{\mathrm{GB}_A}_{\mu \nu}{}^{\rho \sigma'}$, $\mathcal{T}^{\mathrm{GB}_\Phi}_{\mu \nu}$ and $\mathcal{T}^{\mathrm{Gh}}_{\mu \nu}$ are the differential operators constructed by point splitting from the expressions \eqref{SET_v1_cl}, \eqref{SET_v1_GB} and \eqref{SET_v1_Gh}. We have
\begin{subequations}
\allowdisplaybreaks
\label{qSET_v1_do_cl_GB_Gh}
\begin{eqnarray}
\label{qSET_v1_do_clA}
&& \mathcal{T}^{\mathrm{cl}_A}_{\mu \nu}{}^{\rho \sigma'} =
  g^{\phantom{\nu} \alpha'}_{\nu} g^{\rho \sigma'} \nabla_{\mu} \nabla_{\alpha'}
+ g_{\mu}^{\phantom{\mu} \rho} g_{\nu}^{\phantom{\nu} \sigma'} g^{\alpha \beta'} \nabla_{\alpha} \nabla_{\beta'}
\nonumber \\ && \quad
- 2 \, g_{\mu}^{\phantom{\mu} \rho} g_{\nu}^{\phantom{\nu} \alpha'} g^{\beta \sigma'} \nabla_{\beta} \nabla_{\alpha'}
+ m^2 g_{\mu}^{\phantom{\mu} \rho} g_{\nu}^{\phantom{\nu} \sigma'}
\nonumber \\ && \quad
- \frac{1}{2} \, g_{\mu \nu} \left\{
  g^{\rho \sigma'} g^{\alpha \beta'} \nabla_{\alpha} \nabla_{\beta'}
- g^{\rho \alpha'} g^{\beta \sigma'} \nabla_{\beta} \nabla_{\alpha'}
\right. \nonumber \\ && \qquad \left.
+ m^2 g^{\rho \sigma'}
\right\} ,
\end{eqnarray}
\begin{eqnarray}
\label{qSET_v1_do_clPhi}
&& \mathcal{T}^{\mathrm{cl}_\Phi}_{\mu \nu} =
  g_{\nu}^{\phantom{\nu} \nu'} \nabla_{\mu} \nabla_{\nu'}
- \frac{1}{2} \, g_{\mu \nu} \left\{
g^{\alpha \beta'} \nabla_{\alpha} \nabla_{\beta'}
\right\} ,
\end{eqnarray}
\begin{eqnarray}
\label{qSET_v1_do_GBA}
&& \mathcal{T}^{\mathrm{GB}_A}_{\mu \nu}{}^{\rho \sigma'} =
- 2 \, g_{\mu}^{\phantom{\mu} \rho} g_{\nu}^{\phantom{\nu} \alpha'} \nabla_{\alpha'} \nabla^{\sigma'}
\nonumber \\ && \quad
- \frac{1}{2} \, g_{\mu \nu} \left\{
- \nabla^{\rho} \nabla^{\sigma'} - 2 \, g^{\rho \alpha'} \nabla_{\alpha'} \nabla^{\sigma'}
\right\} ,
\end{eqnarray}
\begin{eqnarray}
\label{qSET_v1_do_GBPhi}
&& \mathcal{T}^{\mathrm{GB}_\Phi}_{\mu \nu} =
- \frac{1}{2} \, m^2 \, g_{\mu \nu}
\end{eqnarray}
and
\begin{eqnarray}
\label{qSET_v1_do_Gh}
&& \hspace{-10mm} \mathcal{T}^{\mathrm{Gh}}_{\mu \nu} = - 2 \, g_{\nu}^{\phantom{\nu} \nu'} \nabla_{\mu} \nabla_{\nu'}
+ g_{\mu \nu} \left\{
g^{\alpha \beta'} \nabla_{\alpha} \nabla_{\beta'} + m^2
\right\} .
\end{eqnarray}
\end{subequations}
It should be noted that we have not included in Eqs.~\eqref{qSET_v1_cl} and \eqref{qSET_v1_GB} the contributions which can be obtained by point splitting from the terms coupling $A_\mu$ and $\Phi$ in Eqs.~\eqref{SET_v1_cl} and \eqref{SET_v1_GB}. Such contributions are not present because, due to the absence of coupling between $A_\mu$ and $\Phi$ in the quantum action \eqref{Action_Stueck_Quant_v2}, two-point correlation functions involving both $A_\mu$ and $\Phi$ vanish identically. It should be noted that the absence of these contributions can be also justified in another way: in the quantum stress-energy-tensor operator \eqref{SET_v2}, any coupling between $A_\mu$ and $\Phi$ has disappeared.

Here, some remarks are in order:
\begin{itemize}

  \item[(i)] When we use the point-splitting method, it is more convenient to define the expectation value $\langle \psi | \widehat{T}_{\mu \nu} | \psi \rangle$ from Hadamard Green functions rather than from Feynman propagators. Indeed, this avoids us having to deal with additional singular terms due to the time-ordered product.

  \item[(ii)] Of course, because of the short-distance behavior of the Hadamard Green functions, the expressions \eqref{qSET_v1_cl_GB_Gh} as well as the expectation value $\langle \psi | \widehat{T}_{\mu \nu} | \psi \rangle$ given in Eq.~\eqref{qSET_v1} are divergent and therefore meaningless. In Sec.~\ref{Sec.IVc} we will regularize these quantities.

  \item[(iii)] Even if the formal expression \eqref{qSET_v1} of the expectation value of the stress-energy-tensor operator is divergent, it is interesting to note that
\begin{equation}
\label{qSET_v1_GB+Gh=0}
\langle \psi | \widehat{T}^\mathrm{GB}_{\mu \nu} | \psi \rangle + \langle \psi | \widehat{T}^\mathrm{Gh}_{\mu \nu} | \psi \rangle = 0 .
\end{equation}
Indeed, from the definitions \eqref{qSET_v1_GB} and \eqref{qSET_v1_Gh}, we can obtain Eq.~\eqref{qSET_v1_GB+Gh=0} by using Eqs.~\eqref{WardId_G1A} and \eqref{WardId_G1Phi} as well as the wave equation \eqref{WEQ_G1Gh}. It should be noted that, as a consequence of Eq.~\eqref{qSET_v1_GB+Gh=0},  Eq.~\eqref{qSET_v1} reduces to
\begin{equation}
\label{qSET_v1_simplified}
\langle \psi | \widehat{T}_{\mu \nu} | \psi \rangle =
  \langle \psi | \widehat{T}^{\mathrm{cl}}_{\mu \nu} | \psi \rangle .
\end{equation}

\end{itemize}

We can also give the alternative expression of the expectation value $\langle \psi | \widehat{T}_{\mu \nu} | \psi \rangle$ obtained from Eq.~\eqref{SET_v2}. It takes the following form
\begin{equation}
\label{qSET_v2}
\langle \psi | \widehat{T}_{\mu \nu} | \psi \rangle = \langle \psi | \widehat{T}^A_{\mu \nu} | \psi \rangle + \langle \psi | \widehat{T}^\Phi_{\mu \nu} | \psi \rangle + \langle \psi | \widehat{T}^\mathrm{Gh}_{\mu \nu} | \psi \rangle ,
\end{equation}
where the contributions associated with the massive vector field $A_\mu$ and the auxiliary scalar field $\Phi$ are separated and given by
\begin{subequations}
\allowdisplaybreaks
\label{qSET_v1_A_Phi_Gh}
\begin{eqnarray}
\label{qSET_v1_A}
&& \hspace{-10mm} \langle \psi | \widehat{T}^A_{\mu \nu}(x) | \psi \rangle = \frac{1}{2} \lim_{x' \to x} \mathcal{T}^A_{\mu \nu}{}^{\rho \sigma'} (x,x') \left[ G^{(1)A}_{\rho \sigma'} (x,x') \right]
\end{eqnarray}
and
\begin{eqnarray}
\label{qSET_v1_Phi}
&& \hspace{-10mm} \langle \psi | \widehat{T}^\Phi_{\mu \nu}(x) | \psi \rangle = \frac{1}{2} \lim_{x' \to x} \mathcal{T}^{\Phi}_{\mu \nu} (x,x') \left[ G^{(1)\Phi} (x,x') \right] .
\end{eqnarray}
\end{subequations}
Here, the differential operators $\mathcal{T}^A_{\mu \nu}{}^{\rho \sigma'}$ and $\mathcal{T}^{\Phi}_{\mu \nu}$ are constructed by point splitting from the expressions \eqref{SET_v2_A} and \eqref{SET_v2_Phi}. We have
\begin{subequations}
\label{qSET_v2_do_A_Phi}
\begin{eqnarray}
\label{qSET_v2_do_A}
&& \mathcal{T}^A_{\mu \nu}{}^{\rho \sigma'} =
  g^{\phantom{\nu} \alpha'}_{\nu} g^{\rho \sigma'} \nabla_{\mu} \nabla_{\alpha'}
+ g_{\mu}^{\phantom{\mu} \rho} g_{\nu}^{\phantom{\nu} \sigma'} g^{\alpha \beta'} \nabla_{\alpha} \nabla_{\beta'}
\nonumber \\ && \quad
- 2 \, g_{\mu}^{\phantom{\mu} \rho} g_{\nu}^{\phantom{\nu} \alpha'} g^{\beta \sigma'} \nabla_{\beta} \nabla_{\alpha'}
+ m^2 g_{\mu}^{\phantom{\mu} \rho} g_{\nu}^{\phantom{\nu} \sigma'}
\nonumber \\ && \quad
- 2 \, g_{\mu}^{\phantom{\mu} \rho} g_{\nu}^{\phantom{\nu} \alpha'} \nabla_{\alpha'} \nabla^{\sigma'}
\nonumber \\ && \quad
- \frac{1}{2} \, g_{\mu \nu} \left\{
g^{\rho \sigma'} g^{\alpha \beta'} \nabla_{\alpha} \nabla_{\beta'}
- g^{\rho \alpha'} g^{\beta \sigma'} \nabla_{\beta} \nabla_{\alpha'}
\right. \nonumber \\ && \qquad \left.
+ m^2 g^{\rho \sigma'}
- \nabla^{\rho} \nabla^{\sigma'} - 2 \, g^{\rho \alpha'} \nabla_{\alpha'} \nabla^{\sigma'}
\right\}
\end{eqnarray}
and
\begin{eqnarray}
\label{qSET_v2_do_Phi}
&& \hspace{-7mm} \mathcal{T}^{\Phi}_{\mu \nu} =
  g_{\nu}^{\phantom{\nu} \nu'} \nabla_{\mu} \nabla_{\nu'}
- \frac{1}{2} \, g_{\mu \nu} \left\{
g^{\alpha \beta'} \nabla_{\alpha} \nabla_{\beta'} + m^2
\right\} .
\end{eqnarray}
\end{subequations}
It should be noted that the expressions \eqref{qSET_v1} and \eqref{qSET_v2} of the expectation value $\langle \psi | \widehat{T}_{\mu \nu} | \psi \rangle$ are identical because the various differential operators $\mathcal{T}^{\mathrm{cl}_A}_{\mu \nu}{}^{\rho \sigma'}$, $\mathcal{T}^{\mathrm{cl}_\Phi}_{\mu \nu}$, $\mathcal{T}^{\mathrm{GB}_A}_{\mu \nu}{}^{\rho \sigma'}$ and $\mathcal{T}^{\mathrm{GB}_\Phi}_{\mu \nu}$ appearing in \eqref{qSET_v1_cl_GB_Gh} and $\mathcal{T}^A_{\mu \nu}{}^{\rho \sigma'}$ and $\mathcal{T}^{\Phi}_{\mu \nu}$ appearing in \eqref{qSET_v1_A_Phi_Gh} are related by
\begin{subequations}
\label{qSET_do}
\begin{eqnarray}
\label{qSET_do_A}
&& \mathcal{T}^A_{\mu \nu}{}^{\rho \sigma'} = \mathcal{T}^{\mathrm{cl}_A}_{\mu \nu}{}^{\rho \sigma'} + \mathcal{T}^{\mathrm{GB}_A}_{\mu \nu}{}^{\rho \sigma'} ,
\\
\label{qSET_do_Phi}
&& \mathcal{T}^{\Phi}_{\mu \nu} = \mathcal{T}^{\mathrm{cl}_\Phi}_{\mu \nu} + \mathcal{T}^{\mathrm{GB}_\Phi}_{\mu \nu} .
\end{eqnarray}
\end{subequations}

\subsection{Renormalized stress-energy tensor}
\label{Sec.IVc}

\subsubsection{Definition and conservation}
\label{Sec.IVc1}

As we have already noted, the expectation value $\langle \psi | \widehat{T}_{\mu \nu} | \psi \rangle$ given by Eq.~\eqref{qSET_v1} is divergent due to the short-distance behavior of the Green functions or, more precisely, to the singular purely geometrical part of the Hadamard functions given in Eqs.~\eqref{HadamardRep_G1A_sing} and \eqref{HadamardRep_G1_sing} (see the terms in $1/\sigma$ and $\ln |\sigma|$). It is possible to construct the renormalized expectation value of the stress-energy-tensor operator with respect to the Hadamard quantum state $| \psi \rangle$ by using the prescription proposed by Wald in Refs.~\cite{Wald:1977up,Wald:1978pj,Wald:1995yp}. In Eqs.~\eqref{qSET_v1_cl}--\eqref{qSET_v1_Gh} we first discard the singular contributions, i.e., we make the replacements
\begin{subequations}
\begin{eqnarray}
&& \hspace{-9mm} G^{(1)A}_{\mu \nu'} (x,x') \to G^{(1)A}_\mathrm{reg}{}_{\mu \nu'} (x,x') = \frac{1}{4\pi^2} \, W^A_{\mu \nu'}(x,x') ,
\\
&& \hspace{-9mm} G^{(1)\Phi} (x,x') \to G^{(1)\Phi}_\mathrm{reg} (x,x') = \frac{1}{4\pi^2} \, W^\Phi(x,x')  ,
\\
&& \hspace{-9mm} G^{(1)\mathrm{Gh}} (x,x') \to G^{(1)\mathrm{Gh}}_\mathrm{reg} (x,x') = \frac{1}{4\pi^2} \, W^\mathrm{Gh}(x,x') ,
\end{eqnarray}
\end{subequations}
and we add to the result a state-independent tensor $\widetilde{\Theta}_{\mu \nu}$ which only depends on the mass parameter and on the local geometry and which ensures the conservation of the final expression. In other words, we consider that the renormalized expectation value of the stress-energy-tensor operator is given by
\begin{eqnarray}
\label{RqSET_v1}
&& \langle \psi | \widehat{T}_{\mu \nu} | \psi \rangle_\mathrm{ren} =
  \frac{1}{8 \pi^2} \left\{
  \mathcal{T}^{\mathrm{cl}_A}_{\mu \nu} \left[ W^A \right]
+ \mathcal{T}^{\mathrm{cl}_\Phi}_{\mu \nu} \left[ W^\Phi \right]
\right\}
\nonumber \\ && \phantom{\langle \psi | \widehat{T}_{\mu \nu} | \psi \rangle_\mathrm{ren}}
+ \frac{1}{8 \pi^2} \left\{
  \mathcal{T}^{\mathrm{GB}_A}_{\mu \nu} \left[ W^A \right]
+ \mathcal{T}^{\mathrm{GB}_\Phi}_{\mu \nu} \left[ W^\Phi \right]
\right\}
\nonumber \\ && \phantom{\langle \psi | \widehat{T}_{\mu \nu} | \psi \rangle_\mathrm{ren}}
+ \frac{1}{8 \pi^2} \,
  \mathcal{T}^\mathrm{Gh}_{\mu \nu} \left[ W^\mathrm{Gh} \right]
\nonumber \\ && \phantom{\langle \psi | \widehat{T}_{\mu \nu} | \psi \rangle_\mathrm{ren}}
+ \widetilde{\Theta}_{\mu \nu}
\end{eqnarray}
with
\begin{subequations}
\label{RqSET_v1_def_cl_GB_Gh}
\begin{eqnarray}
\label{RqSET_v1_def_clA}
&& \mathcal{T}^{\mathrm{cl}_A}_{\mu \nu} \left[ W^A \right](x) = \lim_{x' \to x} \mathcal{T}^{\mathrm{cl}_A}_{\mu \nu}{}^{\rho \sigma'} (x,x') \left[ W^A_{\rho \sigma'} (x,x') \right] ,
\nonumber \\
\end{eqnarray}
\begin{eqnarray}
\label{RqSET_v1_def_clPhi}
&& \mathcal{T}^{\mathrm{cl}_\Phi}_{\mu \nu} \left[ W^\Phi \right](x) = \lim_{x' \to x} \mathcal{T}^{\mathrm{cl}_\Phi}_{\mu \nu} (x,x') \left[ W^\Phi (x,x') \right] ,
\nonumber \\
\end{eqnarray}
\begin{eqnarray}
\label{RqSET_v1_def_GBA}
&& \mathcal{T}^{\mathrm{GB}_A}_{\mu \nu} \left[ W^A \right](x) = \lim_{x' \to x} \mathcal{T}^{\mathrm{GB}_A}_{\mu \nu}{}^{\rho \sigma'} (x,x') \left[ W^A_{\rho \sigma'} (x,x') \right] ,
\nonumber \\
\end{eqnarray}
\begin{eqnarray}
\label{RqSET_v1_def_GBPhi}
&& \mathcal{T}^{\mathrm{GB}_\Phi}_{\mu \nu} \left[ W^\Phi \right](x) = \lim_{x' \to x} \mathcal{T}^{\mathrm{GB}_\Phi}_{\mu \nu} (x,x') \left[ W^\Phi (x,x') \right]
\nonumber \\
\end{eqnarray}
and
\begin{eqnarray}
\label{RqSET_v1_def_Gh}
&& \mathcal{T}^\mathrm{Gh}_{\mu \nu} \left[ W^\mathrm{Gh} \right](x) = \lim_{x' \to x} \mathcal{T}^\mathrm{Gh}_{\mu \nu} (x,x') \left[ W^\mathrm{Gh} (x,x') \right] .
\nonumber \\
\end{eqnarray}
\end{subequations}
Here, the differential operators $\mathcal{T}^{\mathrm{cl}_A}_{\mu \nu}{}^{\rho \sigma'}$, $\mathcal{T}^{\mathrm{cl}_\Phi}_{\mu \nu}$, $\mathcal{T}^{\mathrm{GB}_A}_{\mu \nu}{}^{\rho \sigma'}$, $\mathcal{T}^{\mathrm{GB}_\Phi}_{\mu \nu}$ and $\mathcal{T}^{\mathrm{Gh}}_{\mu \nu}$ are given by Eqs.~\eqref{qSET_v1_do_clA}--\eqref{qSET_v1_do_Gh}. In Eqs.~\eqref{RqSET_v1_def_clA}--\eqref{RqSET_v1_def_Gh}, the coincidence limits $x' \to x$ are obtained from the covariant Taylor series expansions \eqref{covTaylorSeries_WA} and \eqref{covTaylorSeries_W} by using extensively some of the results displayed in the Appendix. The final expressions can be simplified by using the relations \eqref{Relation_coefTaylorSeries_WA_1}, \eqref{Relation_coefTaylorSeries_WA_2} and \eqref{Relation_coefTaylorSeries_W_1} we have previously obtained from the wave equations. We have
\begin{subequations}
\allowdisplaybreaks
\label{RqSET_v1_cl_GB_Gh}
\begin{eqnarray}
\label{RqSET_v1_clA}
&& \mathcal{T}^{\mathrm{cl}_A}_{\mu \nu} \left[ W^A \right] =
  (1/2) \, s^{\phantom{\rho} \rho}_{\rho \phantom{\rho} ;\mu \nu} + (1/2) \, \Box s_{\mu \nu} - s^{\phantom{\rho (\mu ;\nu)} \rho}_{\rho (\mu ;\nu)}
\nonumber \\ && \quad
- a^{\phantom{\mu} \rho}_{\mu \phantom{\rho} [\nu ;\rho]} - a^{\phantom{\nu} \rho}_{\nu \phantom{\rho} [\mu ;\rho]} - s^{\phantom{\rho} \rho}_{\rho \phantom{\rho} \mu \nu} + 2 \, s^{\phantom{\rho (\mu \nu)} \rho}_{\rho (\mu \nu)}
\nonumber \\ && \quad
- (1/2) \, g_{\mu \nu} \left\{
(1/2) \, \Box s^{\phantom{\rho} \rho}_{\rho} - (1/2) \, s^{\phantom{\rho \tau} ;\rho \tau}_{\rho \tau}
\right. \nonumber \\ && \left. \qquad
- (1/2) \, R^{\rho \tau} s_{\rho \tau} - a^{\phantom{\rho \tau} \rho ;\tau}_{\rho \tau} + s^{\phantom{\rho \tau} \rho \tau}_{\rho \tau}
\right\}
\nonumber \\ && \quad
+ 6 \, v^A_1{}_{\mu \nu} - 3 \, g_{\mu \nu} \, v^A_1{}^{\phantom{\rho} \rho}_{\rho} ,
\end{eqnarray}
\begin{eqnarray}
\label{RqSET_v1_clPhi}
&& \hspace{-7mm} \mathcal{T}^{\mathrm{cl}_\Phi}_{\mu \nu} \left[ W^\Phi \right] = (1/2) \, w^\Phi_{;\mu \nu} - w^\Phi_{\mu \nu}
\nonumber \\ && \hspace{-7mm} \quad
- (1/2) \, g_{\mu \nu} \left\{ (1/2) \, \Box w^\Phi - m^2 w^\Phi \right\}
- 3 \, g_{\mu \nu} \, v_1 ,
\end{eqnarray}
\begin{eqnarray}
\label{RqSET_v1_GBA}
&& \mathcal{T}^{\mathrm{GB}_A}_{\mu \nu} \left[ W^A \right] =
  R^{\rho}_{\phantom{\rho} (\mu} s_{\nu) \rho}
- a^{\phantom{\mu} \rho}_{\mu \phantom{\rho} (\nu ;\rho)} - a^{\phantom{\nu} \rho}_{\nu \phantom{\rho} (\mu ;\rho)}
- 2 \, s^{\phantom{\rho (\mu \nu)} \rho}_{\rho (\mu \nu)}
\nonumber \\ && \quad
- (1/2) \, g_{\mu \nu} \left\{
- (1/2) \, s^{\phantom{\rho \tau} ;\rho \tau}_{\rho \tau}
+ (1/2) \, R^{\rho \tau} s_{\rho \tau}
\right. \nonumber \\ && \left. \qquad
+ a^{\phantom{\rho \tau} \rho ;\tau}_{\rho \tau}
- s^{\phantom{\rho \tau} \rho \tau}_{\rho \tau}
\right\} ,
\end{eqnarray}
\begin{eqnarray}
\label{RqSET_v1_GBPhi}
&& \mathcal{T}^{\mathrm{GB}_\Phi}_{\mu \nu} \left[ W^\Phi \right] =
- (1/2) \, g_{\mu \nu} \, m^2 w^\Phi
\end{eqnarray}
and
\begin{eqnarray}
\label{RqSET_v1_Gh}
&& \mathcal{T}^\mathrm{Gh}_{\mu \nu} \left[ W^\mathrm{Gh} \right] = - w^\mathrm{Gh}_{;\mu \nu} + 2 \, w^\mathrm{Gh}_{\mu \nu}
\nonumber \\ && \quad
+ (1/2) \, g_{\mu \nu} \, \Box w^\mathrm{Gh}
+ 6 \, g_{\mu \nu} \, v_1 .
\end{eqnarray}
\end{subequations}

Let us now consider the divergence of the terms given by Eqs.~\eqref{RqSET_v1_clA}--\eqref{RqSET_v1_Gh}. By taking into account Eqs.~\eqref{Relation_coefTaylorSeries_WA} and \eqref{Relation_coefTaylorSeries_W}, we obtain
\begin{subequations}
\label{divRqSET_v1_A_Phi_Gh}
\begin{eqnarray}
\label{divRqSET_v1_A}
&& \left( \mathcal{T}^{\mathrm{cl}_A}_{\mu \nu} \left[ W^A \right] + \mathcal{T}^{\mathrm{GB}_A}_{\mu \nu} \left[ W^A \right] \right)^{;\nu} =
6 \, v^A_1{}^{\phantom{\mu \nu} ;\nu}_{\mu \nu} - 2 \, v^A_1{}^{\phantom{\rho} \rho }_{\rho \phantom{\rho} ;\mu} ,
\nonumber \\
\end{eqnarray}
\begin{eqnarray}
\label{divRqSET_v1_Phi}
&& \left( \mathcal{T}^{\mathrm{cl}_\Phi}_{\mu \nu} \left[ W^\Phi \right] + \mathcal{T}^{\mathrm{GB}_\Phi}_{\mu \nu} \left[ W^\Phi \right] \right)^{;\nu} =
- 2 \, v_1{}_{;\mu}
\end{eqnarray}
and
\begin{eqnarray}
\label{divRqSET_v1_Gh}
&& \left( \mathcal{T}^\mathrm{Gh}_{\mu \nu} \left[ W^\mathrm{Gh} \right] \right)^{;\nu} =
4 \, v_1{}_{;\mu} ,
\end{eqnarray}
\end{subequations}
and we then have
\begin{eqnarray}
\label{divRqSET_v1}
&& \left( \langle \psi | \widehat{T}_{\mu \nu} | \psi \rangle_\mathrm{ren} \right)^{;\nu} = \frac{1}{8 \pi^2} \left\{
  6 \, v^A_1{}_{\mu \nu}
- 2 \, g_{\mu \nu} \, v^A_1{}^{\phantom{\rho} \rho}_{\rho}
\right. \nonumber \\ && \left. \quad
+ 2 \, g_{\mu \nu} \, v_1
\right\}^{;\nu} + \widetilde{\Theta}^{\phantom{\mu \nu} ;\nu}_{\mu \nu}
= 0 .
\end{eqnarray}
It is therefore suitable to redefine the purely geometrical tensor $\widetilde{\Theta}_{\mu \nu}$ by
\begin{eqnarray}
\label{Theta}
&& \widetilde{\Theta}_{\mu \nu} \to \Theta_{\mu \nu} - \frac{1}{8 \pi^2} \left\{
  6 \, v^A_1{}_{\mu \nu}
- 2 \, g_{\mu \nu} \, v^A_1{}^{\phantom{\rho} \rho}_{\rho}
+ 2 \, g_{\mu \nu} \, v_1
\right\} ,
\nonumber \\
\end{eqnarray}
where the new local tensor $\Theta_{\mu \nu}$ is assumed to be conserved, i.e.,
\begin{equation}
\label{divTheta}
\Theta^{\phantom{\mu \nu} ;\nu}_{\mu \nu} = 0 .
\end{equation}
As a consequence, the renormalized expectation value of the stress-energy-tensor operator takes the following form:
\begin{eqnarray}
\label{RqSET_v2}
&& \langle \psi | \widehat{T}_{\mu \nu} | \psi \rangle_\mathrm{ren} =
\frac{1}{8 \pi^2} \left\{
  \mathcal{T}^{\mathrm{cl}_A}_{\mu \nu} \left[ W^A \right] + \mathcal{T}^{\mathrm{GB}_A}_{\mu \nu} \left[ W^A \right]
\right. \nonumber \\ && \left. \hphantom{\langle \psi | \widehat{T}_{\mu \nu} | \psi \rangle_\mathrm{ren} = \frac{1}{8 \pi^2}} \quad
- 6 \, v^A_1{}_{\mu \nu}
+ 2 \, g_{\mu \nu} \, v^A_1{}^{\phantom{\rho} \rho}_{\rho}
\right\}
\nonumber \\ && \phantom{\langle \psi | \widehat{T}_{\mu \nu} | \psi \rangle_\mathrm{ren}}
+ \frac{1}{8 \pi^2} \left\{
  \mathcal{T}^{\mathrm{cl}_\Phi}_{\mu \nu} \left[ W^\Phi \right] + \mathcal{T}^{\mathrm{GB}_\Phi}_{\mu \nu} \left[ W^\Phi \right]
\right. \nonumber \\ && \left. \hphantom{\langle \psi | \widehat{T}_{\mu \nu} | \psi \rangle_\mathrm{ren} = \frac{1}{8 \pi^2}} \quad
+ 2 \, g_{\mu \nu} \, v_1
\right\}
\nonumber \\ && \phantom{\langle \psi | \widehat{T}_{\mu \nu} | \psi \rangle_\mathrm{ren}}
+ \frac{1}{8 \pi^2} \left\{
  \mathcal{T}^\mathrm{Gh}_{\mu \nu} \left[ W^\mathrm{Gh} \right]
- 4 \, g_{\mu \nu} \, v_1
\right\}
\nonumber \\ && \phantom{\langle \psi | \widehat{T}_{\mu \nu} | \psi \rangle_\mathrm{ren}}
+ \Theta_{\mu \nu} ,
\end{eqnarray}
where the various state-dependent contributions are given by Eqs.~\eqref{RqSET_v1_clA}--\eqref{RqSET_v1_Gh}.

\subsubsection{Cancellation of the gauge-breaking and ghost contributions}
\label{Sec.IVc2}

In Sec.~\ref{Sec.IVb} we have mentioned that the formal contributions of the gauge-breaking term $\langle \psi | \widehat{T}^\mathrm{GB}_{\mu \nu} | \psi \rangle$ and the ghost term $\langle \psi | \widehat{T}^\mathrm{Gh}_{\mu \nu} | \psi \rangle$ cancel each other out [see Eq.~\eqref{qSET_v1_GB+Gh=0}]. This still remains valid for the corresponding regularized expectation values up to purely geometrical terms. Indeed, by using the first Ward identity in the form \eqref{Relation_coefTaylorSeries_WA_W} as well as the second Ward identity in the form \eqref{WardId_WPhi_WGh}, we obtain
\begin{eqnarray}
\label{RqSET_v1_Gh_GB}
&& \mathcal{T}^{\mathrm{GB}_A}_{\mu \nu} \left[ W^A \right] + \mathcal{T}^{\mathrm{GB}_\Phi}_{\mu \nu} \left[ W^\Phi \right] + \mathcal{T}^\mathrm{Gh}_{\mu \nu} \left[ W^\mathrm{Gh} \right] =
\nonumber \\ && \quad
  2 \, v^A_1{}_{\mu \nu}
+ g_{\mu \nu} \left\{ -(1/2) \, v^A_1{}^{\phantom{\rho} \rho}_{\rho} + 3 \, v_1 \right\} .
\end{eqnarray}
Now, by using this relation in connection with Eqs.~\eqref{RqSET_v1_clA} and \eqref{RqSET_v1_clPhi}, we can rewrite the renormalized expectation value of the stress-energy-tensor operator given by Eq.~\eqref{RqSET_v2} in the form
\begin{eqnarray}
\label{RqSET_v3}
&& \langle \psi | \widehat{T}_{\mu \nu} | \psi \rangle_\mathrm{ren} =
\nonumber \\ && \phantom{+} \quad
\frac{1}{8\pi^2} \left\{
  (1/2) \, s^{\phantom{\rho} \rho}_{\rho \phantom{\rho} ;\mu \nu} + (1/2) \, \Box s_{\mu \nu} - s^{\phantom{\rho (\mu ;\nu)} \rho}_{\rho (\mu ;\nu)}
\right. \nonumber \\ && \left. \qquad
- a^{\phantom{\mu} \rho}_{\mu \phantom{\rho} [\nu ;\rho]} - a^{\phantom{\nu} \rho}_{\nu \phantom{\rho} [\mu ;\rho]} - s^{\phantom{\rho} \rho}_{\rho \phantom{\rho} \mu \nu} + 2 \, s^{\phantom{\rho (\mu \nu)} \rho}_{\rho (\mu \nu)}
\right. \nonumber \\ && \left. \qquad
- (1/2) \, g_{\mu \nu} \left[
  (1/2) \, \Box s^{\phantom{\rho} \rho}_{\rho} - (1/2) \, s^{\phantom{\rho \tau} ;\rho \tau}_{\rho \tau}
\right.\right. \nonumber \\ && \left.\left. \qquad \quad
- (1/2) \, R^{\rho \tau} s_{\rho \tau} - a^{\phantom{\rho \tau} \rho ;\tau}_{\rho \tau} + s^{\phantom{\rho \tau} \rho \tau}_{\rho \tau}
\right]
\right. \nonumber \\ && \left. \qquad
+ (1/2) \, w^\Phi_{;\mu \nu} - w^\Phi_{\mu \nu}
\right. \nonumber \\ && \left. \qquad
- (1/2) \, g_{\mu \nu} \left[ (1/2) \, \Box w^\Phi - m^2 w^\Phi \right]
\right. \nonumber \\ && \left. \qquad
+ 2 \, v^A_1{}_{\mu \nu} - (3/2) \, g_{\mu \nu} v^A_1{}^{\phantom{\rho} \rho}_{\rho}
- 2 \, g_{\mu \nu} v_1
\vphantom{s^{\phantom{\rho (\mu ;\nu)} \rho}_{\rho (\mu ;\nu)}}
\right\}
\nonumber \\ && \quad
+ \, \Theta_{\mu \nu} .
\end{eqnarray}
This expression only involves state-dependent and geometrical quantities associated with the quantum fields $A_\mu$ and $\Phi$. We could consider it as our final result, but, in fact, it is very important here to note that, due to the first Ward identity, the decomposition into a part involving the massive vector field and another part involving the auxiliary scalar field is not unique. In the next sections, we shall provide two alternative expressions which, in our opinion, are much more interesting from the physical point of view.

From now, in order to simplify the notations and because this does not lead to any ambiguity, we shall omit the label $\Phi$ for the Taylor coefficients $w^\Phi$ and $w^\Phi_{\mu \nu}$.

\subsubsection{Substitution of the auxiliary scalar field contribution and final result}
\label{Sec.IVc3}

It is possible to remove in Eq.~\eqref{RqSET_v3} any reference to the auxiliary scalar field $\Phi$. In some sense, it plays the role of a kind of ghost field (the so-called Stueckelberg ghost \cite{vanHees:2003dk}), but its contribution must be carefully taken into account. By using Eqs.~\eqref{Relation_coefTaylorSeries_W_1}, \eqref{Relation_coefTaylorSeries_WA_W_1} and \eqref{Relation_coefTaylorSeries_WA_W_2} in the form
\begin{subequations}
\begin{eqnarray}
\label{Relation_coefTaylorSeries_RqSET_1}
&& m^2 w = w^{\phantom{\rho} \rho}_{\rho} + 6 \, v_1
\nonumber \\ && \phantom{m^2 w}
= - (1/2) \, s^{\phantom{\rho \tau} ;\rho \tau}_{\rho \tau}
- (1/2) \, R^{\rho \tau} s_{\rho \tau}
\nonumber \\ && \phantom{m^2 w =} \quad
+ a^{\phantom{\rho \tau} \rho ;\tau}_{\rho \tau}
+ s^{\phantom{\rho \tau} \rho \tau}_{\rho \tau}
+ v^A_1{}^{\phantom{\rho} \rho}_{\rho} + 2 \, v_1 ,
\end{eqnarray}
\begin{eqnarray}
\label{Relation_coefTaylorSeries_RqSET_2}
&& w_{;\mu \nu} = - s^{\phantom{\rho (\mu|} ;\rho}_{\rho (\mu| \phantom{;\rho} |\nu)} + 2 \, a^{\phantom{\rho (\mu|} \rho}_{\rho (\mu| \phantom{\rho} ;|\nu)}
\end{eqnarray}
and
\begin{eqnarray}
\label{Relation_coefTaylorSeries_RqSET_3}
&& w_{\mu \nu} = - (1/2) \, s^{\phantom{\rho (\mu|} ;\rho}_{\rho (\mu| \phantom{;\rho} |\nu)} - (1/2) \, R^{\rho}_{\phantom{\rho} (\mu} s_{\nu) \rho}
\nonumber \\ && \quad
+ (1/2) \, a^{\phantom{\mu} \rho}_{\mu \phantom{\rho} [\nu ;\rho]}
+ (1/2) \, a^{\phantom{\nu} \rho}_{\nu \phantom{\rho} [\mu ;\rho]}
\nonumber \\ && \quad
+ s^{\phantom{\rho (\mu \nu)} \rho}_{\rho (\mu \nu)}
+ v^A_1{}_{\mu \nu} - g_{\mu \nu} v_1 ,
\end{eqnarray}
\end{subequations}
we obtain
\begin{eqnarray}
\label{RqSET_v5}
&& \langle \psi | \widehat{T}_{\mu \nu} | \psi \rangle_\mathrm{ren} =
\nonumber \\ && \phantom{+} \quad
\frac{1}{8\pi^2} \left\{
  (1/2) \, s^{\phantom{\rho} \rho}_{\rho \phantom{\rho} ;\mu \nu} + (1/2) \, \Box s_{\mu \nu} - s^{\phantom{\rho (\mu ;\nu)} \rho}_{\rho (\mu ;\nu)}
\right. \nonumber \\ && \left. \qquad
+ (1/2) \, R^{\rho}_{\phantom{\rho} (\mu} s_{\nu) \rho} - (1/2) \, a^{\phantom{\mu} \rho}_{\mu \phantom{\rho} (\nu ;\rho)} - (1/2) \, a^{\phantom{\nu} \rho}_{\nu \phantom{\rho} (\mu ;\rho)}
\right. \nonumber \\ && \left. \qquad
- a^{\phantom{\mu} \rho}_{\mu \phantom{\rho} [\nu ;\rho]} - a^{\phantom{\nu} \rho}_{\nu \phantom{\rho} [\mu ;\rho]} - s^{\phantom{\rho} \rho}_{\rho \phantom{\rho} \mu \nu} + s^{\phantom{\rho (\mu \nu)} \rho}_{\rho (\mu \nu)}
\right. \nonumber \\ && \left. \qquad
- (1/2) \, g_{\mu \nu} \left[
(1/2) \, \Box s^{\phantom{\rho} \rho}_{\rho} - (1/2) \, s^{\phantom{\rho \tau} ;\rho \tau}_{\rho \tau} - a^{\phantom{\rho \tau} \rho ;\tau}_{\rho \tau}
\right]
\right. \nonumber \\ && \left. \qquad
+ v^A_1{}_{\mu \nu} - g_{\mu \nu} v^A_1{}^{\phantom{\rho} \rho}_{\rho}
\vphantom{s^{\phantom{\rho (\mu ;\nu)} \rho}_{\rho (\mu ;\nu)}}
\right\}
\nonumber \\ && \quad
+ \Theta_{\mu \nu} .
\end{eqnarray}
We have now at our disposal an expression for the renormalized expectation value of the stress-energy-tensor operator associated with the full Stueckelberg theory which only involves state-dependent and geometrical quantities associated with the massive vector field $A_\mu$. It is the main result of our article.

It is interesting to note that Eq.~\eqref{RqSET_v5} combined with Eq.~\eqref{Relation_coefTaylorSeries_WA_2} leads to
\begin{eqnarray}
\label{trRqSET_v5}
&& \langle \psi | \widehat{T}^{\phantom{\rho} \rho}_{\rho} | \psi \rangle_\mathrm{ren} = \frac{1}{8\pi^2} \left\{
- m^2 s^{\phantom{\rho} \rho}_{\rho}
- (1/2) \, R^{\rho \tau} s_{\rho \tau}
\right. \nonumber \\ && \left. \quad
+ s^{\phantom{\rho \tau} \rho \tau}_{\rho \tau}
+ 3 \, v^A_1{}^{\phantom{\rho} \rho}_{\rho}
\right\}
+ \Theta^{\phantom{\rho} \rho}_{\rho} .
\end{eqnarray}

\subsubsection{Another final expression involving both the vector field $A_\mu$ and the auxiliary scalar field $\Phi$}
\label{Sec.IVc4}

Even if we are satisfied with our final expression \eqref{RqSET_v5}, it is worth nothing that it does not reduce, in the limit $m^2 \to 0$, to the result obtained from Maxwell's theory. This is not really surprising because it involves implicitly the contribution of the auxiliary scalar field. In fact, by replacing in Eq.~\eqref{RqSET_v3} the term $m^2 w$ given by Eq.~\eqref{Relation_coefTaylorSeries_RqSET_1}, it is possible to split the renormalized expectation value of the stress-energy-tensor operator in the form
\begin{equation}
\label{RqSET_v4}
\langle \psi | \widehat{T}_{\mu \nu} | \psi \rangle_\mathrm{ren} = \mathcal{T}^A_{\mu \nu} + \mathcal{T}^\Phi_{\mu \nu} + \Theta_{\mu \nu} ,
\end{equation}
where the terms associated with the vector and scalar fields are given by
\begin{subequations}
\allowdisplaybreaks
\label{RqSET_v4_A_Phi}
\begin{eqnarray}
\label{RqSET_v4_A}
&& \mathcal{T}^A_{\mu \nu} = \frac{1}{8\pi^2} \left\{
  (1/2) \, s^{\phantom{\rho} \rho}_{\rho \phantom{\rho} ;\mu \nu} + (1/2) \, \Box s_{\mu \nu} - s^{\phantom{\rho (\mu ;\nu)} \rho}_{\rho (\mu ;\nu)}
\right. \nonumber \\ && \left. \quad
- a^{\phantom{\mu} \rho}_{\mu \phantom{\rho} [\nu ;\rho]} - a^{\phantom{\nu} \rho}_{\nu \phantom{\rho} [\mu ;\rho]} - s^{\phantom{\rho} \rho}_{\rho \phantom{\rho} \mu \nu} + 2 \, s^{\phantom{\rho (\mu \nu)} \rho}_{\rho (\mu \nu)}
\right. \nonumber \\ && \left. \quad
- (1/2) \, g_{\mu \nu} \left[
  (1/2) \, \Box s^{\phantom{\rho} \rho}_{\rho}
- 2 \, a^{\phantom{\rho \tau} \rho ;\tau}_{\rho \tau}
\right]
\right. \nonumber \\ && \left. \quad
+ 2 \, v^A_1{}_{\mu \nu} - g_{\mu \nu} v^A_1{}^{\phantom{\rho} \rho}_{\rho}
\vphantom{s^{\phantom{\rho (\mu ;\nu)} \rho}_{\rho (\mu ;\nu)}}
\right\}
\end{eqnarray}
and
\begin{eqnarray}
\label{RqSET_v4_Phi}
&& \mathcal{T}^\Phi_{\mu \nu} = \frac{1}{8\pi^2} \left\{
  (1/2) \, w_{;\mu \nu} - w_{\mu \nu}
\right. \nonumber \\ && \left. \quad
- (1/4) \, g_{\mu \nu} \Box w
- g_{\mu \nu} v_1
\right\} .
\end{eqnarray}
\end{subequations}
The stress-energy tensors $\mathcal{T}^A_{\mu \nu}$ and $\mathcal{T}^\Phi_{\mu \nu}$ are separately conserved (this can be checked from relations obtained in Sec.~\ref{Sec.IIId}), and, moreover, the expression of $\mathcal{T}^\Phi_{\mu \nu}$ corresponds exactly to the renormalized expectation value of the stress-energy-tensor operator associated with the quantum action \eqref{Action_Stueck_Quant_v2_Phi} for $\xi$=1 (see, e.g., Refs.~\cite{Brown:1986tj,Bernard:1986vc}). As a consequence, it could be rather natural to consider $\mathcal{T}^A_{\mu \nu}$ given by Eq.~\eqref{RqSET_v4_A} as the renormalized expectation value of the stress-energy-tensor operator associated with the massive vector field $A_\mu$. This physical interpretation is strengthened by noting that, in the limit $m^2 \to 0$, $\mathcal{T}^A_{\mu \nu}$ reduces to the result obtained from Maxwell's theory (see Sec.~\ref{Sec.IVd}). However, despite this, we are not really satisfied by this artificial way to split the contributions of the vector and scalar fields because, as we have already noted, the first Ward identity allows us to move terms from one contribution to the other. So, we consider that the only nonambiguous result is the one given by Eq.~\eqref{RqSET_v5}.

It is interesting to note that Eq.~\eqref{RqSET_v4} combined with Eqs.~\eqref{Relation_coefTaylorSeries_WA_2} and \eqref{Relation_coefTaylorSeries_W_1} leads to
\begin{equation}
\label{trRqSET_v4}
\langle \widehat{T}^{\phantom{\rho} \rho}_{\rho} \rangle_\mathrm{ren} = \mathcal{T}^A{}^{\phantom{\rho} \rho}_{\rho} + \mathcal{T}^\Phi{}^{\phantom{\rho} \rho}_{\rho} + \Theta^{\phantom{\rho} \rho}_{\rho}
\end{equation}
with
\begin{subequations}
\label{trRqSET_v4_A_Phi}
\begin{eqnarray}
\label{trRqSET_v4_A}
&& \mathcal{T}^A{}^{\phantom{\rho} \rho}_{\rho} = \frac{1}{8\pi^2} \left\{
- s^{\phantom{\rho \tau} ;\rho \tau}_{\rho \tau}
- m^2 s^{\phantom{\rho} \rho}_{\rho}
- R^{\rho \tau} s_{\rho \tau}
\right. \nonumber \\ && \left. \quad
+ 2 \, a^{\phantom{\rho \tau} \rho ;\tau}_{\rho \tau}
+ 2 \, s^{\phantom{\rho \tau} \rho \tau}_{\rho \tau}
+ 4 \, v^A_1{}^{\phantom{\rho} \rho}_{\rho}
\right\}
\end{eqnarray}
and
\begin{eqnarray}
\label{trRqSET_v4_Phi}
&& \hspace{-8mm} \mathcal{T}^\Phi{}^{\phantom{\rho} \rho}_{\rho} = \frac{1}{8\pi^2} \left\{ - (1/2) \, \Box w - m^2 w + 2 \, v_1 \right\} .
\end{eqnarray}
\end{subequations}

\subsection{Maxwell's theory}
\label{Sec.IVd}

Let us now consider the limit $m^2 \to 0$ of $\mathcal{T}^A_{\mu \nu}$ given by Eq.~\eqref{RqSET_v4_A}. By using Eq.~\eqref{Relation_coefTaylorSeries_RqSET_1}, it reduces to
\begin{eqnarray}
\label{RqSET_Maxwell}
&& \mathcal{T}^\mathrm{Maxwell}_{\mu \nu} = \frac{1}{8\pi^2} \left\{
  (1/2) \, s^{\phantom{\rho} \rho}_{\rho \phantom{\rho} ;\mu \nu} + (1/2) \, \Box s_{\mu \nu} - s^{\phantom{\rho (\mu ;\nu)} \rho}_{\rho (\mu ;\nu)}
\right. \nonumber \\ && \left. \qquad
- a^{\phantom{\mu} \rho}_{\mu \phantom{\rho} [\nu ;\rho]} - a^{\phantom{\nu} \rho}_{\nu \phantom{\rho} [\mu ;\rho]} - s^{\phantom{\rho} \rho}_{\rho \phantom{\rho} \mu \nu} + 2 \, s^{\phantom{\rho (\mu \nu)} \rho}_{\rho (\mu \nu)}
\right. \nonumber \\ && \left. \qquad
- (1/2) \, g_{\mu \nu} \left[
  (1/2) \, \Box s^{\phantom{\rho} \rho}_{\rho} - (1/2) \, s^{\phantom{\rho \tau} ;\rho \tau}_{\rho \tau}
\right.\right. \nonumber \\ && \left.\left. \qquad \quad
- (1/2) \, R^{\rho \tau} s_{\rho \tau} - a^{\phantom{\rho \tau} \rho ;\tau}_{\rho \tau} + s^{\phantom{\rho \tau} \rho \tau}_{\rho \tau}
\right]
\right. \nonumber \\ && \left. \qquad
+ 2 \, v^A_1{}_{\mu \nu}
-  g_{\mu \nu} \left[ (3/2) \, v^A_1{}^{\phantom{\rho} \rho}_{\rho} + v_1 \right]
\vphantom{s^{\phantom{\rho (\mu ;\nu)} \rho}_{\rho (\mu ;\nu)}}
\right\} .
\end{eqnarray}
This last expression is nothing other than the renormalized expectation value of the stress-energy-tensor operator associated with Maxwell's electromagnetism (see Eq.~(3.41b) of Ref.~\cite{Folacci:1990eb}).

It is interesting to note that Eq.~\eqref{RqSET_Maxwell} combined with Eq.~\eqref{Relation_coefTaylorSeries_WA_2} leads to
\begin{eqnarray}
\label{trRqSET_Maxwell}
&& g^{\mu \nu} \mathcal{T}^\mathrm{Maxwell}_{\mu \nu} = \frac{1}{8\pi^2} \left\{
  2 \, v^A_1{}^{\phantom{\rho} \rho}_{\rho} - 4 \, v_1
\right\}
\nonumber \\ && \quad
= \frac{1}{8\pi^2} \left\{
- (1/20) \, \Box R
- (5/72) \, R^2
+ (11/45) \, R_{p q} \, R^{p q}
\right. \nonumber \\ && \left. \qquad
\vphantom{R^2}
- (13/360) \, R_{p q r s} \, R^{p q r s}
\right\} .
\end{eqnarray}
We recover the trace anomaly for Maxwell's theory.

\subsection{Ambiguities in the renormalized stress-energy tensor}
\label{Sec.IVe}

\subsubsection{General expression of the ambiguities}
\label{Sec.IVe1}

The renormalized expectation value $\langle \psi | \widehat{T}_{\mu \nu}(x) | \psi \rangle_\mathrm{ren}$ is unique up to the addition of a geometrical conserved tensor $\Theta_{\mu \nu}$. In other words, even if it takes perfectly into account the quantum state dependence of the theory, it is ambiguously defined (see, Sec.~III of Ref.~\cite{Wald:1978pj} as well as, e.g., comments in Refs.~\cite{Wald:1995yp,Fulling:1989nb,Tichy:1998ws,Moretti:2001qh,Hollands:2004yh}).

As noted by Wald \cite{Wald:1978pj}, $\Theta_{\mu \nu}$ is a local conserved tensor of dimension $(\mathrm{mass})^4 = (\mathrm{length})^{-4}$ which remains finite in the massless limit. As a consequence, it can be constructed by functional derivation with respect to the metric tensor from a geometrical Lagrangian of dimension $(\mathrm{mass})^4 = (\mathrm{length})^{-4}$. Such a Lagrangian is necessarily  a linear combination of the following four terms: $m^4$, $m^2 R$, $R^2$, $R_{p q} R^{p q}$. It should be noted that we could also take into account the term $R_{p q r s} R^{p q r s}$. But, in fact, we can eliminate this term because, in a four-dimensional background, the Euler number
\begin{equation}
\label{Euler_Gauss-Bonnet}
\chi = \int_\mathcal{M} d^4 x \sqrt{-g} \, \left[ R^2 - 4 \, R_{p q} R^{p q} + R_{p q r s} R^{p q r s} \right]
\end{equation}
associated with the quadratic Gauss-Bonnet action is a topological invariant.

The functional derivation of the action terms previously discussed provides the conserved tensors
\begin{subequations}
\allowdisplaybreaks
\label{ConservedTensors}
\begin{eqnarray}
\label{ConservedTensors_m4}
&& \frac{1}{\sqrt{-g}} \frac{\delta}{\delta g_{\mu \nu}} \int_\mathcal{M} d^4 x \sqrt{-g} \, m^4
= (1/2) \, m^4 g^{\mu \nu} ,
\\
\label{ConservedTensors_m2R}
&& \frac{1}{\sqrt{-g}} \frac{\delta}{\delta g_{\mu \nu}} \int_\mathcal{M} d^4 x \sqrt{-g} \, m^2 R
\nonumber \\ && \quad
= - m^2 \left[ R^{\mu \nu} - (1/2) \, R \, g^{\mu \nu} \right] ,
\\
\label{ConservedTensors_R2}
&& \tensor[^{(1)}]{H}{^{\mu \nu}} \equiv \frac{1}{\sqrt{-g}} \frac{\delta}{\delta g_{\mu \nu}} \int_\mathcal{M} d^4 x \sqrt{-g} \, R^2
\nonumber \\ && \quad
= 2 \, R^{;\mu \nu} - 2 \, R R^{\mu \nu}
\nonumber \\ && \qquad
+ g^{\mu \nu} \left[ -2 \, \Box R + (1/2) \, R^2 \right] ,
\\
\label{ConservedTensors_Rab2}
&& \tensor[^{(2)}]{H}{^{\mu \nu}} \equiv \frac{1}{\sqrt{-g}} \frac{\delta}{\delta g_{\mu \nu}} \int_\mathcal{M} d^4 x \sqrt{-g} \, R_{p q} R^{p q}
\nonumber \\ && \quad
= R^{;\mu \nu} - \Box R^{\mu \nu} - 2 \, R_{p q} R^{\mu p \nu q}
\nonumber \\ && \qquad
+ g^{\mu \nu} \left[ - (1/2) \, \Box R + (1/2) \, R_{p q} R^{p q} \right] .
\end{eqnarray}
\end{subequations}
The general expression of the local conserved tensor $\Theta_{\mu \nu} $ can be therefore written in the form
\begin{eqnarray}
\label{ThetaGeneral}
&& \Theta_{\mu \nu} = \frac{1}{8\pi^2} \left\{ \alpha \, m^4 g_{\mu \nu} + \beta \, m^2 \left[ R_{\mu \nu} - (1/2) \, R \, g_{\mu \nu} \right]
\vphantom{\tensor[^{(1)}]{H}{_{\mu \nu}}}
\right. \nonumber \\ && \left. \quad
+ \gamma_1 \, \tensor[^{(1)}]{H}{_{\mu \nu}} + \gamma_2 \, \tensor[^{(2)}]{H}{_{\mu \nu}} \right\} ,
\end{eqnarray}
where $\alpha$, $\beta$, $\gamma_1$ and $\gamma_2$ are constants which can be fixed by imposing additional physical conditions on the renormalized expectation value of the stress-energy-operator tensor, these conditions being appropriate to the problem treated.

\subsubsection{Ambiguities associated with the renormalization mass}
\label{Sec.IVe2}

So far, in order to simplify the calculations, we have dropped the scale length $\lambda$ (or, equivalently, the mass scale $M=1/\lambda$, i.e., the so-called renormalization mass) that should be introduced in order to make dimensionless the argument of the logarithm in the Hadamard representation of the Green functions. In fact, in Eqs.~\eqref{HadamardRep_GAF} and \eqref{HadamardRep_GF} it is necessary to make the substitution $\ln[ \sigma(x,x') + i \epsilon ] \to \ln[ \sigma(x,x')/\lambda^2 + i \epsilon ]$ which leads in Eqs.~\eqref{HadamardRep_G1A} and \eqref{HadamardRep_G1} to the substitution
\begin{equation}
\label{log_to_logL}
\ln| \sigma(x,x') | \to \ln| \sigma(x,x')/\lambda^2 | .
\end{equation}
This scale length induces an indeterminacy in the bitensors $W^A_{\mu \nu'}(x,x')$ and $W(x,x')$ which corresponds to the replacements
\begin{subequations}
\label{W_to_W_V_L}
\begin{eqnarray}
\label{WA_to_WA_VA_L}
&& \hspace{-7mm} W^A_{\mu \nu'}(x,x') \to W^A_{\mu \nu'}(x,x') - V^A_{\mu \nu'}(x,x') \ln(M^2) ,
\\
\label{WPhi_to_WPhi_VPhi_L}
&& \hspace{-7mm} W^\Phi(x,x') \to W^\Phi(x,x') - V^\Phi(x,x') \ln(M^2) ,
\\
\label{WGh_to_WGh_VGh_L}
&& \hspace{-7mm} W^\mathrm{Gh}(x,x') \to W^\mathrm{Gh}(x,x') - V^\mathrm{Gh}(x,x') \ln(M^2) ,
\end{eqnarray}
\end{subequations}
i.e., in terms of the associated Taylor coefficients, to replacements
\begin{subequations}
\label{wA_to_wA_vA_L}
\begin{eqnarray}
\label{s0_to_s0_v0_L}
&& \hspace{-8mm} s_{\mu \nu} \to s_{\mu \nu} - {v^A_0}_{(\mu \nu)} \ln(M^2) ,
\\
\label{a1_to_a1_v1_L}
&& \hspace{-8mm} a_{\mu \nu a} \to a_{\mu \nu a} - {v^A_0}_{[\mu \nu] a} \ln(M^2) ,
\\
\label{s2_to_s2_v_L}
&& \hspace{-8mm} s_{\mu \nu a b} \to s_{\mu \nu a b} - \left( {v^A_0}_{(\mu \nu) a b} + {v^A_1}_{(\mu \nu)} g_{a b} \right) \ln(M^2)
\end{eqnarray}
\end{subequations}
for the vector field $A_\mu$ and
\begin{subequations}
\label{w_to_w_v_L}
\begin{eqnarray}
\label{w0_to_w0_v0_L}
&& w \to w - v_0 \ln(M^2) ,
\\
\label{w2_to_w2_v2_L}
&& w_{a b} \to w_{a b} - \left( {v_0}_{a b} + v_1 g_{a b} \right) \ln(M^2)
\end{eqnarray}
\end{subequations}
for the scalar field $\Phi$ or the ghost fields. By substituting Eqs.~\eqref{WA_to_WA_VA_L}--\eqref{WGh_to_WGh_VGh_L} into the general expression \eqref{RqSET_v1} of the renormalized expectation value of the stress-energy-tensor operator, we obtain the general form of the ambiguity associated with the scale length (or with the renormalization mass). It is given by
\begin{eqnarray}
\label{ThetaL}
&& \Theta_{\mu \nu}(M) = - \frac{\ln(M^2)}{8\pi^2} \left\{
  \Theta^{\mathrm{cl}_A}_{\mu \nu} \left[ V^A \right]
+ \Theta^{\mathrm{cl}_\Phi}_{\mu \nu} \left[ V^\Phi \right]
\right\}
\nonumber \\ && \phantom{\Theta^{M}_{\mu \nu} =}
- \frac{\ln(M^2)}{8\pi^2} \left\{
  \Theta^{\mathrm{GB}_A}_{\mu \nu} \left[ V^A \right]
+ \Theta^{\mathrm{GB}_\Phi}_{\mu \nu} \left[ V^\Phi \right]
\right\}
\nonumber \\ && \phantom{\Theta^{M}_{\mu \nu} =}
- \frac{\ln(M^2)}{8\pi^2} \,
  \Theta^\mathrm{Gh}_{\mu \nu} \left[ V^\mathrm{Gh} \right]
\end{eqnarray}
with
\begin{subequations}
\label{ThetaL_in_V}
\begin{eqnarray}
\label{ThetaLA_in_VclA}
&& \Theta^{\mathrm{cl}_A}_{\mu \nu} \left[ V^A \right] (x) = \lim_{x' \to x} \mathcal{T}^{\mathrm{cl}_A}_{\mu \nu}{}^{\rho \sigma'} (x,x') \left[ V^A_{\rho \sigma'} (x,x') \right] ,
\nonumber \\
\end{eqnarray}
\begin{eqnarray}
\label{ThetaLA_in_VclPhi}
&& \Theta^{\mathrm{cl}_\Phi}_{\mu \nu} \left[ V^\Phi \right] (x) = \lim_{x' \to x} \mathcal{T}^{\mathrm{cl}_\Phi}_{\mu \nu} (x,x') \left[ V^\Phi (x,x') \right] ,
\nonumber \\
\end{eqnarray}
\begin{eqnarray}
\label{ThetaLPhi_in_VGBA}
&& \Theta^{\mathrm{GB}_A}_{\mu \nu} \left[ V^A \right] (x) = \lim_{x' \to x} \mathcal{T}^{\mathrm{GB}_A}_{\mu \nu}{}^{\rho \sigma'} (x,x') \left[ V^A_{\rho \sigma'} (x,x') \right] ,
\nonumber \\
\end{eqnarray}
\begin{eqnarray}
\label{ThetaLPhi_in_VGBPhi}
&& \Theta^{\mathrm{GB}_\Phi}_{\mu \nu} \left[ V^\Phi \right] (x) = \lim_{x' \to x} \mathcal{T}^{\mathrm{GB}_\Phi}_{\mu \nu} (x,x') \left[ V^\Phi (x,x') \right]
\nonumber \\
\end{eqnarray}
and
\begin{eqnarray}
\label{ThetaLGh_in_VGh}
&& \Theta^\mathrm{Gh}_{\mu \nu} \left[ V^\mathrm{Gh} \right] (x) = \lim_{x' \to x} \mathcal{T}^\mathrm{Gh}_{\mu \nu} (x,x') \left[ V^\mathrm{Gh} (x,x') \right] ,
\nonumber \\
\end{eqnarray}
\end{subequations}
where the differential operators $\mathcal{T}^{\mathrm{cl}_A}_{\mu \nu}{}^{\rho \sigma'}$, $\mathcal{T}^{\mathrm{cl}_\Phi}_{\mu \nu}$, $\mathcal{T}^{\mathrm{GB}_A}_{\mu \nu}{}^{\rho \sigma'}$, $\mathcal{T}^{\mathrm{GB}_\Phi}_{\mu \nu}$ and $\mathcal{T}^{\mathrm{Gh}}_{\mu \nu}$ are given in Eqs.~\eqref{qSET_v1_do_clA}--\eqref{qSET_v1_do_Gh}. It should be noted that $\Theta_{\mu \nu}(M)$ is a purely geometrical object. This is due to the geometrical nature of the Hadamard coefficients $V^A_{\mu \nu'}(x,x')$ and $V(x,x')$.

In order to obtain the explicit expression of the stress-energy tensor $\Theta_{\mu \nu}(M)$, we can repeat the calculations of Sec.~\ref{Sec.IVc} by replacing $W^A_{\mu \nu'}$ by $V^A_{\mu \nu'}$, $W^\Phi$ by $V^\Phi$ and $W^\mathrm{Gh}$ by $V^\mathrm{Gh}$. From Eqs.~\eqref{RqSET_v1_clA}--\eqref{RqSET_v1_Gh} it is straightforward to obtain explicitly  $\Theta^{\mathrm{cl}_A}_{\mu \nu} \left[ V^A \right]$, $\Theta^{\mathrm{cl}_\Phi}_{\mu \nu} \left[ V^\Phi \right]$, $\Theta^{\mathrm{GB}_A}_{\mu \nu} \left[ V^A \right]$, $\Theta^{\mathrm{GB}_\Phi}_{\mu \nu} \left[ V^\Phi \right]$ and $\Theta^\mathrm{Gh}_{\mu \nu} \left[ V^\mathrm{Gh} \right]$ by using the replacements \eqref{wA_to_wA_vA_L} and \eqref{w_to_w_v_L}. We can then show that
\begin{subequations}
\label{divThetaL_A_Phi_Gh}
\begin{eqnarray}
\label{divThetaL_A}
&& \left( \Theta^{\mathrm{cl}_A}_{\mu \nu} \left[ V^A \right] + \Theta^{\mathrm{GB}_A}_{\mu \nu} \left[ V^A \right] \right)^{;\nu} = 0 ,
\end{eqnarray}
\begin{eqnarray}
\label{divThetaL_Phi}
&& \left( \Theta^{\mathrm{cl}_\Phi}_{\mu \nu} \left[ V^\Phi \right] + \Theta^{\mathrm{GB}_\Phi}_{\mu \nu} \left[ V^\Phi \right] \right)^{;\nu} = 0
\end{eqnarray}
and
\begin{eqnarray}
\label{divThetaL_Gh}
&& \left( \Theta^\mathrm{Gh}_{\mu \nu} \left[ V^\mathrm{Gh} \right] \right)^{;\nu} = 0 .
\end{eqnarray}
\end{subequations}
Equations \eqref{divThetaL_A}--\eqref{divThetaL_Gh} are similar to Eqs.~\eqref{divRqSET_v1_A}--\eqref{divRqSET_v1_Gh} but now with the right-hand sides vanishing. This is due to the fact that, unlike the wave equations \eqref{EQ_WA} and \eqref{EQ_W} for $W^A_{\mu \nu'}$, $W^\Phi$ and $W^\mathrm{Gh}$, the wave equations for $V^A_{\mu \nu'}$, $V^\Phi$ and $V^\mathrm{Gh}$ [cf. Eqs.~\eqref{EQ_VA} and \eqref{EQ_V}] have no source terms. As a consequence $\Theta_{\mu \nu}(M)$ is a conserved geometrical tensor. We can also check that
\begin{eqnarray}
\label{ThetaL_Gh_GB}
&& \Theta^{\mathrm{GB}_A}_{\mu \nu} \left[ V^A \right] + \Theta^{\mathrm{GB}_\Phi}_{\mu \nu} \left[ V^\Phi \right] + \Theta^\mathrm{Gh}_{\mu \nu} \left[ V^\mathrm{Gh} \right] = 0 .
\end{eqnarray}
Equation \eqref{ThetaL_Gh_GB} is similar to Eq.~\eqref{RqSET_v1_Gh_GB} but now with the right-hand side vanishing. This is due to the fact that, unlike the Ward identity \eqref{WardId_WA_W} linking $W^A_{\mu \nu'}$ and $W^\mathrm{Gh}$, the Ward identity \eqref{WardId_VA_V} linking $V^A_{\mu \nu'}$ and $V^\mathrm{Gh}$ has no right-hand side. As a consequence, from  Eqs.~\eqref{ThetaL} and \eqref{ThetaL_Gh_GB}, we obtain
\begin{eqnarray}
\label{ThetaL_v1}
&& \hspace{-5mm} \Theta_{\mu \nu}(M) = - \frac{\ln(M^2)}{8\pi^2} \left\{
  \Theta^{\mathrm{cl}_A}_{\mu \nu} \left[ V^A \right]
+ \Theta^{\mathrm{cl}_\Phi}_{\mu \nu} \left[ V^\Phi \right]
\right\} .
\end{eqnarray}

The ambiguities associated with the scale length can now be obtained explicitly from the replacements \eqref{wA_to_wA_vA_L} and \eqref{w_to_w_v_L}. If we use the form \eqref{RqSET_v5} without taking into account the geometrical terms, we obtain
\begin{eqnarray}
\label{ThetaL_v1_vA}
&& \Theta_{\mu \nu}(M) = - \frac{\ln(M^2)}{8\pi^2} \left\{
  (1/2) \, v^A_0{}^{\phantom{\rho} \rho}_{\rho \phantom{\rho} ;\mu \nu}
+ (1/2) \, \Box v^A_0{}_{\mu \nu}
\vphantom{v^A_0{}^{\phantom{[\rho \tau]} \rho ;\tau}_{[\rho \tau]}}
\nonumber \right. \\ && \left. \quad
- v^A_0{}^{\phantom{\rho (\mu ;\nu)} \rho}_{\rho (\mu ;\nu)}
+ (1/2) \, R^{\rho}_{\phantom{\rho} (\mu} v^A_0{}_{\nu) \rho}
- (1/2) \, g^{\rho \tau} v^A_0{}_{[\mu \rho] (\nu ;\tau)}
\nonumber \right. \\ && \left. \quad
- (1/2) \, g^{\rho \tau} v^A_0{}_{[\nu \rho] (\mu ;\tau)}
- g^{\rho \tau} v^A_0{}_{[\mu \rho] [\nu ;\tau]}
- g^{\rho \tau} v^A_0{}_{[\nu \rho] [\mu ;\tau]}
\nonumber \right. \\ && \left. \quad
- v^A_0{}^{\phantom{\rho} \rho}_{\rho \phantom{\rho} \mu \nu}
+ (1/2) \, v^A_0{}^{\phantom{(\rho \mu) \nu} \rho}_{(\rho \mu) \nu}
+ (1/2) \, v^A_0{}^{\phantom{(\rho \nu) \mu} \rho}_{(\rho \nu) \mu}
+ v^A_1{}_{\mu \nu}
\nonumber \right. \\ && \left. \quad
- g_{\mu \nu} \left[
  (1/4) \, \Box v^A_0{}^{\phantom{\rho} \rho}_{\rho \phantom{\rho}}
- (1/4) \, v^A_0{}^{\phantom{\rho \tau} ;\rho \tau}_{\rho \tau \phantom{;\rho \tau}}
- (1/2) \, v^A_0{}^{\phantom{[\rho \tau]} \rho ;\tau}_{[\rho \tau]}
\nonumber \right.\right. \\ && \left.\left. \qquad
+ v^A_1{}^{\phantom{\rho} \rho}_{\rho \phantom{\rho}}
\vphantom{v^A_0{}^{\phantom{[\rho \tau]} \rho ;\tau}_{[\rho \tau]}}
\right]
\right\} .
\end{eqnarray}
Similarly, if we use the alternative form \eqref{RqSET_v4} where the contributions corresponding to the vector field $A_\mu$ and the auxiliary scalar field $\Phi$ are highlighted [see Eqs.~\eqref{RqSET_v4_A} and \eqref{RqSET_v4_Phi}], we obtain
\begin{equation}
\label{ThetaL_v4}
\Theta_{\mu \nu}(M) = \Theta^A_{\mu \nu}(M) + \Theta^\Phi_{\mu \nu}(M)
\end{equation}
with
\begin{subequations}
\label{ThetaL_v0_A_Phi}
\begin{eqnarray}
\label{ThetaL_v0_A}
&& \Theta^A_{\mu \nu}(M) = - \frac{\ln(M^2)}{8\pi^2} \left\{
  (1/2) \, v^A_0{}^{\phantom{\rho} \rho}_{\rho \phantom{\rho} ;\mu \nu}
+ (1/2) \, \Box v^A_0{}_{\mu \nu}
\vphantom{v^A_0{}^{\phantom{[\rho \tau]} \rho ;\tau}_{[\rho \tau]}}
\nonumber \right. \\ && \left. \quad
- v^A_0{}^{\phantom{\rho (\mu ;\nu)} \rho}_{\rho (\mu ;\nu)}
- g^{\rho \tau} v^A_0{}_{[\mu \rho] [\nu ;\tau]}
- g^{\rho \tau} v^A_0{}_{[\nu \rho] [\mu ;\tau]}
\nonumber \right. \\ && \left. \quad
- v^A_0{}^{\phantom{\rho} \rho}_{\rho \phantom{\rho} \mu \nu}
+ v^A_0{}^{\phantom{(\rho \mu) \nu} \rho}_{(\rho \mu) \nu}
+ v^A_0{}^{\phantom{(\rho \nu) \mu} \rho}_{(\rho \nu) \mu}
+ 2 \, v^A_1{}_{\mu \nu}
\nonumber \right. \\ && \left. \quad
- g_{\mu \nu} \left[
  (1/4) \, \Box v^A_0{}^{\phantom{\rho} \rho}_{\rho \phantom{\rho}}
- v^A_0{}^{\phantom{[\rho \tau]} \rho ;\tau}_{[\rho \tau]}
+ v^A_1{}^{\phantom{\rho} \rho}_{\rho \phantom{\rho}}
\right]
\right\}
\end{eqnarray}
and
\begin{eqnarray}
\label{ThetaL_v0_Phi}
&& \Theta^\Phi_{\mu \nu}(M) = - \frac{\ln(M^2)}{8\pi^2} \left\{
  (1/2) \, v_0{}_{;\mu \nu} - v_0{}_{\mu \nu}
\right. \nonumber \\ && \left. \quad
- g_{\mu \nu} \left[ (1/4) \, \Box v_0 + v_1 \right]
\right\} .
\end{eqnarray}
\end{subequations}

Now, by using the explicit expressions \eqref{coefTaylorSeries_vA0_vA1} and \eqref{coefTaylorSeries_v0_v1} of the Taylor coefficients of the purely geometrical Hadamard coefficients, we can show that Eq.~\eqref{ThetaL_v1_vA} reduces to
\begin{eqnarray}
\label{ThetaL_v1_Exp}
&& \Theta_{\mu \nu}(M) = - \frac{\ln(M^2)}{8\pi^2} \left\{
  (1/4) \, m^2 R_{\mu \nu} - (1/20) \, R_{;\mu \nu}
\right. \nonumber \\ && \left. \quad
+ (13/120) \, \Box R_{\mu \nu} - (1/8) \, R R_{\mu \nu} + (2/15) \, R_{\mu p} R^{\phantom{\nu} p}_{\nu}
\right. \nonumber \\ && \left. \quad
+ (7/20) \, R_{p q} R^{\phantom{\mu} p \phantom{\nu} q}_{\mu \phantom{p} \nu} - (1/15) \, R_{\mu p q r} R^{\phantom{\nu} p q r}_{\nu}
\right. \nonumber \\ && \left. \quad
+ g_{\mu \nu} \left[
- (3/8) \, m^4 - (1/8) \, m^2 R - (1/240) \, \Box R
\right.\right. \nonumber \\ && \left.\left. \qquad
+ (1/32) \, R^2 - (29/240) \, R_{p q} R^{p q}
\right.\right. \nonumber \\ && \left.\left. \qquad
+ (1/60) \, R_{p q r s} R^{p q r s}
\vphantom{m^4} \right]
\right\} ,
\end{eqnarray}
while Eqs.~\eqref{ThetaL_v0_A} and \eqref{ThetaL_v0_Phi} provide
\begin{subequations}
\allowdisplaybreaks
\label{ThetaL_v0_A_Phi_Exp}
\begin{eqnarray}
\label{ThetaL_v0_A_Exp}
&& \Theta^A_{\mu \nu}(M) = - \frac{\ln(M^2)}{8\pi^2} \left\{
  (1/3) \, m^2 R_{\mu \nu} - (1/30) \, R_{;\mu \nu}
\right. \nonumber \\ && \left. \quad
+ (1/10) \, \Box R_{\mu \nu} - (5/36) \, R R_{\mu \nu} + (13/90) \, R_{\mu p} R^{\phantom{\nu} p}_{\nu}
\right. \nonumber \\ && \left. \quad
+ (31/90) \, R_{p q} R^{\phantom{\mu} p \phantom{\nu} q}_{\mu \phantom{p} \nu} - (13/180) \, R_{\mu p q r} R^{\phantom{\nu} p q r}_{\nu}
\right. \nonumber \\ && \left. \quad
+ g_{\mu \nu} \left[
- (1/4) \, m^4 - (1/6) \, m^2 R - (1/60) \, \Box R
\right.\right. \nonumber \\ && \left.\left. \qquad
+ (5/144) \, R^2 - (11/90) \, R_{p q} R^{p q}
\right.\right. \nonumber \\ && \left.\left. \qquad
+ (13/720) \, R_{p q r s} R^{p q r s}
\vphantom{m^4} \right]
\right\}
\end{eqnarray}
and
\begin{eqnarray}
\label{ThetaL_v0_Phi_Exp}
&& \Theta^\Phi_{\mu \nu}(M) = - \frac{\ln(M^2)}{8\pi^2} \left\{
- (1/12) \, m^2 R_{\mu \nu} - (1/60) \, R_{;\mu \nu}
\right. \nonumber \\ && \left. \quad
+ (1/120) \, \Box R_{\mu \nu} + (1/72) \, R R_{\mu \nu} - (1/90) \, R_{\mu p} R^{\phantom{\nu} p}_{\nu}
\right. \nonumber \\ && \left. \quad
+ (1/180) \, R_{p q} R^{\phantom{\mu} p \phantom{\nu} q}_{\mu \phantom{p} \nu} + (1/180) \, R_{\mu p q r} R^{\phantom{\nu} p q r}_{\nu}
\right. \nonumber \\ && \left. \quad
+ g_{\mu \nu} \left[
- (1/8) \, m^4 + (1/24) \, m^2 R + (1/80) \, \Box R
\right.\right. \nonumber \\ && \left.\left. \qquad
- (1/288) \, R^2 + (1/720) \, R_{p q} R^{p q}
\right.\right. \nonumber \\ && \left.\left. \qquad
- (1/720) \, R_{p q r s} R^{p q r s}
\vphantom{m^4} \right]
\right\} .
\end{eqnarray}
\end{subequations}
Of course, it is easy to check that the sum of $\Theta^A_{\mu \nu}(M)$ and $\Theta^\Phi_{\mu \nu}(M)$ is equal to $\Theta_{\mu \nu}(M)$.

It is possible to obtain a more compact form for the stress-energy tensors \eqref{ThetaL_v1_Exp}, \eqref{ThetaL_v0_A_Exp} and \eqref{ThetaL_v0_Phi_Exp} by using the conserved tensors \eqref{ConservedTensors_m4}--\eqref{ConservedTensors_Rab2}. It should be noted that the terms in $R_{\mu p} R^{\phantom{\nu} p}_{\nu}$, $R_{\mu p q r} R^{\phantom{\nu} p q r}_{\nu}$ and $R_{p q r s} R^{p q r s}$ which are not present in $\tensor[^{(1)}]{H}{_{\mu \nu}}$ and $\tensor[^{(2)}]{H}{_{\mu \nu}}$ can be eliminated by introducing
\begin{eqnarray}
\label{GonservedTensors_Rabcd2}
&& \tensor[^{(3)}]{H}{^{\mu \nu}} \equiv \frac{1}{\sqrt{-g}} \frac{\delta}{\delta g_{\mu \nu}} \int_\mathcal{M} d^4 x \sqrt{-g} \, R_{p q r s} R^{p q r s}
\nonumber \\ && \quad
= 2 \, R^{;\mu \nu} - 4 \, \Box R^{\mu \nu} + 4 \, R^{\mu}_{\phantom{\mu} p} R^{\nu p} - 4 \, R_{p q} R^{\mu p \nu q}
\nonumber \\ && \qquad
- 2 \, R^{\mu}_{\phantom{\mu} p q r} R^{\nu p q r}
+ g^{\mu \nu} \left[ (1/2) \, R_{p q r s} R^{p q r s} \right]
\end{eqnarray}
and by noting that, due to Eq.~\eqref{Euler_Gauss-Bonnet},
\begin{equation}
\label{DerivFunct_Euler_Gauss-Bonnet}
\tensor[^{(1)}]{H}{_{\mu \nu}} - 4 \, \tensor[^{(2)}]{H}{_{\mu \nu}} + \tensor[^{(3)}]{H}{_{\mu \nu}} = 0 .
\end{equation}
We then have
\begin{eqnarray}
\label{ThetaL_v1_Simp}
&& \Theta_{\mu \nu}(M) = \frac{\ln(M^2)}{8\pi^2} \left\{ \vphantom{H^{(3)}_{\mu \nu}}
  (3/8) \, m^4 g_{\mu \nu}
\nonumber \right. \\ && \left. \quad
- (1/4) \, m^2 \left[ R_{\mu \nu} - (1/2) \, R \, g_{\mu \nu} \right]
\nonumber \right. \\ && \left. \quad
- (7/240) \, \tensor[^{(1)}]{H}{_{\mu \nu}}
+ (13/120) \, \tensor[^{(2)}]{H}{_{\mu \nu}}
\right\} ,
\end{eqnarray}
\begin{subequations}
\label{ThetaL_v0_A_Phi_Simp}
\begin{eqnarray}
\label{ThetaL_v0_A_Simp}
&& \Theta^A_{\mu \nu}(M) = \frac{\ln(M^2)}{8\pi^2} \left\{ \vphantom{H^{(3)}_{\mu \nu}}
  (1/4) \, m^4 g_{\mu \nu}
\nonumber \right. \\ && \left. \quad
- (1/3) \, m^2 \left[ R_{\mu \nu} - (1/2) \, R \, g_{\mu \nu} \right]
\nonumber \right. \\ && \left. \quad
- (1/30) \, \tensor[^{(1)}]{H}{_{\mu \nu}}
+ (1/10) \, \tensor[^{(2)}]{H}{_{\mu \nu}}
\right\}
\end{eqnarray}
and
\begin{eqnarray}
\label{ThetaL_v0_Phi_Simp}
&& \Theta^\Phi_{\mu \nu}(M) = \frac{\ln(M^2)}{8\pi^2} \left\{ \vphantom{H^{(3)}_{\mu \nu}}
  (1/8) \, m^4 g_{\mu \nu}
\nonumber \right. \\ && \left. \quad
+ (1/12) \, m^2 \left[ R_{\mu \nu} - (1/2) \, R \, g_{\mu \nu} \right]
\nonumber \right. \\ && \left. \quad
+ (1/240) \, \tensor[^{(1)}]{H}{_{\mu \nu}}
+ (1/120) \, \tensor[^{(2)}]{H}{_{\mu \nu}}
\right\} .
\end{eqnarray}
\end{subequations}
As expected, we can note that the ambiguities associated with the scale length (or with the renormalization mass) are of the form \eqref{ThetaGeneral}.

\section{Casimir effect}
\label{Sec.V}

\subsection{General considerations}
\label{Sec.Va}

\begin{figure}[t]
  \centering
  \includegraphics[width=\columnwidth]{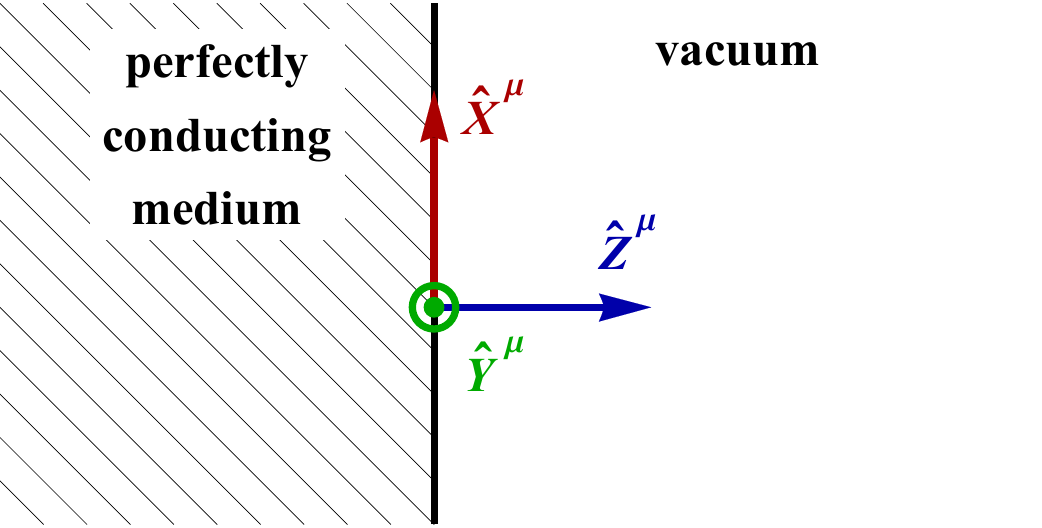}
  \caption{\textbf{Geometry of the Casimir effect}}
  \label{fig_1}
\end{figure}

In this section, we shall consider the Casimir effect for Stueckelberg massive electromagnetism in the Minkowski spacetime $(\mathbb{R}^4, \eta_{\mu \nu})$ with $\eta_{\mu \nu} = \mathrm{diag}(-1,+1,+1,+1)$. We denote by $(T, X, Y, Z)$ the coordinates of an event in this spacetime. We shall provide the renormalized vacuum expectation value of the stress-energy-tensor operator outside of a perfectly conducting medium with a plane boundary wall at $Z=0$ separating it from free space (see Fig.~\ref{fig_1}). It is worth pointing out that this problem has been studied a long time ago by Davies and Toms in the framework of de Broglie-Proca electromagnetism \cite{Davies:1984vm}. We shall revisit this problem in order to compare, at the quantum level and in the case of a simple example, de Broglie-Proca and  Stueckelberg theories and to discuss their limit for $m^2 \to 0$. It should be noted that the Casimir effect in connection with a massive photon has been considered for various geometries (see, e.g., Refs.~\cite{Davies:1980ji,Barton:1984ic,Barton:1984im,Teo:2010hr,Teo:2010sb}).

From symmetries and physical considerations, we can observe that, outside of the perfectly conducting medium, the renormalized stress-energy tensor takes the form (see Chap.~4 of Ref.~\cite{Birrell:1982ix})
\begin{equation}
\label{SET_Minkowski+Wall}
\langle 0 | \widehat{T}_{\mu \nu} | 0 \rangle_\mathrm{ren} = \frac{1}{3} \, \langle 0 | \widehat{T}^{\phantom{\rho} \rho}_{\rho} | 0 \rangle_\mathrm{ren} \, \left( \eta_{\mu \nu} - \hat{Z}_\mu \hat{Z}_\nu \right) ,
\end{equation}
where $\hat{Z}^\mu$ is the spacelike unit vector orthogonal to the wall. As a consequence, it is sufficient to determine the trace of the renormalized stress-energy tensor. From Eq.~\eqref{trRqSET_v5}, we have
\begin{eqnarray}
\label{trRqSET_v5_Minkowski}
&& \langle 0 | \widehat{T}^{\phantom{\rho} \rho}_{\rho} | 0 \rangle_\mathrm{ren} = \frac{1}{8\pi^2} \left\{
- m^2 s^{\phantom{\rho} \rho}_{\rho}
+ s^{\phantom{\rho \tau} \rho \tau}_{\rho \tau}
+ (3/2) \, m^4
\right\}
\nonumber \\ && \phantom{\langle 0 | \widehat{T}^{\phantom{\rho} \rho}_{\rho} | 0 \rangle_\mathrm{ren} =}
+ \Theta^{\phantom{\rho} \rho}_{\rho} .
\end{eqnarray}
The term $\Theta^{\phantom{\rho} \rho}_{\rho}$ encodes the usual ambiguities discussed in Sec.~\ref{Sec.IVe}. In the Minkowski spacetime it reduces to
\begin{eqnarray}
\label{trRqSET_Theta_Minkowski}
&& \Theta^{\phantom{\rho} \rho}_{\rho} = \frac{1}{8\pi^2} \left\{ \alpha \, m^4 \right\} ,
\end{eqnarray}
where $\alpha$ is a constant. From Eq.~\eqref{trRqSET_v5_Minkowski} it is clear that in order to evaluate $\langle 0 | \widehat{T}^{\phantom{\rho} \rho}_{\rho} | 0 \rangle_\mathrm{ren}$, it is sufficient to take the coincidence limit $x' \to x$ of $W^A_{\mu \nu}(x,x')$ and $W^A_{\mu \nu; (a b)}(x,x')$ [see Eqs.~\eqref{coefTaylorSeries_WA_1} and \eqref{coefTaylorSeries_WA_3} and note that, in the Minkowski spacetime, the bivector of parallel transport $g^{\phantom{\mu} \nu'}_\mu (x,x')$ is equal to the unit matrix $\delta^{\phantom{\mu} \nu'}_\mu (x,x')$], where $W^A_{\mu \nu}(x,x')$ is the regular part of the Feynman propagator $G^A_{\mu \nu}(x, x')$ corresponding to the geometry of the Casimir effect.

\subsection{Stress-energy tensor in the Minkowski spacetime}
\label{Sec.Vb}

Let us first consider the vacuum expectation value of the stress-energy-tensor operator in the ordinary Minkowski spacetime (i.e., without the boundary wall). This will permit us to establish some notations and, moreover, to fix the constant $\alpha$ appearing in Eq.~\eqref{trRqSET_Theta_Minkowski}. Due to symmetry considerations, we have
\begin{equation}
\label{SET_Minkowski}
\langle 0 | \widehat{T}_{\mu \nu} | 0 \rangle_\mathrm{ren} = \frac{1}{4} \, \langle 0 | \widehat{T}^{\phantom{\rho} \rho}_{\rho} | 0 \rangle_\mathrm{ren} \, \eta_{\mu \nu} ,
\end{equation}
where $\langle 0 | \widehat{T}^{\phantom{\rho} \rho}_{\rho} | 0 \rangle_\mathrm{ren}$ is still given by Eqs.~\eqref{trRqSET_v5_Minkowski} and \eqref{trRqSET_Theta_Minkowski}. Of course, we must have $\langle 0 | \widehat{T}_{\mu \nu} | 0 \rangle_\mathrm{ren} = 0$, and we have therefore
\begin{equation}
\label{trSET_Minkowski}
\langle 0 | \widehat{T}^{\phantom{\rho} \rho}_{\rho} | 0 \rangle_\mathrm{ren} = 0
\end{equation}
which plays the role of a constraint for $\alpha$.

In the Minkowski spacetime, the Feynman propagator $G^A_{\mu \nu}(x, x')$ associated with the vector field $A_\mu$ satisfies the wave equation \eqref{WEQ_GFA}, i.e.,
\begin{equation}
\label{WEQ_GFA_Minkowski}
\left[ \Box_x - m^2 \right] G^{A}_{\mu \nu} (x,x') = - \eta_{\mu \nu} \delta^4(x,x') ,
\end{equation}
and its explicit expression is given in terms of a Hankel function of the second kind by (see, e.g., Chap.~27 of Ref.~\cite{DeWitt:2003pm})
\begin{equation}
\label{GFA_Expression_Minkowski}
G^A_{\mu \nu} (x, x') =
- \frac{m^2}{8\pi} \, \frac{1}{\mathcal{Z}(x,x')} H^{(2)}_{1} \left[ \mathcal{Z}(x,x') \right] \, \eta_{\mu \nu} .
\end{equation}
Here, $\mathcal{Z}(x,x') = \sqrt{- 2 \, m^2 [\sigma(x,x') + i \epsilon]}$ with $2 \, \sigma(x,x') = - (T-T')^2 + (X-X')^2 + (Y-Y')^2 + (Z-Z')^2$.

We have (see Chap.~9 of Ref.~\cite{AS65})
\begin{align}
\label{fn_Hankel2_1}
H^{(2)}_{1}(z) = J_1(z) - i Y_1(z) ,
\end{align}
where $J_1(z)$ and $Y_1(z)$ are the Bessel functions of the first and second kinds. By using the series expansions for $z \to 0$ (see Eqs.~(9.1.10) and (9.1.11) of Ref.~\cite{AS65})
\begin{subequations}
\begin{eqnarray}
\label{fn_Bessel1_1}
&& J_1(z) = \dfrac{z}{2} \sum_{k=0}^{\infty} \dfrac{(-z^2/4)^{k}}{k!(k+1)!}
\end{eqnarray}
and
\begin{eqnarray}
\label{fn_Bessel2_1}
&& Y_1(z) = - \dfrac{2}{\pi z} + \dfrac{2}{\pi} \ln\left( \dfrac{z}{2} \right) J_1(z)
\nonumber \\ && \quad \hspace{-2mm}
- \dfrac{z}{2 \pi} \sum_{k=0}^{\infty} [\psi(k+1) + \psi(k+2)] \, \dfrac{(-z^2/4)^{k}}{k!(k+1)!}
\end{eqnarray}
\end{subequations}
[we note that Eq.~\eqref{fn_Bessel2_1} is valid for $|\arg(z)| < \pi$, and we recall that the digamma function $\psi$ is defined by the recursion relation $\psi(z+1) = \psi(z) + 1/z$ with $\psi(1) = - \gamma$], we can provide the Hadamard representation of $G^A_{\mu \nu} (x, x')$ given by Eq.~\eqref{GFA_Expression_Minkowski}. We can write
\begin{eqnarray}
\label{HadamardForm}
&& - \frac{m^2}{8\pi} \, \frac{1}{\mathcal{Z}(x,x')} H^{(2)}_{1} \left[ \mathcal{Z}(x,x') \right] = \dfrac{i}{8\pi^2} \left[ \dfrac{\Delta^{1/2}(x,x')}{\sigma(x,x')+i\epsilon}
\right. \nonumber \\ && \left. \quad \vphantom{\dfrac{\Delta^{(1/2)}(x,x')}{\sigma(x,x')+i\epsilon}}
+ V(x,x') \ln[\sigma(x,x')+i\epsilon] + W(x,x') \right] ,
\end{eqnarray}
where
\begin{subequations}
\label{HadamardCoefs}
\begin{eqnarray}
&& \Delta^{1/2}(x,x') = 1 ,
\end{eqnarray}
\begin{eqnarray}
\label{HadamardCoef_V}
&& V(x,x') =  \sum_{k=0}^{\infty} V_k \, \sigma^k(x,x')
\end{eqnarray}
with
\begin{eqnarray}
\label{HadamardCoef_V_k}
&& V_k = \dfrac{(m^2/2)^{k+1}}{k! (k+1)!}
\end{eqnarray}
and
\begin{eqnarray}
\label{HadamardCoef_W}
&& W(x,x') =  \sum_{k=0}^{\infty} W_k \sigma^k(x,x')
\end{eqnarray}
with
\begin{eqnarray}
\label{HadamardCoef_W_k}
&& W_k = - \dfrac{(m^2/2)^{k+1}}{k! (k+1)!} \, \left[ \psi(k+1) + \psi(k+2) - \ln\left( \dfrac{m^2}{2} \right) \right] .
\nonumber \\
\end{eqnarray}
\end{subequations}

By noting that
\begin{equation}
\label{HadamardCoef_WA=etaW}
W^A_{\mu \nu}(x,x') = W(x,x') \, \eta_{\mu \nu} ,
\end{equation}
where $W(x,x')$ is given by Eqs.~\eqref{HadamardCoef_W} and \eqref{HadamardCoef_W_k}, we are now able to express $\langle 0 | \widehat{T}^{\phantom{\rho} \rho}_{\rho} | 0 \rangle_\mathrm{ren}$. From Eqs.~\eqref{coefTaylorSeries_WA_1} and \eqref{coefTaylorSeries_WA_3} we have, respectively,
\begin{eqnarray}
\label{s1_Minkowski}
&& s_{\mu \nu} = m^2 \left[ - 1/2 + \gamma + (1/2) \, \ln(m^2/2) \right] \eta_{\mu \nu}
\end{eqnarray}
and
\begin{eqnarray}
\label{s2_Minkowski}
&& s_{\mu \nu a b} = m^4 \left[ - 5/16 + (1/4) \, \gamma + (1/8) \ln(m^2/2) \right] \eta_{\mu \nu} \eta_{a b} .
\nonumber \\
\end{eqnarray}
Then, from Eq.~\eqref{trRqSET_v5_Minkowski}, we obtain
\begin{eqnarray}
\label{trRqSET_v5_Expression_Minkowski}
&& \langle 0 | \widehat{T}^{\phantom{\rho} \rho}_{\rho} | 0 \rangle_\mathrm{ren} = \frac{m^4}{8\pi^2} \left\{ \alpha + 9/4 - 3 \, \gamma - (3/2) \, \ln(m^2/2) \right\} ,
\nonumber \\
\end{eqnarray}
and, necessarily, by using Eq.~\eqref{trSET_Minkowski}, we have the constraint
\begin{eqnarray}
\label{Theta_alpha}
&& \alpha = - 9/4 + 3 \, \gamma + (3/2) \, \ln(m^2/2) .
\end{eqnarray}

\subsection{Stress-energy tensor for the Casimir effect}
\label{Sec.Vc}

Let us now come back to our initial problem. The Feynman propagator previously considered is modified by the presence of the plane boundary wall. The new Feynman propagator $\widetilde{G}^A_{\mu \nu}(x, x')$ can be constructed by the method of images if we assume, in order to simplify our problem, a perfectly reflecting wall. It should be noted that this particular boundary condition is questionable from the physical point of view. It is logical for the transverse components of the electromagnetic field but much less natural for its longitudinal component. Indeed, for this component, we could also consider perfect transmission instead of complete reflection (see Refs.~\cite{BassSchrodinger,Davies:1980ji,Davies:1984vm}). We shall now consider that the Feynman propagator is given by
\begin{eqnarray}
\label{GFA_Minkowski+Wall}
&& \widetilde{G}^A_{\mu \nu}(x, x') = G^A_{\mu \nu}(x, x') - q_\nu \, G^A_{\mu \nu}(x, \tilde{x}') .
\end{eqnarray}
Here, $x'^\mu = (T', X', Y', Z')$ and $\tilde{x}'^\mu = (T', X', Y', -Z')$, while $q_\nu = (1 - 2 \, \delta_{3 \nu})$. It is important to note that, in Eq.~\eqref{GFA_Minkowski+Wall}, the index $\nu$ is not summed. Furthermore, we remark that the term $G^A_{\mu \nu}(x, \tilde{x}')$ which is obtained by replacing $x'$ by $\tilde{x}'$ in Eq.~\eqref{GFA_Expression_Minkowski} as well as its derivatives are regular in the limit $x' \to x$.

By following the steps of Sec.~\ref{Sec.Vb} and using the relation
\begin{eqnarray}
&& K_\nu (z) = - (1/2) \, i \pi \, e^{-i\pi\nu/2} \, H^{(2)}_\nu (z \, e^{-i\pi/2})
\end{eqnarray}
which is valid for $- \pi/2 \leq \arg(z) \leq \pi$ as well as the properties of the modified Bessel functions of the second kind $K_1$, $K_2$ and $K_3$ (see Chap.~9 of Ref.~\cite{AS65}), it is easy to show that the Taylor coefficients $s_{\mu \nu}$ and $s_{\mu \nu a b}$ involved in $\langle 0 | \widehat{T}^{\phantom{\rho} \rho}_{\rho} | 0 \rangle_\mathrm{ren}$ are now given by
\begin{eqnarray}
\label{s1_Minkowski+Wall}
&& s_{\mu \nu} = m^2 \left[ - 1/2 + \gamma + (1/2) \, \ln(m^2/2) \right] \eta_{\mu \nu}
\nonumber \\ && \quad
- (m/Z) \, K_1(2 m Z) \, q_\nu \, \eta_{\mu \nu}
\end{eqnarray}
and
\begin{eqnarray}
\label{s2_Minkowski+Wall}
&& s_{\mu \nu a b} = m^4 \left[ - 5/16 + (1/4) \, \gamma + (1/8) \ln(m^2/2) \right] \eta_{\mu \nu} \eta_{a b}
\nonumber \\ && \quad
- \left[
  (m^2/Z^2) \, K_2(2 m Z) \, q_\nu \, \eta_{\mu \nu} \left( 2 \, \eta_{3 a} \, \eta_{3 b} - (1/2) \, \eta_{a b} \right)
\right. \nonumber \\ && \left. \qquad
+ (m^3/Z) \, K_1(2 m Z) \, q_\nu \, \eta_{\mu \nu} \eta_{3 a} \eta_{3 b}
\right] .
\end{eqnarray}
By inserting Eqs.~\eqref{s1_Minkowski+Wall} and \eqref{s2_Minkowski+Wall} in the expression \eqref{trRqSET_v5_Minkowski} and using the value of $\alpha$ fixed by Eq.~\eqref{Theta_alpha}, we obtain
\begin{eqnarray}
\label{trSET_Expression_Minkowski+Wall}
&& \langle 0 | \widehat{T}^{\phantom{\rho} \rho}_\rho | 0 \rangle_\mathrm{ren} = \frac{3}{8\pi^2} \left\{ \frac{m^2}{Z^2} \, K_2(2mZ) + \frac{m^3}{Z} \, K_1(2mZ) \right\} ,
\nonumber \\
\end{eqnarray}
and from Eq.~\eqref{SET_Minkowski+Wall} we have
\begin{eqnarray}
\label{SET_Expression_Minkowski+Wall}
&& \langle 0 | \widehat{T}_{\mu \nu} | 0 \rangle_\mathrm{ren} = \frac{1}{8\pi^2} \left\{ \frac{m^2}{Z^2} \, K_2(2mZ) + \frac{m^3}{Z} \, K_1(2mZ) \right\}
\nonumber \\ && \hphantom{\langle \widehat{T}^{\phantom{\rho} \rho}_\rho \rangle_\mathrm{ren} =}
\times \left( \eta_{\mu \nu} - \hat{Z}_\mu \hat{Z}_\nu \right) .
\end{eqnarray}
It is very important to note that this result coincides exactly with the result obtained by Davies and Toms in the framework of de Broglie-Proca electromagnetism \cite{Davies:1984vm}.

In the limit $m^2 \to 0$ and for $Z \neq 0$, we obtain
\begin{eqnarray}
\label{SET_Expression_Minkowski+Wall_m=0}
&& \langle 0 | \widehat{T}_{\mu \nu} | 0 \rangle_\mathrm{ren} = \frac{1}{16\pi^2} \, \frac{1}{Z^4} \left( \eta_{\mu \nu} - \hat{Z}_\mu \hat{Z}_\nu \right) .
\end{eqnarray}
In the massless limit, the vacuum expectation value of the renormalized stress-energy tensor associated with the Stueckelberg theory diverges like $Z^{-4}$ as the boundary surface is approached. This result contrasts with that obtained from Maxwell's theory (see also Ref.~\cite{Davies:1984vm}). Indeed, for this theory, the renormalized stress-energy-tensor operator vanishes identically. In order to extract that result from the Stuckelberg theory, we will now repeat the previous calculations from the expressions \eqref{RqSET_v4} and \eqref{RqSET_v4_A_Phi} [as well as \eqref{trRqSET_v4} and \eqref{trRqSET_v4_A_Phi}] given in Sec.~\ref{Sec.IVc4}, where we have proposed an artificial separation of the contributions associated with the vector field $A_\mu$ and the auxiliary scalar field $\Phi$.

\subsection{Separation of the contributions associated with the vector field $A_\mu$ and the auxiliary scalar field $\Phi$}
\label{Sec.Vd}

In the Minkowski spacetime, Eqs.~\eqref{trRqSET_v4} and \eqref{trRqSET_v4_A_Phi} reduce to
\begin{equation}
\label{trRqSET_v4_Minkowski}
\langle 0 | \widehat{T}^{\phantom{\rho} \rho}_{\rho} | 0 \rangle_\mathrm{ren} = \mathcal{T}^A{}^{\phantom{\rho} \rho}_{\rho} + \mathcal{T}^\Phi{}^{\phantom{\rho} \rho}_{\rho} + \Theta^{\phantom{\rho} \rho}_{\rho}
\end{equation}
with
\begin{eqnarray}
\label{trRqSET_v4_A_Minkowski}
&& \mathcal{T}^A{}^{\phantom{\rho} \rho}_{\rho} = \frac{1}{8\pi^2} \left\{
- s^{\phantom{\rho \tau} ;\rho \tau}_{\rho \tau}
- m^2 s^{\phantom{\rho} \rho}_{\rho}
\right. \nonumber \\ && \left. \quad
+ 2 \, a^{\phantom{\rho \tau} \rho ;\tau}_{\rho \tau}
+ 2 \, s^{\phantom{\rho \tau} \rho \tau}_{\rho \tau}
+ 2 \, m^4
\right\}
\end{eqnarray}
and
\begin{eqnarray}
\label{trRqSET_v4_Phi_Minkowski}
&& \hspace{-8mm} \mathcal{T}^\Phi{}^{\phantom{\rho} \rho}_{\rho} = \frac{1}{8\pi^2} \left\{ - (1/2) \, \Box w - m^2 w + (1/4) \, m^4 \right\} .
\end{eqnarray}
The term $\Theta^{\phantom{\rho} \rho}_{\rho}$ encodes the usual ambiguities discussed in Sec.~\ref{Sec.IVe}. We can split it in the form
\begin{subequations}
\label{trRqSET_Theta_v4_Minkowski}
\begin{eqnarray}
\label{trRqSET_Theta_v4_Minkowski_A_Phi}
&& \Theta^{\phantom{\rho} \rho}_{\rho} = \Theta^A{}^{\phantom{\rho} \rho}_{\rho} + \Theta^\Phi{}^{\phantom{\rho} \rho}_{\rho}
\end{eqnarray}
with
\begin{eqnarray}
\label{trRqSET_Theta_v4_Minkowski_A}
&& \Theta^A{}^{\phantom{\rho} \rho}_{\rho} = \frac{1}{8\pi^2} \left\{ \alpha^A \, m^4 \right\}
\end{eqnarray}
and
\begin{eqnarray}
\label{trRqSET_Theta_v4_Minkowski_Phi}
&& \Theta^\Phi{}^{\phantom{\rho} \rho}_{\rho} = \frac{1}{8\pi^2} \left\{ \alpha^\Phi \, m^4 \right\} ,
\end{eqnarray}
\end{subequations}
where $\alpha^A$ and $\alpha^\Phi$ are two constants associated, respectively, with the contributions of the vector field $A_\mu$ and the auxiliary scalar field $\Phi$. We can then replace Eq.~\eqref{trRqSET_v4_Minkowski} by
\begin{eqnarray}
\label{trRqSET_v4_Expression_Minkowski_bis}
&& \langle 0 | \widehat{T}^{\phantom{\rho} \rho}_{\rho} | 0 \rangle_\mathrm{ren} =
  \langle 0 | \widehat{T}^A{}^{\phantom{\rho} \rho}_{\rho} | 0 \rangle_\mathrm{ren}
+ \langle 0 | \widehat{T}^\Phi{}^{\phantom{\rho} \rho}_{\rho} | 0 \rangle_\mathrm{ren}
\end{eqnarray}
with
\begin{eqnarray}
\label{trRqSET_v4_A_Expression_Minkowski_bis}
&& \langle 0 | \widehat{T}^A{}^{\phantom{\rho} \rho}_{\rho} | 0 \rangle_\mathrm{ren} = \mathcal{T}^A{}^{\phantom{\rho} \rho}_{\rho} + \Theta^A{}^{\phantom{\rho} \rho}_{\rho}
\end{eqnarray}
and
\begin{eqnarray}
\label{trRqSET_v4_Phi_Expression_Minkowski_bis}
&& \langle 0 | \widehat{T}^\Phi{}^{\phantom{\rho} \rho}_{\rho} | 0 \rangle_\mathrm{ren} = \mathcal{T}^\Phi{}^{\phantom{\rho} \rho}_{\rho} + \Theta^\Phi{}^{\phantom{\rho} \rho}_{\rho} ,
\end{eqnarray}
where the contributions associated with the vector field $A_\mu$ and the auxiliary scalar field $\Phi$ are separated. At first sight, $\mathcal{T}^A{}^{\phantom{\rho} \rho}_{\rho}$ seems complicated because it involves Taylor coefficients of orders $\sigma^{1/2}$ and $\sigma^1$ of $W^A_{\mu \nu}(x,x')$. In fact, its expression can be simplified by replacing the sum $a^{\phantom{\rho \tau} \rho ;\tau}_{\rho \tau} + s^{\phantom{\rho \tau} \rho \tau}_{\rho \tau}$ from the relation \eqref{Relation_coefTaylorSeries_WA_W_2}, and we obtain
\begin{eqnarray}
\label{trRqSET_v4_A_Minkowski_bis}
&& \mathcal{T}^A{}^{\phantom{\rho} \rho}_{\rho} = \frac{1}{8\pi^2} \left\{
- m^2 s^{\phantom{\rho} \rho}_{\rho}
+ 2 \, m^2 w
+ (1/2) \, m^4
\right\}
\end{eqnarray}
which only involves the first Taylor coefficients $s_{\mu\nu}$ and $w$ of order $\sigma^0$. So, in order to evaluate $\langle 0 | \widehat{T}^{\phantom{\rho} \rho}_{\rho} | 0 \rangle_\mathrm{ren}$ given by Eq.~\eqref{trRqSET_v4_Minkowski}, it is sufficient to take the coincidence limit $x' \to x$ of the state-dependent Hadamard coefficients $W^A_{\mu \nu}(x,x')$ and $W(x,x')$ associated with the Feynman propagators $G^A_{\mu \nu}(x, x')$ and $G^\Phi(x, x')$ corresponding to the geometry of the Casimir effect.

At first, we must fix the constants $\alpha^A$ and $\alpha^\Phi$ appearing in Eq.~\eqref{trRqSET_Theta_v4_Minkowski}. This can be achieved by imposing, in the Minkowski spacetime without boundary, the vanishing of $\langle 0 | \widehat{T}^A{}^{\phantom{\rho} \rho}_{\rho} | 0 \rangle_\mathrm{ren}$ given by Eq.~\eqref{trRqSET_v4_A_Expression_Minkowski_bis} and $\langle 0 | \widehat{T}^\Phi{}^{\phantom{\rho} \rho}_{\rho} | 0 \rangle_\mathrm{ren}$ given by Eq.~\eqref{trRqSET_v4_Phi_Expression_Minkowski_bis}. In this spacetime, everything related to the Feynman propagator $G^A_{\mu \nu}(x, x')$ has been already given in Sec.~\ref{Sec.Vb}, while the Feynman propagator $G^\Phi(x, x')$ associated with the scalar field $\Phi$ satisfies the wave equation \eqref{WEQ_GFPhi} and is explicitly given by
\begin{equation}
\label{GFPhi_Expression_Minkowski}
G^\Phi (x, x') =
- \frac{m^2}{8\pi} \, \frac{1}{\mathcal{Z}(x,x')} H^{(2)}_{1} \left[ \mathcal{Z}(x,x') \right] .
\end{equation}
By using Eqs.~\eqref{HadamardForm} and \eqref{HadamardCoefs}, it is easy to see that this propagator can be represented in the Hadamard form and to obtain
\begin{eqnarray}
\label{w1_Minkowski}
&& w = m^2 \left[ - 1/2 + \gamma + (1/2) \, \ln(m^2/2) \right] .
\end{eqnarray}
We are now able to express $\langle 0 | \widehat{T}^{\phantom{\rho} \rho}_\rho | 0 \rangle_\mathrm{ren}$. From Eqs.~\eqref{trRqSET_v4_A_Expression_Minkowski_bis}, \eqref{trRqSET_v4_Phi_Expression_Minkowski_bis}, \eqref{trRqSET_v4_A_Minkowski_bis}, \eqref{trRqSET_v4_Phi_Minkowski} and \eqref{trRqSET_Theta_v4_Minkowski}, we obtain
\begin{eqnarray}
\label{trSET_Expression_A_Minkowski}
&& \langle 0 | \widehat{T}^A{}^{\phantom{\rho} \rho}_{\rho} | 0 \rangle_\mathrm{ren} = \frac{m^4}{8\pi^2} \left\{ \alpha^A + 3/2 - 2 \, \gamma - \ln(m^2/2) \right\}
\nonumber \\
\end{eqnarray}
and
\begin{eqnarray}
\label{trSET_Expression_Phi_Minkowski}
&& \langle 0 | \widehat{T}^\Phi{}^{\phantom{\rho} \rho}_{\rho} | 0 \rangle_\mathrm{ren} = \frac{m^4}{8\pi^2} \left\{ \alpha^\Phi + 3/4 - \gamma - (1/2) \, \ln(m^2/2) \right\} ,
\nonumber \\
\end{eqnarray}
and, necessarily, the vanishing of these traces provides the two constraints
\begin{subequations}
\begin{eqnarray}
\label{Theta_alpha_A}
\alpha^A = - 3/2 + 2 \, \gamma + \ln(m^2/2)
\end{eqnarray}
and
\begin{eqnarray}
\label{Theta_alpha_Phi}
\alpha^\Phi = - 3/4 + \gamma + (1/2) \, \ln(m^2/2) .
\end{eqnarray}
\end{subequations}

We now come back to the Casimir effect. The two Feynman propagators previously considered are modified by the presence of the plane boundary wall. The new Feynman propagators can be constructed by the method of images. Of course, the propagator of the vector field $A_\mu$ is still given by Eq.~\eqref{GFA_Minkowski+Wall}, while we have
\begin{eqnarray}
\label{GFPhi_Minkowski+Wall}
&& \widetilde{G}^\Phi(x, x') = G^\Phi(x, x') - G^\Phi(x, \tilde{x}')
\end{eqnarray}
for the propagator of the scalar field $\Phi$. In the context of the Casimir effect, Eq.~\eqref{w1_Minkowski} must be replaced by
\begin{eqnarray}
\label{w1_Minkowski+Wall}
&& w = m^2 \left[ - 1/2 + \gamma + (1/2) \, \ln(m^2/2) \right]
\nonumber \\ && \quad
- (m/Z) \, K_1(2 m Z) ,
\end{eqnarray}
and $s_{\mu \nu}$ is given by Eq.~\eqref{s1_Minkowski+Wall}. By inserting Eqs.~\eqref{s1_Minkowski+Wall} and \eqref{w1_Minkowski+Wall} in Eqs.~\eqref{trRqSET_v4_A_Minkowski_bis} and \eqref{trRqSET_v4_Phi_Minkowski} and taking into account the constraints \eqref{Theta_alpha_A} and \eqref{Theta_alpha_Phi}, we obtain from Eq.~\eqref{trRqSET_v4_A_Expression_Minkowski_bis}
\begin{eqnarray}
\label{trRqSET_v4_A_Expression_Minkowski+Wall_bis_Expr}
&& \langle 0 | \widehat{T}^A{}^{\phantom{\rho} \rho}_{\rho} | 0 \rangle_\mathrm{ren} = 0
\end{eqnarray}
and from Eq.~\eqref{trRqSET_v4_Phi_Expression_Minkowski_bis}
\begin{eqnarray}
\label{trRqSET_v4_Phi_Expression_Minkowski+Wall_bis_Expr}
&& \langle 0 | \widehat{T}^\Phi{}^{\phantom{\rho} \rho}_{\rho} | 0 \rangle_\mathrm{ren} = \frac{3}{8\pi^2} \left\{ \frac{m^2}{Z^2} \, K_2(2mZ) + \frac{m^3}{Z} \, K_1(2mZ) \right\} .
\nonumber \\
\end{eqnarray}
From Eq.~\eqref{SET_Minkowski+Wall} we can then see that the vacuum expectation value of the stress-energy-tensor operator associated with the vector field $A_\mu$ is such that
\begin{eqnarray}
\label{SET_A_Expression_Minkowski+Wall}
&& \langle 0 | \widehat{T}^A_{\mu \nu} | 0 \rangle_\mathrm{ren} = 0 ,
\end{eqnarray}
while the vacuum expectation value of the stress-energy-tensor operator associated with the auxiliary scalar field $\Phi$ is given by
\begin{eqnarray}
\label{SET_Phi_Expression_Minkowski+Wall}
&& \langle 0 | \widehat{T}^\Phi_{\mu \nu} | 0 \rangle_\mathrm{ren} = \frac{1}{8\pi^2} \left\{ \frac{m^2}{Z^2} \, K_2(2mZ) + \frac{m^3}{Z} \, K_1(2mZ) \right\}
\nonumber \\ && \hphantom{\langle \widehat{T}^{\phantom{\rho} \rho}_\rho \rangle_\mathrm{ren} =}
\times \left( \eta_{\mu \nu} - \hat{Z}_\mu \hat{Z}_\nu \right) .
\end{eqnarray}
Of course, the sum of these two contributions permits us to recover the result \eqref{SET_Expression_Minkowski+Wall} of Sec.~\ref{Sec.Vc} which is also the result obtained by Davies and Toms in the framework of de Broglie-Proca electromagnetism \cite{Davies:1984vm}. It is moreover interesting to note that the contribution \eqref{SET_A_Expression_Minkowski+Wall} associated with the vector field $A_\mu$ and which has been artificially separated from the scalar field contribution (see Sec.~\ref{Sec.IVc4}) vanishes identically for any value of the mass parameter $m$. This result coincides exactly with that obtained from Maxwell's theory (see also Ref.~\cite{Davies:1984vm}).

\section{Conclusion}
\label{Sec.VI}

In the context of quantum field theory in curved spacetime and with possible applications to cosmology and to black hole physics in mind, the massive vector field is frequently studied. It should be, however, noted that, in this particular domain, it is its description via the de Broglie-Proca theory which is mostly considered and that there are very few works dealing with the Stueckelberg point of view (see, e.g., Refs.~\cite{Janssen:1986fz,Chimento:1990dk,Tsamis:2006gj,Frob:2013qsa,Akarsu:2014eaa,Kouwn:2015cdw}, but remark that these papers are restricted to de Sitter and anti-de Sitter spacetimes or to Roberstson-Walker backgrounds with spatially flat sections). In this article, in order to fill a void, we have developed the general formalism of the Stueckelberg theory on an arbitrary four-dimensional spacetime (quantum action, Feynman propagators, Ward identities, Hadamard representation of the Green functions), and we have particularly focussed on the aspects linked with the construction, for a Hadamard quantum state, of the expectation value of the renormalized stress-energy-tensor operator. It is important to note that we have given two alternative but equivalent expressions for this result. The first one has been obtained by eliminating from a Ward identity the contribution of the auxiliary scalar field $\Phi$ (the so-called Stueckelberg ghost \cite{vanHees:2003dk}) and only involves state-dependent and geometrical quantities associated with the massive vector field $A_\mu$ [see Eq.~\eqref{RqSET_v5}]. The other one involves contributions coming from both the massive vector field and the auxiliary Stueckelberg scalar field [see Eqs.~\eqref{RqSET_v4}--\eqref{RqSET_v4_A_Phi}], and it has been constructed artificially in such a way that these two contributions are independently conserved and that, in the zero-mass limit, the massive vector field contribution reduces smoothly to the result obtained from Maxwell's electromagnetism. It is also important to note that, in Sec.~\ref{Sec.IVe}, we have discussed the geometrical ambiguities of the expectation value of the renormalized stress-energy-tensor operator. They are of fundamental importance (see, e.g., in Sec.~\ref{Sec.V}, their role in the context of the Casimir effet).

We intend to use our results in the near future in cosmology of the very early universe, but we hope they will be useful for other authors. This is why we shall now provide a step-by-step guide for the reader who is not especially interested by the technical details of our work but who wishes to calculate the expectation value of the renormalized stress-energy tensor from the expression \eqref{RqSET_v5}, i.e., from the expression where any reference to the Stueckelberg auxiliary scalar field $\Phi$ has disappeared. We shall describe the calculation from the Feynman propagator as well as from the anticommutator function :

\begin{itemize}
  \item[(i)] We assume that the Feynman propagator $G^{A}_{\mu \nu'} (x,x')$ which is given by Eq.~\eqref{FeynmanProp_A} and satisfies the wave equation \eqref{WEQ_GFA_1} [or that the anticommutator $G^{(1)A}_{\mu \nu'} (x,x')$ which is given by Eq.~\eqref{HadamardGreenFn_A} and satisfies the wave equation \eqref{WEQ_G1A}] has been determined in a particular gravitational background and for a Hadamard quantum state. In other words, we consider that the Feynman propagator $G^{A}_{\mu \nu'} (x,x')$ can be represented in the Hadamard form \eqref{HadamardRep_GAF} [or that the anticommutator $G^{(1)A}_{\mu \nu'} (x,x')$ can be represented in the Hadamard form \eqref{HadamardRep_G1A}].
  \item[(ii)] We need the regular part of the Feynman propagator $G^{A}_{\mu \nu'} (x,x')$ [or that of the anticommutator $G^{(1)A}_{\mu \nu'} (x,x')$] at order $\sigma$. To extract it, we subtract from the Feynman propagator $G^{A}_{\mu \nu'} (x,x')$ its singular part \eqref{HadamardRep_GFA_sing} in order to obtain its regular part \eqref{HadamardRep_GFA_reg} [or we subtract from the anticommutator $G^{(1)A}_{\mu \nu'} (x,x')$ its singular part \eqref{HadamardRep_G1A_sing} in order to obtain its regular part \eqref{HadamardRep_G1A_reg}]. We have then at our disposal the state-dependent Hadamard bivector $W^{A}_{\mu \nu'} (x,x')$. Here, it is important to note that we do not need the full expression of the singular part of the Green function considered, but we can truncate it by neglecting the terms vanishing faster than $\sigma (x,x')$ for $x' \to x$. As a consequence, we can construct the singular part \eqref{HadamardRep_GFA_sing} [or the singular part \eqref{HadamardRep_G1A_sing}] by using the covariant Taylor series expansion \eqref{CovTaylorSeries_1} of $\Delta^{1/2}$ up to order $\sigma^2$, the covariant Taylor series expansion \eqref{covTaylorSeries_VA0} of $V^A_0{}_{\mu \nu}$ up to order $\sigma^1$ [see Eqs.~\eqref{coefTaylorSeries_vA0()}--\eqref{coefTaylorSeries_vA0()ab}] and the covariant Taylor series expansion \eqref{covTaylorSeries_VA1} of $V^A_1{}_{\mu \nu}$ up to order $\sigma^0$ [see Eq.~\eqref{coefTaylorSeries_vA1()}].
  \item[(iii)] Finally, by using Eqs.~\eqref{coefTaylorSeries_WA_1}--\eqref{coefTaylorSeries_WA_3}, we can construct the expectation value of the renormalized stress-energy tensor given by Eq.~\eqref{RqSET_v5}.
\end{itemize}

It is interesting to note that, in the literature concerning Stueckelberg electromagnetism, some authors only focus on the part of the action associated with the massive vector field $A_\mu$ and which is given by Eq.~\eqref{Action_Stueck_Quant_v2_A_1} (see, e.g., Refs.~\cite{ItzyksonZuber,Janssen:1986fz,Frob:2013qsa}). Of course, this is sufficient because they are mainly interested, in the context of canonical quantization, by the determination of the Feynman propagator associated with this field. However, in order to calculate physical quantities, it is necessary to take into account the contribution of the auxiliary scalar field $\Phi$. It cannot be discarded. This is very clear in the context of the construction of the renormalized stress-energy-tensor operator as we have shown in our article and remains true for any other physical quantity.

To conclude this article, we shall briefly compare the de Broglie-Proca and Stueckelberg formulations of massive electomagnetism and discuss the advantages of the Stueckelberg formulation over the de Broglie-Proca one. It is interesting to note the existence of a nice paper by Pitts \cite{Pitts:2009dj}, where de Broglie-Proca and Stueckelberg approaches of massive electromagnetism are discussed from a philosophical point of view based on the machinery of the Hamiltonian formalism (primary and secondary constraints, Poisson brackets, \dots). Here, we adopt a more pragmatic point of view. We discuss the two formulations in light of the results obtained in our article. In our opinion:

\begin{itemize}
  \item[(i)] De Broglie-Proca and Stueckelberg approaches of massive electromagnetism are two faces of the same theory. Indeed, the transition from de Broglie-Proca to Stueckelberg theory is achieved via the Stueckelberg trick \eqref{Action_dBProca_to_Stueck} which permits us, by introducing an auxiliary scalar field $\Phi$, to artificially restore Maxwell's gauge symmetry in massive electromagnetism, but, reciprocally, the transition from Stueckelberg to de Broglie-Proca theory is achieved by imposing the gauge choice $\Phi =0$ [see Eq.~\eqref{Action_Stueck_to_dBProca}]. As a consequence, it is not really surprising to obtain the same result for the renormalized stress-energy-tensor operator associated with the Casimir effect (see Sec.~\ref{Sec.V}) when we consider this problem in the framework of the de Broglie-Proca and Stueckelberg formulations of massive electromagnetism. Indeed, we can expect that this remains true for any other quantum quantity.
  \item[(ii)] However, we can note that with regularization and renormalization in mind, it is much more interesting to work in the framework of the Stueckelberg formulation of massive electromagnetism. Indeed, this permits us to have at our disposal the machinery of the Hadamard formalism which is not the case in the framework of the de Broglie-Proca formulation. Indeed, due to the constraint \eqref{LorenzCondition_dBProca}, the Feynman propagator $G^A_{\mu \nu'} (x,x')$ associated with the vector field $A_\mu$ cannot be represented in the Hadamard form \eqref{HadamardRep_GAF}.
\end{itemize}

\begin{acknowledgments}

We wish to thank Yves D\'ecanini, Mohamed Ould El Hadj and Julien Queva for various discussions and the ``Collectivit\'e Territoriale de Corse" for its support through the COMPA project.

\end{acknowledgments}

\bigskip

\appendix*

\section{Biscalars, bivectors and their covariant Taylor series expansions}
\label{Appendix}

Regularization and renormalization of quantum field theories in the Minkowski spacetime are most times based on the representation of Green functions in momentum space, and, in general, this greatly simplifies reasoning and calculations. The use of such a representation is not possible in an arbitrary gravitational background where the lack of symmetries as well as spacetime curvature prevent us from working within the framework of the Fourier transform. As a consequence, regularization and renormalization in curved spacetime are necessarily based on representations of Green functions in coordinate space, and, moreover, they require extensively the concepts of biscalars, bivectors and, more generally, bitensors. Thanks to the work of some mathematicians \cite{Hadamard,Lichnerowicz:1961,Garabedian,Friedlander} and of DeWitt \cite{DeWitt:1960fc,DeWitt65,DeWitt:1975ys,DeWitt:2003pm} and coworkers \cite{Christensen:1976vb,Christensen:1978yd}, we have at our disposal all the tools necessary to deal with this subject.

In this short Appendix, in order to make a self-consistent paper (i.e., to avoid the reader needing to consult the references mentioned above), we have gathered some important results which are directly related with the representations of Green functions in coordinate space and, more particularly, with the Hadamard representations of the Green functions appearing in Stueckelberg electromagnetism [see Eqs.~\eqref{HadamardRep_GAF}, \eqref{HadamardRep_GF}, \eqref{HadamardRep_G1A} and \eqref{HadamardRep_G1}] which is the main subject of Sec.~\ref{Sec.III} and which plays a crucial role in Sec.~\ref{Sec.IV}. In particular, we define the geodetic interval $\sigma(x,x')$, the Van Vleck-Morette determinant $\Delta(x,x')$ and the bivector of parallel transport from $x$ to $x'$ denoted by $g_{\mu \nu'} (x,x')$ (see, e.g., Ref.~\cite{DeWitt:1960fc}), and we moreover discuss the concept of covariant Taylor series expansions for biscalars and bivectors.

We first recall that $2\sigma(x,x')$ is the square of the geodesic distance between $x$ and $x'$ which satisfies
\begin{equation}
\label{Def_GeodeticInterval}
2 \sigma = \sigma^{;\mu} \sigma_{;\mu} .
\end{equation}
We have $\sigma(x,x') < 0$ if $x$ and $x'$ are timelike related, $\sigma(x,x') = 0$ if $x$ and $x'$ are null related and $\sigma(x,x') > 0$ if $x$ and $x'$ are spacelike related. We furthermore recall that $\Delta(x,x')$ is given by
\begin{equation}
\label{Def_DetVanVleckMorette}
\Delta (x,x') = - [-g(x)]^{-1/2} \mathrm{det} ( - \sigma_{;\mu \nu'}(x,x')) [-g(x')]^{-1/2}
\end{equation}
and satisfies the partial differential equation
\begin{subequations}
\label{Def_DetVanVleckMorette_DiffEq_Boundary}
\begin{equation}
\label{Def_DetVanVleckMorette_DiffEq}
\Box_x \sigma = 4 - 2 \Delta ^{-1/2}{\Delta ^{1/2}}_{;\mu} \sigma^{;\mu}
\end{equation}
as well as the boundary condition
\begin{equation}
\label{Def_DetVanVleckMorette_Boundary}
\lim_{x' \to x} \Delta (x,x') = 1 .
\end{equation}
\end{subequations}
The bivector of parallel transport from $x$ to $x'$ is defined by the partial differential equation
\begin{subequations}
\label{Def_ParallelTransport_EqDef_Boundary}
\begin{equation}
\label{Def_ParallelTransport_EqDef}
g_{\mu \nu' ;\rho} \sigma^{;\rho} = 0
\end{equation}
and the boundary condition
\begin{equation}
\label{Def_ParallelTransport_Boundary}
\lim_{x' \to x} g_{\mu \nu'} (x,x') = g_{\mu \nu} (x) .
\end{equation}
\end{subequations}

The Hadamard coefficients $V^A_n{}_{\mu \nu'}(x,x')$ and $W^A_n{}_{\mu \nu'}(x,x')$ introduced in Eq.~\eqref{Expansion_VA_WA} and which are bivectors involved in the Hadamard representation of the Green functions \eqref{HadamardRep_GAF} and \eqref{HadamardRep_G1A} or the Hadamard coefficients $V_n(x,x')$ and $W_n(x,x')$ introduced in Eq.~\eqref{Expansion_V_W} and which are biscalars involved in the Hadamard representation of the Green functions \eqref{HadamardRep_GF} and \eqref{HadamardRep_G1} cannot in general be determined exactly. They are solutions of the recursion relations \eqref{Recursion_VAn}, \eqref{Recursion_VA0} and \eqref{Recursion_WAn} or \eqref{Recursion_Vn}, \eqref{Recursion_V0} and \eqref{Recursion_Wn}, and, following DeWitt \cite{DeWitt:1960fc,DeWitt65}, we can look for the solutions of these equations in the form of covariant Taylor series expansions for $x'$ in the neighborhood of $x$. This is the method we use in Sec.~\ref{Sec.III}. The series defining the biscalars $V_n(x,x')$ and $W_n(x,x')$ can be written in the form

\begin{widetext}
\begin{eqnarray}
\label{CovTaylorSeries_Scalar}
&& T(x,x') = t(x) - t_{a_1}(x) \, \sigma^{;a_1}(x,x')
+ \frac{1}{2!} \, t_{a_1 a_2}(x) \, \sigma^{;a_1}(x,x') \, \sigma^{;a_2}(x,x')
\nonumber \\ && \qquad
- \frac{1}{3!} \, t_{a_1 a_2 a_3}(x) \, \sigma^{;a_1}(x,x') \, \sigma^{;a_2}(x,x')  \, \sigma^{;a_3}(x,x')
\nonumber \\ && \qquad
+ \frac{1}{4!} \, t_{a_1 a_2 a_3 a_4}(x) \, \sigma^{;a_1}(x,x') \, \sigma^{;a_2}(x,x') \, \sigma^{;a_3}(x,x') \, \sigma^{;a_4}(x,x')
+ \cdots .
\end{eqnarray}
\end{widetext}

\noindent By construction, the coefficients $t_{a_1 \dots a_p}(x)$ are symmetric in the exchange of the indices $a_1 \dots a_p$, i.e., $t_{a_1 \dots a_p}(x) = t_{(a_1 \dots a_p)}(x)$, and, moreover, by requiring the symmetry of $T(x,x')$ in the exchange of $x$ and $x'$, i.e., $T(x,x') = T(x',x)$, the coefficients $t(x)$ and $t_{a_1 \dots a_p}(x)$ with $p=1,2, \dots $ are constrained. The symmetry of $T(x,x')$ permits us to express the odd coefficients of the covariant Taylor series
expansion of $T(x,x')$ in terms of the even ones. We have for the odd coefficients of lowest orders (see, e.g., Ref.~\cite{Decanini:2005gt})
\begin{subequations}
\label{ScalarTaylorCoef_Sym}
\begin{eqnarray}
\label{ScalarTaylorCoef_Sym_1}
&& t_{a_1}=(1/2) \, t_{;a_1} ,
\\
\label{ScalarTaylorCoef_Sym_2}
&& t_{a_1 a_2 a_3}= (3/2) \, t_{(a_1 a_2 ;a_3)} - (1/4) \, t_{;(a_1 a_2 a_3)} .
\end{eqnarray}
\end{subequations}
Similarly, the series defining the bivectors $V^A_n{}_{\mu \nu'}(x,x')$ and $W^A_n{}_{\mu \nu'}(x,x')$ can be written in the form

\begin{eqnarray}
\label{CovTaylorSeries_Vector}
&& T_{\mu \nu}(x,x') = g^{\phantom{\nu} \nu' }_{\nu}(x,x') \, T_{\mu \nu'}(x,x')
\nonumber \\ && \quad
= t_{\mu \nu}(x) - t_{\mu \nu a_1}(x) \, \sigma^{;a_1}(x,x')
\nonumber \\ && \qquad
+ \frac{1}{2!} \, t_{\mu \nu a_1 a_2}(x) \, \sigma^{;a_1}(x,x') \, \sigma^{;a_2}(x,x')
\nonumber \\ && \qquad
- \frac{1}{3!} \, t_{\mu \nu a_1 a_2 a_3}(x) \, \sigma^{;a_1}(x,x') \, \sigma^{;a_2}(x,x') \, \sigma^{;a_3}(x,x')
\nonumber \\ && \qquad
+ \cdots .
\end{eqnarray}

\noindent By construction, the coefficients $t_{\mu \nu a_1 \dots a_p}(x)$ are symmetric in the exchange of indices $a_1 \dots a_p$, i.e., $t_{\mu \nu a_1 \dots a_p}(x) = t_{\mu \nu (a_1 \dots a_p)}(x)$, and by requiring the symmetry of $T_{\mu \nu'}(x,x')$ in the exchange of $x$ and $x'$, i.e., $T_{\mu \nu'}(x,x') = T_{\nu' \mu}(x',x)$, the coefficients $t_{\mu \nu }(x)$ and $t_{\mu \nu a_1 \dots a_p}(x)$ with $p=1,2, \dots$ are constrained. The symmetry of $T_{\mu \nu'}(x,x')$ permits us to express the coefficients of the covariant Taylor series expansion of $T_{\mu \nu'}(x,x')$ in terms of their symmetric and antisymmetric parts in $\mu$ and $\nu$. We have for the coefficients of lowest orders (see, e.g., Refs.~\cite{Brown:1986tj,Folacci:1990eb})
\begin{subequations}
\label{VectorTaylorCoef_Sym}
\begin{eqnarray}
\label{VectorTaylorCoef_Sym_1}
&& t_{\mu \nu} = t_{(\mu \nu)} ,
\\
\label{VectorTaylorCoef_Sym_2}
&& t_{\mu \nu a_1} = (1/2) \, t_{(\mu \nu) ;a_1} + t_{[\mu \nu] a_1} ,
\\
\label{VectorTaylorCoef_Sym_3}
&& t_{\mu \nu a_1 a_2} = t_{(\mu \nu) a_1 a_2} + t_{[\mu \nu] (a_1 ;a_2)} ,
\\
\label{VectorTaylorCoef_Sym_4}
&& t_{\mu \nu a_1 a_2 a_3} = (3/2) \, t_{(\mu \nu) (a_1 a_2 ;a_3)}
\nonumber \\ && \quad
- (1/4) \, t_{(\mu \nu) ;(a_1 a_2 a_3)} + t_{[\mu \nu] a_1 a_2 a_3} .
\end{eqnarray}
\end{subequations}

In order to solve the recursion relations \eqref{Recursion_VAn}, \eqref{Recursion_VA0}, \eqref{Recursion_WAn}, \eqref{Recursion_Vn}, \eqref{Recursion_V0} and \eqref{Recursion_Wn} but also to do most of the calculations in Secs.~\ref{Sec.III} and \ref{Sec.IV} and, in particular, to obtain the explicit expression of the renormalized stress-energy-tensor operator, it is necessary to have at our disposal the covariant Taylor series expansions of the biscalars $\Delta^{1/2}$, $\Delta^{-1/2} {\Delta^{1/2}}_{;\mu} \sigma^{;\mu}$ and $\Box \Delta^{1/2}$ and of the bivectors $\sigma_{;\mu \nu'}$ and $\Box g_{\mu \nu'}$ but also of some bitensors such as $\sigma_{;\mu \nu}$, $g_{\mu \nu' ;\rho}$ and $g_{\mu \nu' ;\rho'}$. Here, we provide these expansions up to the orders necessary in this article (for higher orders, see Refs.~\cite{Decanini:2005gt,Ottewill:2009uj}). We have
\begin{widetext}
\begin{eqnarray}
\label{CovTaylorSeries_1}
&& \Delta^{1/2} = 1 + \frac{1}{12} \, R_{a_1 a_2} \sigma^{;a_1} \sigma^{;a_2}
- \frac{1}{24} \, R_{a_1 a_2 ;a_3} \sigma^{;a_1} \sigma^{;a_2} \sigma^{;a_3}
\nonumber \\ && \quad
+ \left[
  \frac{1}{80} \, R_{a_1 a_2 ;a_3 a_4}
+ \frac{1}{360} \, R^{p}_{\phantom{p} a_1 q a_2} R^{q}_{\phantom{q} a_3 p a_4}
+ \frac{1}{288} \, R_{a_1 a_2} R_{a_3 a_4}
\right] \sigma^{;a_1} \sigma^{;a_2} \sigma^{;a_3} \sigma^{;a_4}
\nonumber \\ && \quad
- \left[
  \frac{1}{360} \, R_{a_1 a_2 ;a_3 a_4 a_5}
+ \frac{1}{360} \, R^{p}_{\phantom{p} a_1 q a_2} R^{q}_{\phantom{q} a_3 p a_4 ;a_5}
+ \frac{1}{288} \, R_{a_1 a_2} R_{a_3 a_4 ;a_5}
\right] \sigma^{;a_1} \sigma^{;a_2} \sigma^{;a_3} \sigma^{;a_4} \sigma^{;a_5}
+ O( \sigma^{3} ) , \nonumber \\ &&
\end{eqnarray}
\begin{eqnarray}
\label{CovTaylorSeries_2}
&& \Box \Delta^{1/2} = \frac{1}{6} \, R
\nonumber \\ && \quad
+ \left[
  \frac{1}{40} \, \Box R_{a_1 a_2}
- \frac{1}{120} \, R_{;a_1 a_2}
+ \frac{1}{72} \, R R_{a_1 a_2}
- \frac{1}{30} \, R^{p}_{\phantom{p} a_1} R_{p a_2}
+ \frac{1}{60} \,  R^{p q} R_{p a_1 q a_2}
+ \frac{1}{60} \,  R^{p q r}_{\phantom{p q r} a_1} R_{p q r a_2}
\right] \sigma^{;a_1} \sigma^{;a_2}
\nonumber \\ && \quad
- \left[
- \frac{1}{360} \, R _{;a_1 a_2 a_3}
+ \frac{1}{120} \, ( \Box R_{a_1 a_2} )_{;a_3}
+ \frac{1}{144} \, R R_{ a_1 a_2 ;a_3}
- \frac{1}{45} \, R^{p}_{\phantom{p} a_1} R_{p a_2 ;a_3}
+ \frac{1}{180} \, R^{p}_{\phantom{p} q ;a_1} R^{q}_{\phantom{q} a_2 p a_3}
\right. \nonumber \\ && \left. \quad \qquad
+ \frac{1}{180} \, R^{p}_{\phantom{p} q} R^{q}_{\phantom{q} a_1 p a_2 ;a_3}
+ \frac{1}{90} \, R^{p q r}_{\phantom{p q r} a_1} R_{p q r a_2 ;a_3}
\right] \sigma^{;a_1} \sigma^{;a_2} \sigma^{;a_3}
+ O( \sigma^{2} ) ,
\end{eqnarray}
\end{widetext}
\begin{eqnarray}
\label{CovTaylorSeries_3}
&& \Delta^{-1/2} {\Delta^{1/2}}_{;\mu} \sigma^{;\mu} =
  \frac{1}{6} \, R_{a_1 a_2} \sigma^{;a_1} \sigma^{;a_2}
+ O( \sigma^{3/2} ) ,
\nonumber \\
\end{eqnarray}
\begin{eqnarray}
\label{CovTaylorSeries_4}
&& \sigma_{;\mu \nu} = g_{\mu \nu}
- \frac{1}{3} \, R_{\mu a_1 \nu a_2} \sigma^{;a_1} \sigma^{;a_2}
+ O( \sigma^{3/2} ) ,
\end{eqnarray}
\begin{eqnarray}
\label{CovTaylorSeries_5}
&& g_{\nu}^{\phantom{\nu} \nu'} \sigma_{;\mu \nu'} = - g_{\mu \nu}
- \frac{1}{6} \, R_{\mu a_1 \nu a_2} \sigma^{;a_1} \sigma^{;a_2}
+ O( \sigma^{3/2} ) ,
\nonumber \\
\end{eqnarray}
\begin{eqnarray}
\label{CovTaylorSeries_6}
&& g_{\nu}^{\phantom{\nu} \nu'} g_{\mu \nu';\rho} =
- \frac{1}{2} \, R_{\mu \nu \rho a_1} \sigma^{;a_1}
\nonumber \\ && \quad
+ \frac{1}{6} \, R_{\mu \nu \rho a_1 ;a_2} \sigma^{;a_1} \sigma^{;a_2}
+ O( \sigma^{3/2} ) ,
\end{eqnarray}
\begin{eqnarray}
\label{CovTaylorSeries_7}
&& g_{\nu}^{\phantom{\nu} \nu'} g_{\rho}^{\phantom{\rho} \rho'} g_{\mu \nu';\rho'} =
- \frac{1}{2} \, R_{\mu \nu \rho a_1} \sigma^{;a_1}
\nonumber \\ && \quad
+ \frac{1}{3} \, R_{\mu \nu \rho a_1 ;a_2} \sigma^{;a_1} \sigma^{;a_2}
+ O( \sigma^{3/2} )
\end{eqnarray}
and
\begin{eqnarray}
\label{CovTaylorSeries_8}
&& g_{\nu}^{\phantom{\nu} \nu'} \Box g_{\mu \nu'} =
\frac{2}{3} \, R_{a_1 [\mu ;\nu]} \, \sigma^{a_1}
+ \left[
- \frac{1}{6} \, R_{a_1 [\mu ;\nu] a_2}
\right. \nonumber \\ && \left. \quad \qquad
+ \frac{1}{6} \, R_{\mu \nu p a_1} R_{\phantom{p} a_2}^{p}
- \frac{1}{4} \, R_{\mu p q a_1} R_{\nu \phantom{p} \phantom{q} a_2}^{\phantom{\nu} p q}
\right] \sigma^{;a_1} \sigma^{;a_2}
\nonumber \\ && \quad
\vphantom{+} + O( \sigma^{3/2} ) .
\end{eqnarray}

\bibliography{Stueckelberg_SET_regH}

\end{document}